%% file: bpic13.tex
\documentclass[emulateapj,natbib]{emulateapj}

\begin{document}
\title{A Combined VLT and Gemini Study of the Atmosphere of the Directly-Imaged Planet, $\beta$ Pictoris \lowercase{b}}
\author{
Thayne Currie\altaffilmark{1}, 
Adam Burrows\altaffilmark{2},
Nikku Madhusudhan\altaffilmark{3},
 Misato Fukagawa\altaffilmark{4},  
Julien H. Girard\altaffilmark{5},
Rebekah Dawson\altaffilmark{6},
Ruth Murray-Clay\altaffilmark{6},
Scott Kenyon\altaffilmark{6},
Marc Kuchner\altaffilmark{7}, 
Soko Matsumura\altaffilmark{8}, 
Ray Jayawardhana\altaffilmark{1},
John Chambers\altaffilmark{9},
Ben Bromley\altaffilmark{10}
}
\altaffiltext{1}{Department of Astronomy and Astrophysics, University of Toronto}
\altaffiltext{2}{Department of Astrophysical Sciences, Princeton University}
\altaffiltext{3}{Department of Astronomy, Yale University}
\altaffiltext{4}{Osaka University}
\altaffiltext{5}{European Southern Observatory}
\altaffiltext{6}{Harvard-Smithsonian Center for Astrophysics}
\altaffiltext{7}{NASA-Goddard Space Flight Center}
\altaffiltext{8}{Department of Astronomy, University of Maryland-College Park}
\altaffiltext{9}{Department of Terrestrial Magnetism, Carnegie Institution of Washington}
\altaffiltext{10}{Department of Physics, University of Utah}
\begin{abstract}
We analyze new/archival VLT/NaCo and Gemini/NICI high-contrast imaging of the
young, self-luminous planet $\beta$ Pictoris b in seven near-to-mid IR 
photometric filters, using advanced image processing methods 
to achieve high signal-to-noise, high precision measurements.
While $\beta$ Pic b's near-IR colors mimick that of a standard, cloudy early-to-mid L dwarf, 
it is overluminous in the mid-infrared compared to the 
field L/T dwarf sequence.  Few substellar/planet-mass objects -- i.e. $\kappa$ And b and 
1RXJ 1609B -- match $\beta$ Pic b's $JHK_{s}L^\prime$ photometry, and
its 3.1 $\mu m$ and 5 $\mu m$ photometry are particularly difficult to reproduce.  Atmosphere models adopting cloud 
prescriptions and large ($\sim$ 60 $\mu m$) dust grains fail to reproduce the $\beta$ Pic b spectrum. 
However, models incorporating thick clouds similar to those found for HR 8799 bcde but also with small (a few microns)
modal particle sizes yield fits consistent with the data within uncertainties.
Assuming solar abundance models, thick clouds, and small dust particles ($<a>$ = 4 $\mu m$)
we derive atmosphere parameters of log(g) = 3.8 $\pm$ 0.2 and $T_{eff}$ = 1575--1650 $K$,  
an inferred mass of 7$^{+4}_{-3}$ $M_{J}$, and a luminosity of log(L/L$_{\odot}$) $\sim$ -3.80 $\pm$ 0.02.  
The best-estimated planet radius, $\approx$ 1.65 $\pm$ 0.06 $R_{J}$, is 
near the upper end of allowable planet radii for hot-start models given the host star's age and 
likely reflects challenges with constructing accurate atmospheric models.  Alternatively, 
these radii are comfortably consistent with hot-start model predictions if $\beta$ 
Pic b is younger than $\approx$ 7 Myr, consistent with a late formation, 
well after its host star's birth $\sim$ 12$^{+8}_{-4}$ Myr ago.

\end{abstract}
\section{Introduction}
The method of detecting extrasolar planets by 
\textit{direct imaging}, even in its current early stage, fills in an important gap in our knowledge of the 
diversity of planetary systems around nearby stars.  Direct imaging searches with the best 
conventional AO systems (e.g. Keck/NIRC2, VLT/NaCo, Subaru/HiCIAO) are sensitive to very massive planets 
($M$ $\gtrsim$ 5--10 $M_{J}$) at wide separation ($a$ $\sim$ 10-30 $AU$ 
to 100 $AU$) and young ages ($t$ $\lesssim$ 100 Myr), which are not detectable by the radial 
velocity and transit methods \citep[e.g.][]{Lafreniere2007b,Vigan2012,Rameau2013,Galicher2013}.
Planets with these masses and orbital separations pose a stiff challenge to planet formation theories 
 \citep[e.g.][]{Kratter2010,Rafikov2011}.
Young self-luminous directly-imageable planets provide a critical 
probe of planet atmospheric evolution \citep{Fortney2008,Currie2011a,Spiegel2012,Konopacky2013}.

The directly-imaged planet around the nearby star $\beta$ Pictoris ($\beta$ Pictoris b) is a particularly 
clear, crucial test for understanding the formation and atmospheric evolution of gas giant planets 
\citep{Lagrange2009,Lagrange2010}. 
At 12$^{+8}_{-4}$ Myr old \citep{Zuckerman2001}, the $\beta$ Pictoris system provides a way to probe
 planet atmospheric properties only $\approx$ 5--10 Myr after 
the disks from which planets form dissipate \citep[$\approx$ 3--10 Myr, e.g.][]{Pascucci2006,Currie2009}.  
Similar to the case for the HR 8799 planets \citep{Marois2011,Fabrycky2010,Currie2011a,SudolHaghighipour2012}, 
$\beta$ Pic b's mass can be constrained without 
depending on highly-uncertain planet cooling models: in this case, RV-derived 
dynamical mass upper limits when coupled with the range of plausible orbits ($a$ $\sim$ 8--10 AU)
 imply masses less than $\sim$ 10--15 $M_{J}$ \citep{Lagrange2012a,Currie2011b,Chauvin2012,Bonnefoy2013}, 
a mass range consistent with estimates derived from the planet's interaction with the secondary 
disk \citep{Lagrange2009,Dawson2011}.

Furthermore, while other likely/candidate planets such as Fomalhaut b and LkCa 15 b are probably made detectable by
circumplanetary emission in some poorly constrained geometery \citep{Currie2012a,Kraus2012}, $\beta$ Pic b's emission 
appears to be consistent with that from a self-luminous planet's atmosphere \citep{Currie2011b,Bonnefoy2013}.
Other objects of comparable mass appear to have formed more like low-mass binary companions.
Thus, combined with the planets HR 8799 bcde, $\beta$ Pic b provides a crucial reference point with 
which to interpret the properties of many soon-to-be imaged planets with upcoming extreme AO systems like 
$GPI$, $SCExAO$, and $SPHERE$ \citep{Macintosh2008,Martinache2009,Beuzit2008}.

However, investigations into $\beta$ Pic b's atmosphere are still in an early stage compared to
 those for the atmospheres of the HR 8799 planets and other very low-mass, young substellar 
objects \citep[e.g.][]{Currie2011a,Skemer2011,Konopacky2013,Bailey2013}.  Of the current 
published photometry, only $K_{s}$ (2.18 $\mu m$) and $L^\prime$ (3.78 $\mu m$) have photometric 
errors smaller than $\sim$ 0.1 mag \citep{Bonnefoy2011,Currie2011b}.  Other high SNR detections such 
as at $M^\prime$ were obtained without reliable flux calibration 
\citep{Currie2011b} or with additional, large photometric uncertainties due to processing \citep{Bonnefoy2013}. 
   As a result, the best-fit models admit a wide range of temperatures, surface gravities, and cloud structures 
\citep[e.g.][]{Currie2011b}.  Thus, new higher signal-to-noise/precision and flux-calibrated photometry at 
1--5 $\mu m$ should provide a clearer picture of the clouds, chemistry, temperature, and gravity of 
$\beta$ Pic b.  Moreover, new near-to-mid IR data may identify distinguishing characteristics of $\beta$ Pic b's atmosphere, 
much like clouds and non-equilibrium carbon chemistry for HR 8799 bcde \citep{Currie2011a,Galicher2011,Skemer2012,Konopacky2013}.

In this study, we present new 1.5--5 $\mu m$ observations for $\beta$ Pic b obtained with $NaCo$ on the 
\textit{Very Large Telescope} and $NICI$ on \textit{Gemini-South}.  We extract the first detection at the 3.09 $\mu m$ water-ice filter; the first 
high signal-noise, well calibrated H, [4.05], and $M^\prime$ detections; and higher 
signal-to-noise detections at $K_{s}$ and $L^\prime$ (2.18 and 3.8 $\mu m$).  To our new data, we 
add rereduced $\beta$ Pic data obtained in $J$ (1.25 $\mu m$) and $H$ (1.65 $\mu m$) bands and first presented in 
\citet{Bonnefoy2013}, recovering $\beta$ Pic b at a slightly higher signal-to-noise and deriving 
its photometry with smaller errors.

We compare the colors derived from broadband photometry to that for field substellar objects with a range of spectral types to 
assess whether $\beta$ Pic b's colors appear anomalous/redder than the field sequence like those for planets around HR 8799 
and $\kappa$ And; planet-mass companions like 2M 1207 B, GSC 06214 B, and 1RXJ 1609 B \citep{Chauvin2004,IrelandKraus2011,Lafreniere2008a}; 
and other substellar objects 
like Luhman 16B \citep{Luhman2013}.  We use atmosphere modeling to constrain the range of temperatures, 
surface gravities, and cloud structures plausible for the planet.  While previous studies have shown the importance 
of clouds and non-equilibrium carbon chemistry in fitting the spectra/photometry of directly-imaged planets 
\citep{Bowler2010,Currie2011a,Madhusudhan2011,Galicher2011,Skemer2012,Konopacky2013}, here the assumed sizes of dust 
particles entrained in the clouds plays a critical role.
 
\section{Observations and Data Reduction}
\subsection{VLT/NaCo Data and Basic Processing}
We observed $\beta$ Pictoris under photometric conditions on 14 December to 17 December 2012 with the NAOS-CONICA instrument \citep[NaCo;][]{Rousset2003} 
on the \textit{Very Large Telescope} UT4/Yepun at Paranal Observatory (Program ID 090.C-0396).  
All data were taken in pupil-tracking/angular differential imaging \citep{Marois2006} and data cube mode.  
Table \ref{bpiclog} summarizes the basic properties of these observations.  Our full complement of data during the run includes  
imaging at 1.04 $\mu m$, 2.12 $\mu m$, $K_{s}$/2.18 $\mu m$, 2.32 $\mu m$, 3.74 $\mu m$, $L^\prime$/3.78 $\mu m$, 
Br-$\alpha$/4.05 $\mu m$, and $M^\prime$.  Here, we focus only on the $L^\prime$, [4.05], and $M^\prime$ data, 
deferring the rest to a later study.  Each observation was centered on $\beta$ Pictoris's transit for a total 
field rotation of $\sim$ 50--70 degrees and a total observing times ranging between $\sim$ 30 minutes and 59 minutes.

To these new observations, we rereduce $J$-band and $H$-band data first presented in \citet{Bonnefoy2013} and 
taken on 16 December 2011 and 11 January 2012, respectively.
The saturated $J$ band science images are bracketed by two sequences of unsaturated images obtained in neutral density filter 
for flux calibration.  While there were additional frames taken but not analyzed in \citeauthor{Bonnefoy2013}, we found these 
to be of significantly poorer quality and thus do not consider them here.  In total, the $J$-band data we consider 
covers 40 minutes of integration time and $\sim$ 23$^{o}$ of field rotation.   The $H$-band data cover $\sim$ 92 minutes 
of integration time and $\sim$ 36$^{o}$ of field rotation.
 
Basic NaCo image processing steps were performed as in \citet{Currie2010,Currie2011b}.
The thermal IR data at $L^\prime$ and [4.05] ($M^\prime$) were obtained in a dither pattern with offsets every 2 (1) images to remove 
the sky background.  
As all data were obtained in data cube mode, we increased our PSF quality by realigning each individual exposure in the 
cube to a common center position and clipping out frames with low encircled energy (i.e. those with a 
core/halo ratio $<$ max(core/halo) - 3$\times$$\sigma$(core-to-halo ratio)).

\subsection{Gemini/NICI Data and Basic Processing}
We obtained Gemini imaging for $\beta$ Pic b using the Near-Infrared Coronagraphic Imager (NICI) on 23 December 2012 and 
26 December 2012 in the 
H$_{2}$O filter ($\lambda_{o}$ = 3.09 $\mu m$) and 9 January 2013 in the $H$ and $K_{s}$ filters (dual-channel 
imaging), both under photometric conditions (Program GS-2012B-Q-40).  
These observations were also executed in \textit{angular differential imaging} mode.
For the $H_{2}$$O$ data, we dithered each 38 s exposure for sky subtraction for a total of $\sim$ 38 minutes of integration 
time over a field rotation of $\sim$ 30 degrees.
For the $H/K_{s}$ data, we placed the star behind the $r$ = 0\farcs{}22 partially-transmissive coronagraphic 
mask to suppress the stellar halo.  Here, we took shorter exposures of $\beta$ Pic ($t_{int}$ $\sim$ 11.4 s) to better 
identify and filter out frames with bad AO correction.  Our observing sequence consists of $\sim$ 22 minutes of usable data 
centered on transit with a field rotation of $\sim$ 41 degrees.

Basic image processing follows steps described above for NaCo data.  The PSF halo was saturated 
out to $r$ $\sim$ 0\farcs{}32--0\farcs{}36 in $H$ during most of the observations and our sequence suffered periodic seeing bubbles that 
saturated the halo out to angular separations overlapping with the $\beta$ Pic b PSF.  Thus, we focus on reducing only 
those $H$-band frames with less severe halo saturation ($r_{sat}$ $<$ 0\farcs{}36).  The $K_{s}$ observations, obtained 
at a higher Strehl ratio, never suffered halo saturation.  The first of the two $H_{2}0$ sets, suffered from severe 
periodic seeing bubbles and thus generally poor AO performance.  We identify and remove from 
analysis frames whose halo flux exceeded the $F_{min}$+3$\sigma$, where $F_{min}$ is the minimum flux within an 
aperture covering $\beta$ Pic b and $\sigma$ is the dispersion in this flux: about 10-25\% of the 
frames, depending on the data set in question.  

\subsection{PSF Subtraction}
To remove the noisy stellar halo and reveal $\beta$ Pic b, we process the data with 
our ``adaptive" LOCI (A-LOCI) pipeline \citep[][ T. Currie 2013 in prep.]{Currie2012a,Currie2012b}.
This approach adopts ``locally optimized 
combination of images" (LOCI) formalism \citep{Lafreniere2007}, where we perform PSF subtraction in small annular regions (the ``subtraction zone")
at a time over each image.  
Previously-described A-LOCI components we use here include ``subtraction zone centering" 
\citep{Currie2012b}; ``speckle filtering" to identify and remove images with 
noise structure poorly correlated with that from the science image we are wanting to subtract \citep{Currie2012b}; 
a moving pixel mask to increase point source throughput and normalize it as a function of azimuthal angle \citep{Currie2012a}.
We do not consider a PSF reference library \citep{Currie2012a} since $\beta$ Pictoris is our only target.

Into A-LOCI as recently utilized in \citet{Currie2012a}, we incorporate a component different from but complementary 
to our ``speckle filtering", using \textit{singular value decomposition} (SVD) to limit the number of images used 
in a given annular region (i.e. for a given optimization zone) to construct and subtract a reference.  
Briefly, in the (A-)LOCI formalism a matrix inversion yields the set of coefficients $c^{k}$ applied to each image making up the 
reference ``image":  \textbf{$c^{k}$} = \textbf{A$^{-1}$}\textbf{b}.  Here, \textbf{A} is the covariance matrix 
and \textbf{b} is a column matrix defined from $i$ pixels in the ``optimization zones" of the $j$-th reference image section 
\textit{O$^{j}$} and the science image, 
\textit{O$^{T}$}: \textit{b$_{j}$} = $\sum\limits_{i}$ \textit{O$^{j}_{i}$}\textit{O$^{T}_{i}$} \citep[see][]{Lafreniere2007}.
In the previous versions of our codes, we use a simple double-precision 
matrix inversion to invert the covariance matrix and then solve for $c^{k}$ after multiplying by \textbf{b}.  

In this work, we instead use SVD to rewrite \textbf{A} as \textbf{U$\Sigma$V$^{T}$} such that 
\textbf{A$^{-1}$} = \textbf{V$\Sigma^{-1}$U$^{T}$}, where the \textit{T} superscript 
stands for the transpose of the matrix.   Prior to inversion, we truncate the number 
of singular values at a predefined cutoff, $svd_{lim}$.  This eigenvalue truncation is very 
similar to and functions the same as the truncation of principle components, $N_{pca}$, in the 
Karhunen-Loeve image projection (KLIP) \citep{Soummer2012} and has been 
successfully incorporated before \citep{Marois2010b}. 
We found that both speckle filtering and SVD truncation within our 
formalism can yield significant contrast gains over LOCI and KLIP/Principle Component Analysis (PCA), although 
in this study at the angular separation of $\beta$ Pic b ($\approx$ 0\farcs{}45) the gains over LOCI are typically about a 
factor of 1.5, albeit with substantially higher throughput\footnote{Recently, \citet{Amara2012} claimed a contrast gain of
 $\sim$ 5$\times$ over LOCI using PCA.
However, optimal set-ups even \textit{within} a given formalism like LOCI or PCA/KLIP are very dataset-specific 
\citep[cf.][]{Lafreniere2007,Currie2012a,Currie2012b}.
  With LOCI, we obtained roughly equivalent SNRs for $\beta$ Pic b obtained during 
the same observing run but on a night with poorer observing conditions (29 December 2009) than their test data set \citep{Currie2011b}.  
Implementing some A-LOCI filtering and pixel masking yields SNR $\approx$ 30--35.}.

\subsection{Planet Detections}
\begin{figure}
\centering
\includegraphics[scale=.75,clip]{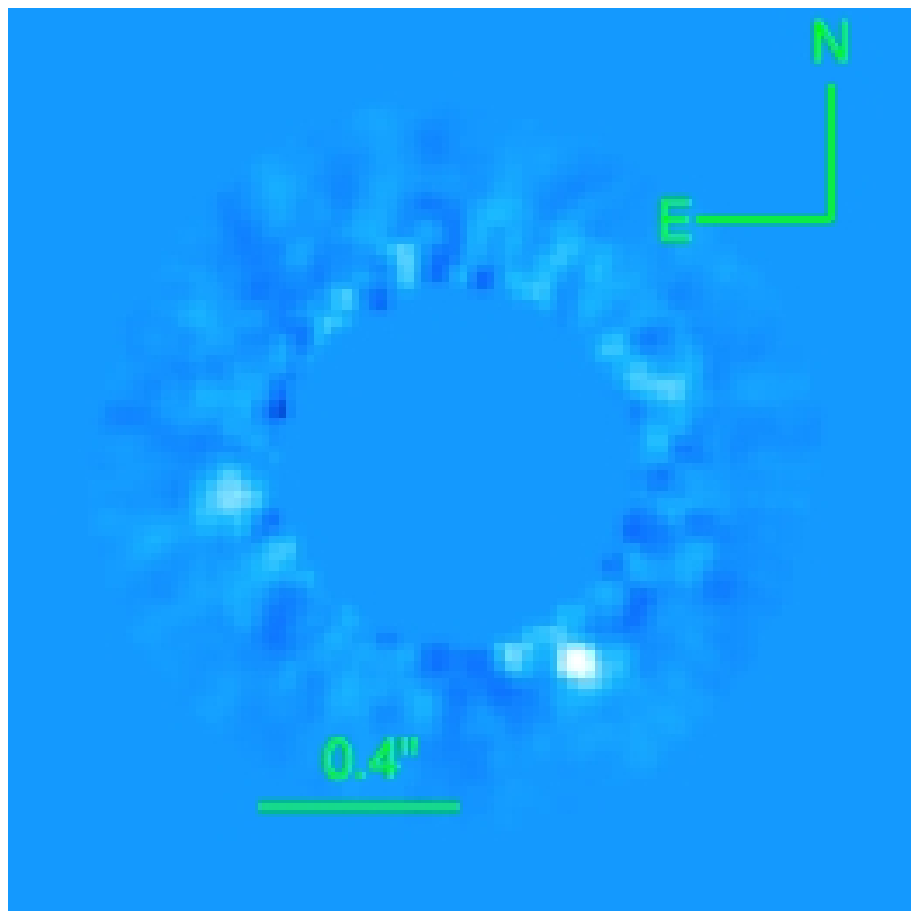}
\includegraphics[scale=.75,clip]{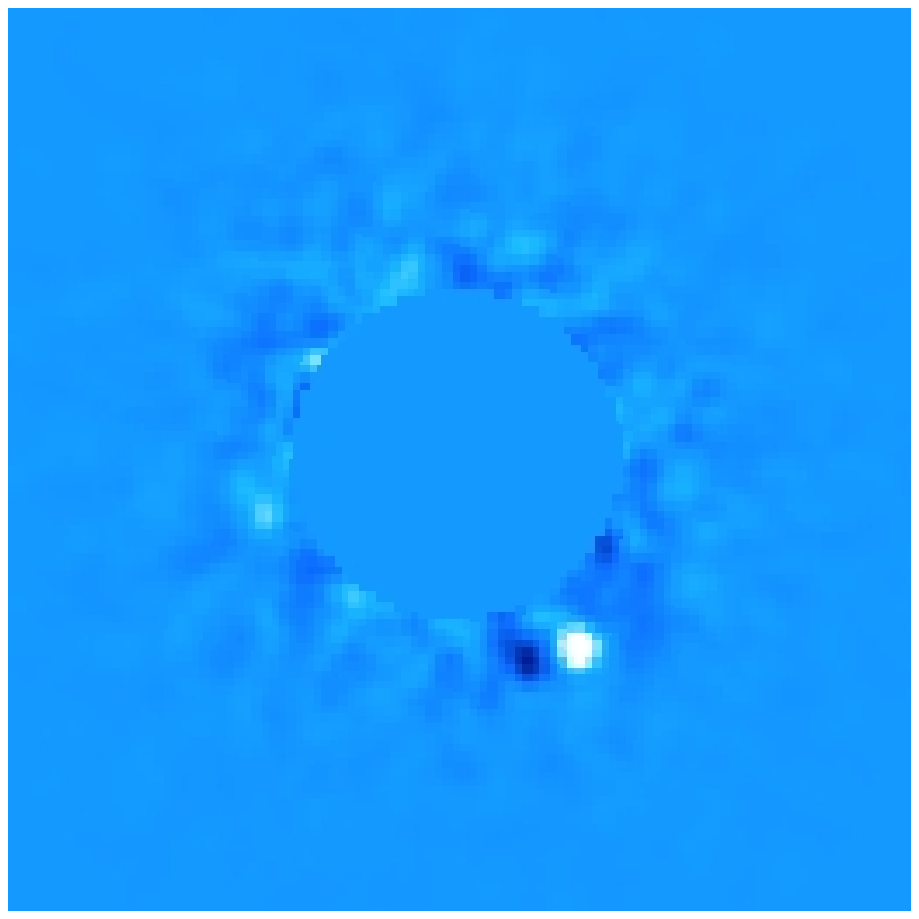}
\\
\includegraphics[scale=.75,trim=0mm 0mm 0mm 0mm,clip]{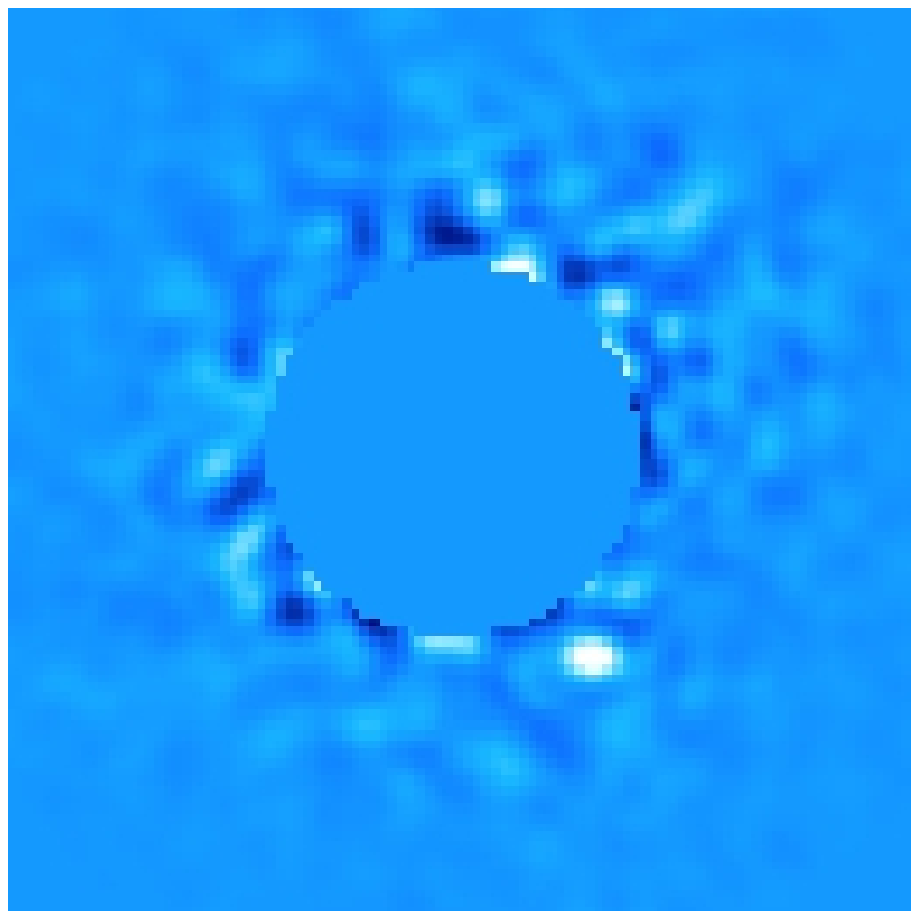}
\includegraphics[scale=.75,trim=0mm 0mm 0mm 0mm,clip]{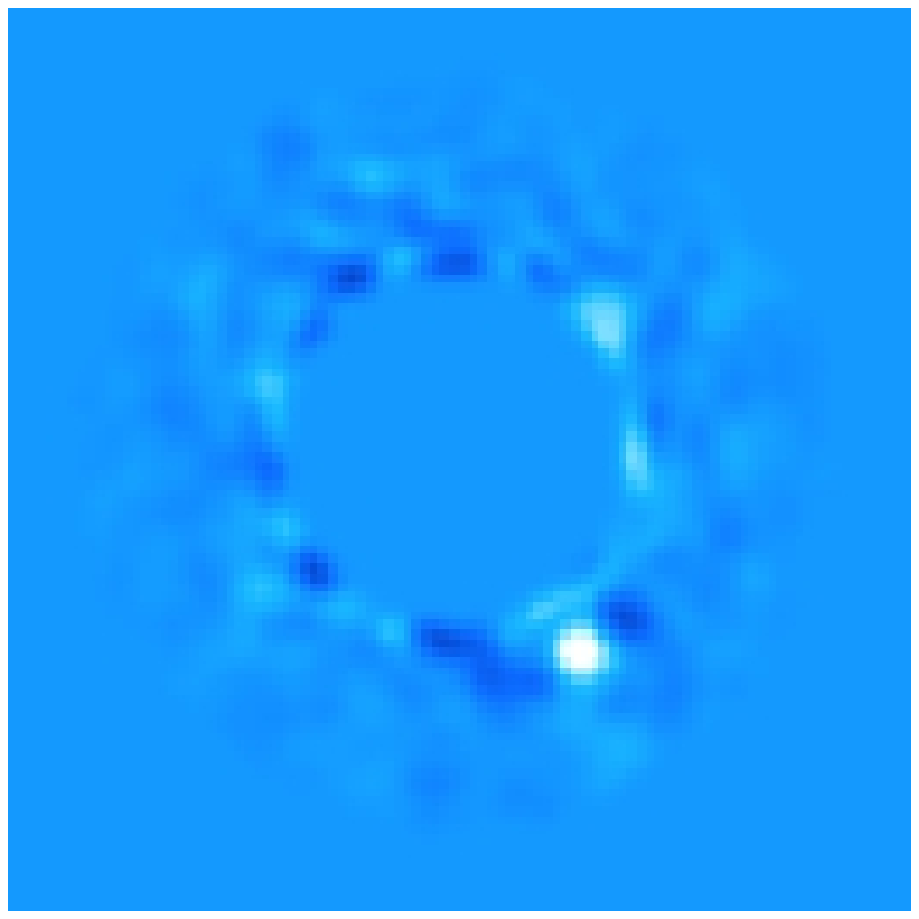}
\caption{Processed Gemini-NICI images obtained at $H$ band (top-left), $K_{s}$ band (top-right), and the 
$H_{2}0$ filter centered on $\sim$ 3.1 $\mu m$ (bottom panels: 23 Dec 2012 data on bottom-left, 26 Dec 2012 data on the bottom 
right).  For clarity, we mask the region interior to $\sim$ 0\farcs{}4 and convolve the image with a gaussian equal 
to the image FWHM.  The planet $\beta$ Pic b is in the lower-right at a position angle of $\sim$ 210$^{o}$ and a 
separation of $\sim$ 0\farcs{}46.  The color scale is set such that over the planet's FWHM the pixel color is white.  }
\label{niciimages}
\end{figure}

\begin{figure}
\centering
\includegraphics[scale=.75,clip]{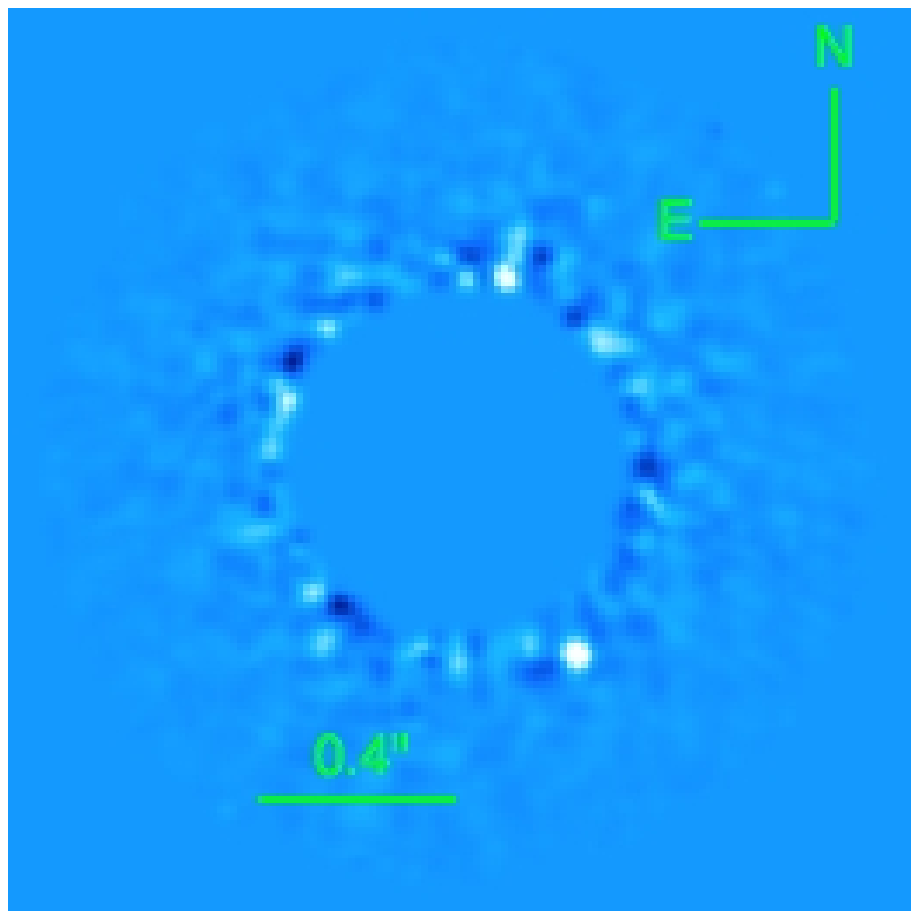}
\includegraphics[scale=.75,clip]{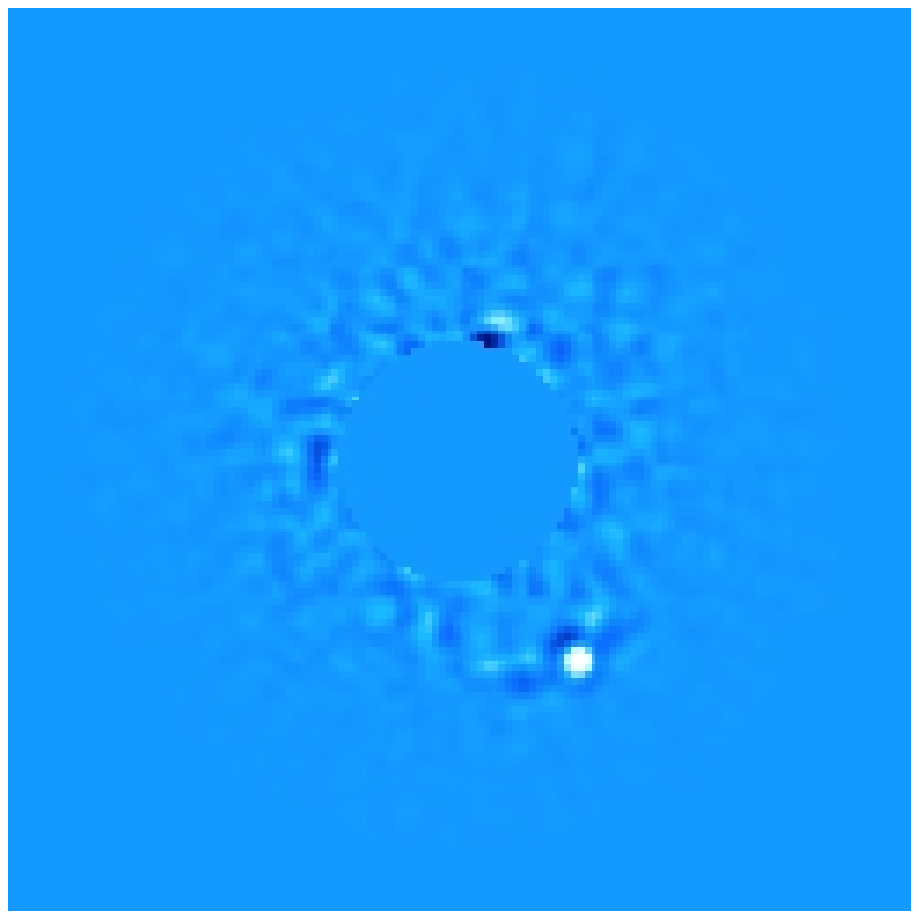}
\caption{Processed VLT/NaCo $J$ (left) and $H$ (right) band images presented in the same manner as Figure \ref{niciimages}.  
Owing to very good speckle suppression, the $H$-band data's effective inner working angle beyond which we 
are sensitive to $\beta$ Pic b-brightness companions is significantly smaller than for the $J$ band data and the preceding 
NICI images ($r_{IWA}$ = 0\farcs{}2).}
\label{jhnacoimages}
\end{figure}

\begin{figure}
\centering
\includegraphics[scale=.55,clip]{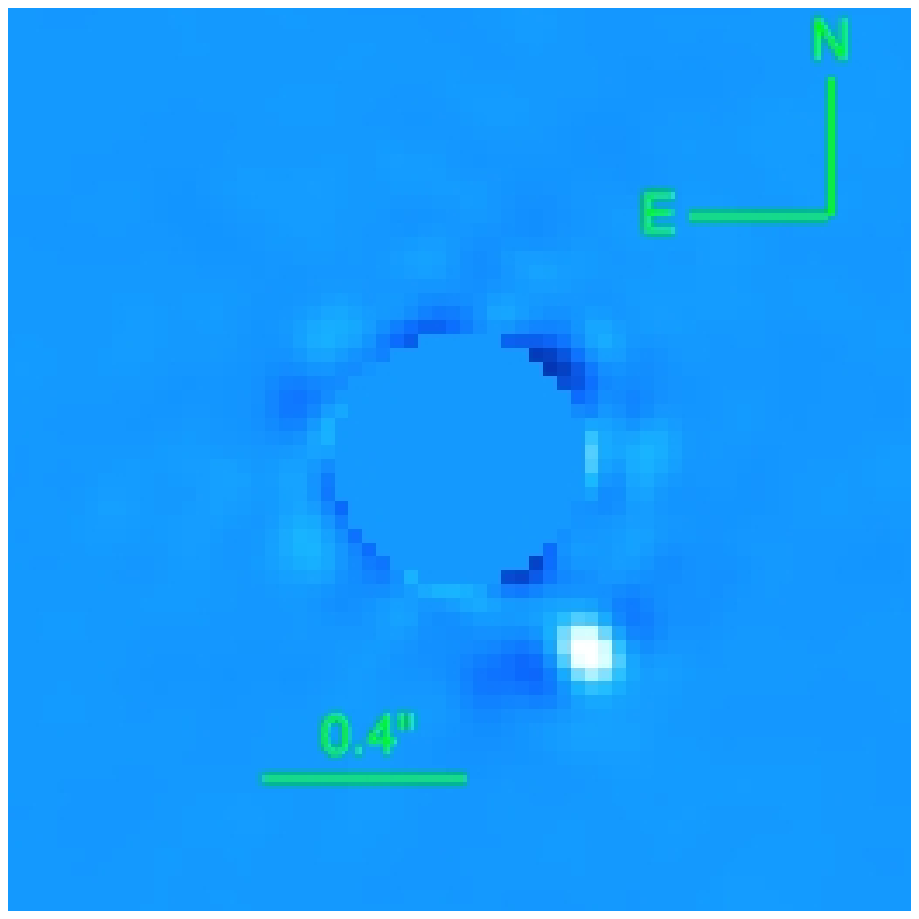}
\includegraphics[scale=.55,clip]{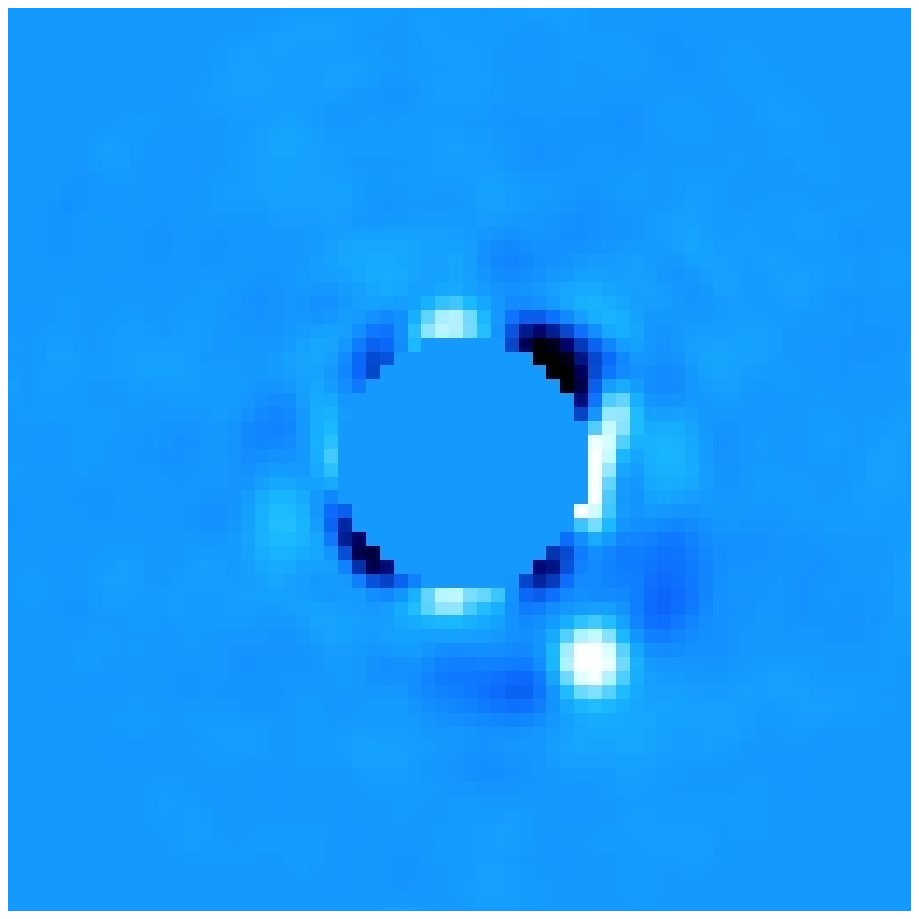}
\includegraphics[scale=.55,clip]{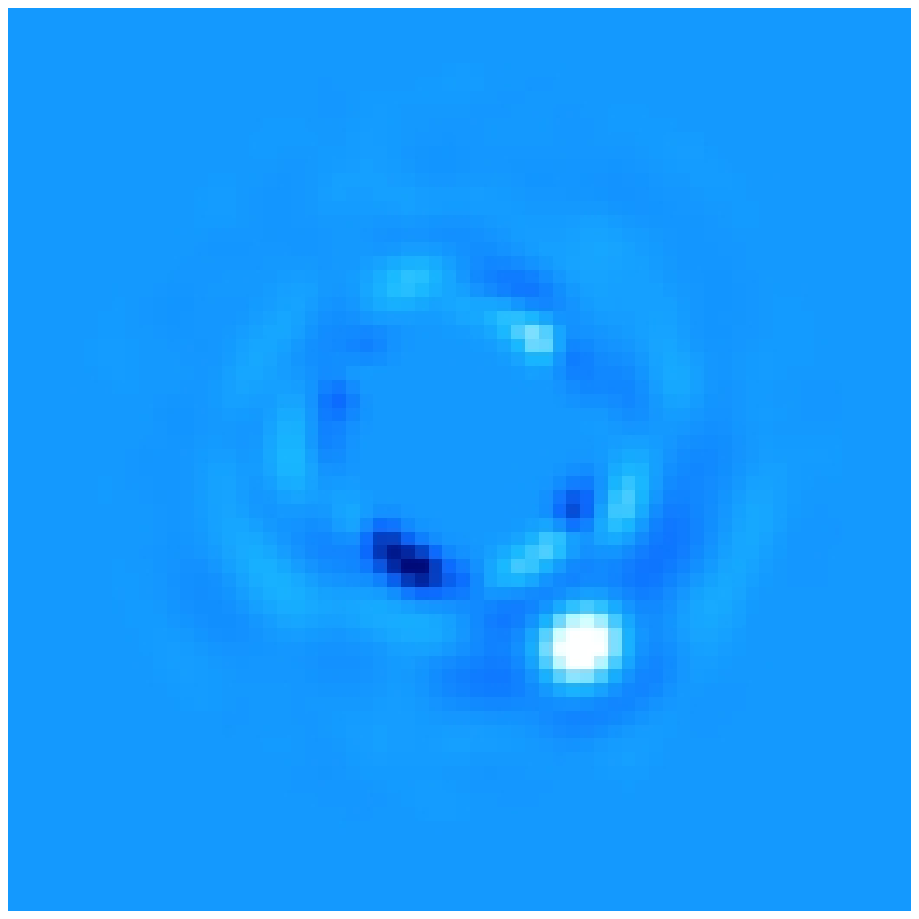}
\\
\caption{Processed VLT/NaCo $L^\prime$ (left), [4.05] (middle), and $M^\prime$ (right) images presented as in Figures \ref{niciimages} 
and \ref{jhnacoimages}.} 
\label{midirnacoimages}
\end{figure}

Figures \ref{niciimages}, \ref{jhnacoimages}, and \ref{midirnacoimages} display reduced NaCo and NICI images 
of $\beta$ Pic.  We detect $\beta$ Pic b in all datasets (summarized in Table \ref{bpicphot}).  To compute the signal-to-noise ratio (SNR) for $\beta$ Pic b, 
we determine the dispersion, $\sigma$, in pixel values of our final image convolved with a gaussian along a ring with width of 
1 FWHM at the same angular separation as $\beta$ Pic b but excluding the planet \citep[e.g.][]{Thalmann2009}, and average the 
SNR/pixel over the aperture area.  For the Gemini-NICI $H$, $K_{s}$, and two [3.1] datasets, the SNRs are thus 6.4, 11, 4.6, and 10, respectively.
  For the $J$ and $H$-band NaCo data 
previously presented in \citet{Bonnefoy2013}, we achieve SNR $\sim$ 9 and SNR $\sim$ 30, respectively.  Generally speaking, our 3.8--5 $\mu m$ 
NaCo data are deeper than the near-IR NaCo and especially near-IR NICI data, where we detect $\beta$ Pic b at SNR = 40 
in $L^\prime$ and 22 at $M^\prime$, roughly a factor of two higher than previously reported \citep{Currie2011b,Bonnefoy2013}, 
gains due to $\beta$ Pic b now being at a wider projected separation ($L^\prime$) or post-processing and slightly better observing conditions 
($M^\prime$).  The high SNR detections obtained with NaCo also leverage on recent engineering upgrades that substantially 
improved the instrument's image quality and the stability of its PSF \citep{Girard2012}.

The optimal A-LOCI algorithm parameters vary significantly from dataset to dataset.  The rotation gap ($\Delta$PA in units of the image 
full-width half maximum) criterion used to produce most of the images is $\delta$ $\sim$ 0.6--0.65, although it is 
significantly larger for the $J$ and $H$ data sets ($\delta$ = 0.75--0.95).   Generally speaking, the optimization areas we use $N_{A}$ are 
significantly smaller ($N_{A}$ = 50-150) than those typically adopted \citep[i.e. $N_{A}$=300;][]{Lafreniere2007}.  We speculate that 
the pixel masking component of A-LOCI drives the optimal $N_{A}$ settings toward these smaller values since the planet flux 
(ostensibly within the subtraction zone) no longer significantly biases the coefficient determinations to the point of reducing 
the planet's SNR.  Filtering parameters 
$r_{corr}$ and $svd_{lim}$ likewise vary wildly from $r_{corr}$ = 0 and $svd_{lim}$ = 2.5$\times$10$^{-7}$ at $J$ to $r_{corr}$ = 0.9 for the NICI 
$H$-band data or $svd_{lim}$ = 2.5$\times$10$^{-2}$ for the $M^\prime$ NaCo data.  

While the many algorithm free parameters make finding an optimal 
combination difficult and computationally expensive, our final image quality is nevertheless \textit{extremely} sensitive to some 
values, in particular $svd_{lim}$ and $r_{corr}$.  As a test, we explored other image processing methods -- ADI-based 
classical PSF subtraction and LOCI.  While A-LOCI always yields deeper contrasts, we easily detect $\beta$ Pic b in the mid-IR NaCo data 
using any method and only the poorer of the two [3.1] data sets requires A-LOCI to yield a better than 4-$\sigma$ detection (i.e. where $\sigma_{det}$ = 
1.0857/SNR = 0.27 mags).
We will present a detailed analysis of image processing methods and algorithm parameters in an upcoming study (T. Currie, 2013 in prep.).

\begin{figure}
\centering
\includegraphics[scale=0.6,trim=0mm 0mm 0mm 0mm,clip]{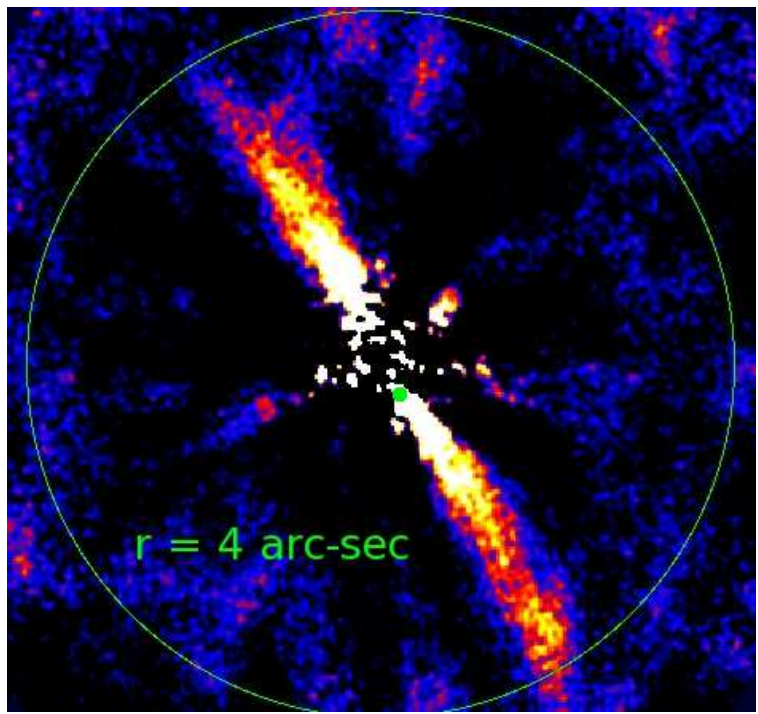}
\caption{Classical ADI reduction of our $L^\prime$ data showing a clear detection of the $\beta$ Pic debris disk.  
The green dot denotes the position of $\beta$ Pic b in the disk.}
\label{diskimage}
\end{figure}

Adopting the pixel scales listed in Table \ref{bpiclog}, $\beta$ Pic b is detected at an angular separation of $r$ $\sim$ 0\farcs{}46 in each data set.
The position angle of $\beta$ Pic b is consistent with previously-listed values (PA $\approx$ 210$^{o}$) and in between
 values for the main disk and the warp, intermediate between the results presented in \citet{Currie2011b} and \citet{Lagrange2012b}.  
While the NICI north position angle on the detector is precisely known and determined from facilty observations, 
we have not yet used our astrometric standard observations to derive the NaCo position angle offset,
which changes every time NaCo is removed from the telescope.
To dissuade others from using the poorly calibrated NaCo data and precisely calibrated data \citep{Lagrange2012b} together, 
we reserve a detailed determination of $\beta$ Pic b's astrometry and a study of its orbit for a future study. 
 We also detect the $\beta$ Pic debris disk in each new broadband data set and at [4.05] (Figure \ref{diskimage}).  
We will analyze its properties at a later time as well.

\subsection{Planet Photometry}
To derive $\beta$ Pic b photometry, we first measured its brightness within an aperture roughly equal to the image FWHM in each case, 
which was known since we either had AO-corrected standard star observations (NICI $H$, $K_{s}$, and [3.1]), unsaturated 
images of the primary as seen through the coronagraphic mask (NICI $K_{s}$), unsaturated neutral density filter observations (NaCo $J$, $H$, 
$L^\prime$, and $M^\prime$), or unsaturated images of the primary (NaCo $L^\prime$ and [4.05]).  We assessed and corrected for 
planet throughput losses due to processing by comparing the flux of synthetic point sources within this aperture 
implanted into registered images at the same angular separation as
$\beta$ Pic b before and after processing.  To derive $\beta$ Pic b's throughput and uncertainty in the throughput ($\sigma_{atten}$), we 
repeat these measurements at 15 different position angles and adopt the clipped mean of the throughput as our throughput
and standard deviation of this mean as its uncertainty.  The planet throughput ranges from 0.38 for the $J$-band data to 0.82 for the [4.05] 
data and 0.96 for the NICI $H$-band data, even with aggressive algorithm parameters (i.e. $\delta$ $\sim$ 0.6), due to the throughput gains yielded 
by our pixel masking and the SVD cutoff.  

For photometric calibration, we followed several different approaches.  For the NICI data, we used TYC 7594-1689-1 and HD 38921 as photometric standards.  
We were only able to obtain photometric calibrations for the first of the two [3.1] datasets.
For all other data we used the primary star, $\beta$ Pic, for flux calibration adopting the measurements listed in \citet{Bonnefoy2013}.  
For the $J$ and $H$ NaCo data, we used images of the primary as viewed through the neutral density filter.  For the $M^\prime$ 
NaCo data, we obtained neutral density filter observations \textit{and} very short exposures.  While the latter were close to 
saturation and were probably in the non-linear regime, the implied photometry for $\beta$ Pic was consistent to within errors.  The primary was unsaturated 
in the [4.05].  Finally, for the $L^\prime$ data, we took 8.372 ms unsaturated images of $\beta$ Pic for flux calibration.  In all cases, 
we again adopt the clipped mean of individual measurements as our photometric calibration uncertainty, $\sigma_{fluxcal}$.  To compute the 
photometric uncertainty for each data set, we considered the SNR of our detection, the uncertainty in the planet throughput, and the uncertainty 
in absolute flux calibration: $\sigma$ = $\sqrt{\sigma_{det}^{2}+\sigma_{atten}^{2}+\sigma_{fluxcal}^{2}}$.

Table \ref{bpicphot} reports our photometry and Table \ref{bpicphoterror} lists sample error budgets for two NICI photometric measurements and 
two NaCo measurements.  The relative contributions from each source of photometric uncertainty to the total uncertainty are 
representative of our combined data set.  For the [3.09] data, residual speckle noise/sky fluctuations greatly 
limit the planet’s SNR and thus $\sigma_{det}$ is the primary source of photometric uncertainty.  
For the $K_{s}$ data, the intrinsic SNR and the two other sources of photometric uncertainty contribute in 
a more equal proportion.   The $L^\prime$ and $M^\prime$ data error budgets are characteristic of most of 
our other data, where the photometric uncertainty is primarily due to the absolute photometric calibration 
and throughput.   With the exception of the [3.09] NICI data, the intrinsic SNR of the detection does not 
dominate the error budget.   
For the best-quality (mid-IR NaCo) data, the throughput uncertainty was small ($\approx$ 5\%) and was 
never any larger than 15\% ($J$ band data) in any data set\footnote{In principle, tuning the algorithm parameters to maximize the SNR of $\beta$ Pic b could 
introduce additional photometric uncertainties if the planet is in significant residual speckle contamination.  In such a case, the 
algorithm parameters maximizing the SNR could instead be the set that maximizes the residual speckle contamination within the
the planet aperture while minimizing it elsewhere, especially as the pixel masking technique normalizes the point source throughput 
but not the noise as a function of azimuthal angle.  However, we do not find substantial differences in 
the derived photometry if we adopt a default set of algorithm parameters.   Furthermore, the parameters maximizing the SNR are 
never the ones maximizing the planet throughput, and our tuning is not just finding the parameter set 
making pixels within the planet aperture 'noisiest'.  Adopting slightly different parameters from the 'optimized' case yields nearly identical photometry.  
Moreover, residual speckle contamination in most data sets is extremely low, and for the mid-IR data the intrinsic SNR is limited 
by sky background fluctuations in addition to speckles.}.  

In general, we find fair agreement with previously published photometry, where 
our measurements are usually consistent within photometric errors with those reported previously (e.g. $m_{H}$= 13.32 $\pm$ 0.14 and 
13.25 $\pm$ 0.18 vs. 13.5 $\pm$ 0.2 in Bonnefoy et al. 2013).
Our $L^\prime$ photometry is more consistent with \citeauthor{Currie2011b}'s measurement of $m_{L^\prime}$=9.73 $\pm$ 0.06
than with that listed in \citet{Bonnefoy2013} ($m_{L^\prime}$=9.5 $\pm$ 0.2), though it is nearly identical to that derived 
for some $\beta$ Pic b data sets listed in \citet{Lagrange2010}.  Our [4.05] photometry implies that $\beta$ Pic b 
is $\sim$ 15-20\% brighter there than previously assumed \citep{Quanz2010} and may have a slightly red $L^\prime$-[4.05] color.
The major difference from previous studies, though, is that our photometric errors are consistently much smaller.  For 
example, the uncertainty in the [4.05] photometry is reduced to 0.08 mag from 0.23 mag due both to higher SNR detections 
and lower uncertainty in our derived photometry (e.g. throughput corrections).
NICI photometry is also substantially less uncertain than in \citet{Boccaletti2013} because $\beta$ Pic b is not occulted by the focal plane mask.
These lower uncertainties should allow more robust comparisons between $\beta$ Pic b and other substellar objects and, from 
modeling, more precise limits on the best-fitting planet atmosphere properties.

\section{Empirical Comparisons to $\beta$ Pic b}
Our new data allows us to compare the spectral energy distribution of $\beta$ Pic b to 
that for the many field L/T-type brown dwarfs as well that for directly-imaged low-surface gravity, low-mass 
brown dwarf companions and directly-imaged planets.  Our goal here is to place 
$\beta$ Pic b within the general L/T type spectral sequence, identify departures from this sequence 
such as those seen for low surface gravity objects like HR 8799 bcde, and identify the substellar 
object(s) with the best-matched SED.  Some bona fide directly-imaged planets like 
HR 8799 bcde and at least some of the lowest-mass brown dwarfs like 2M 1207 B appear redder/cloudier 
than their field dwarf counterparts at comparable temperatures ($T_{eff}$ $\approx$ 900-1100 $K$).  
However, it is unclear whether hotter imaged exoplanets appear different from their (already cloudy) field 
L dwarf counterparts, and $\beta$ Pic b provides a test of any such differences.  We will use 
our comparisons to the L/T dwarf sequence and the SEDs of other substellar objects to inform our atmosphere model 
comparisons later to derive planet physical parameters (e.g. $T_{eff}$ and log(g)).

\subsection{Infrared Colors of $\beta$ Pic b}
To compare the near-to-mid IR properties of $\beta$ Pic b with those
for other cool, substellar objects, we primarily use the sample of L/T dwarfs compiled by
\citet{Leggett2010}, which include field dwarfs spectral classes between $\sim$ M7 and T5, 
corresponding to a range of temperatures between $\sim$ 2500 $K$ and 700 $K$.
To explore how the $\beta$ Pic b SED compares to those with other directly-imaged planets/planet candidates 
and very low-mass brown dwarf companions within this temperature range, we include objects listed in 
Table \ref{photcomptable}.  These include the directly-imaged planets around 
HR 8799 \citep{Marois2008,Marois2011,Currie2011a} and the directly-imaged planet candidate around 
$\kappa$ And \citep{Carson2013}.
Additionally, we include high mass ratio brown dwarf companions with masses less than the 
deuterium-burning limit ($\sim$ 13--14 $M_{J}$) and higher-mass companions whose
youth likely favors a lower surface gravity than for field brown dwarfs, a difference that affect the 
objects' spectra \citep[e.g.][]{Luhman2007}.  Among these objects are 
1RXJ 1609B, AB Pic B, and Luhman 16 B \citep{Lafreniere2008a,Chauvin2005,Luhman2013}.
Table \ref{photcomptable2} compiles photometry for all of these low surface gravity objects.

\begin{figure}
\centering
\includegraphics[scale=0.5,trim = 8mm 0mm 8mm 0mm,clip]{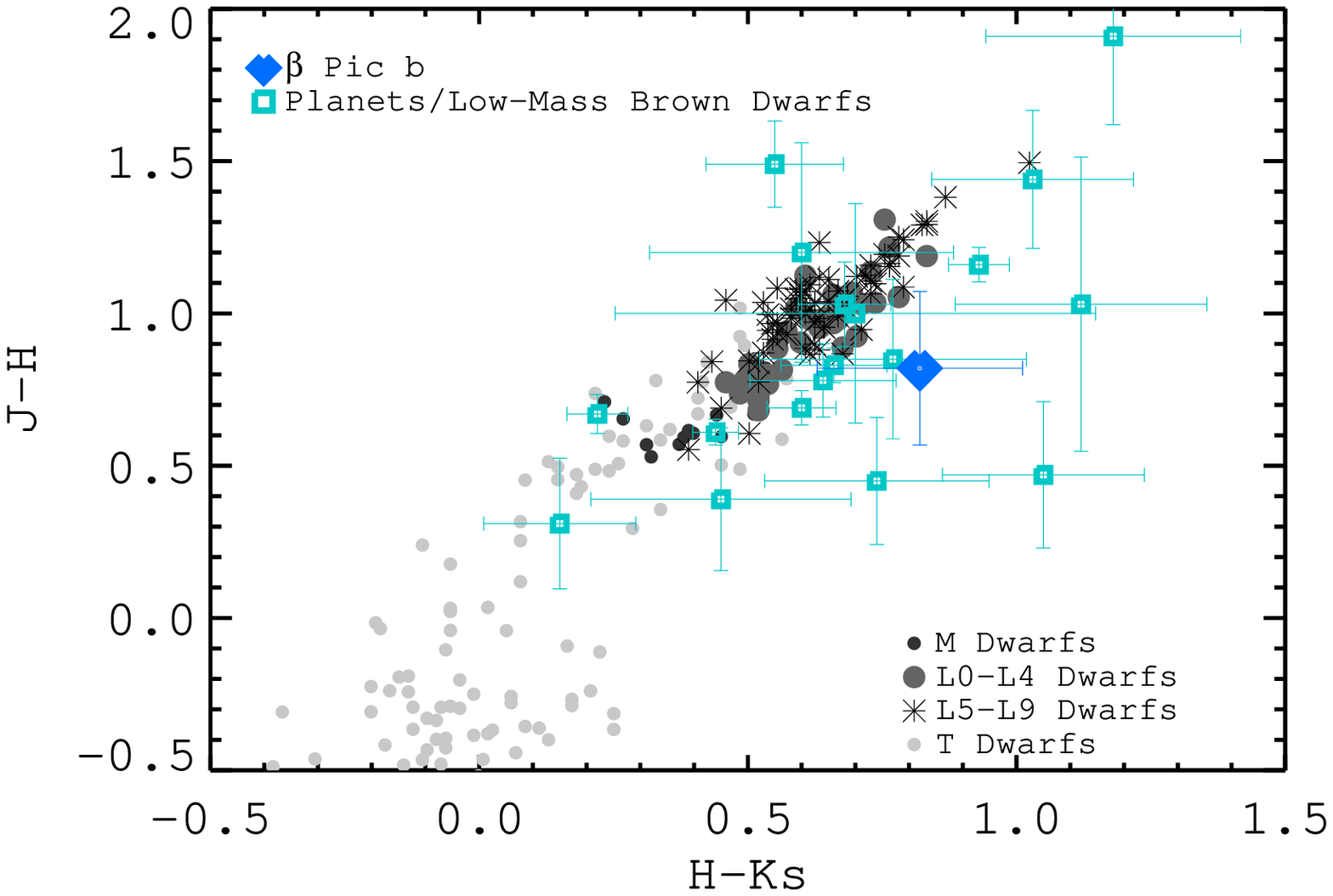}
\includegraphics[scale=0.5,trim = 8mm 0mm 8mm 0mm,clip]{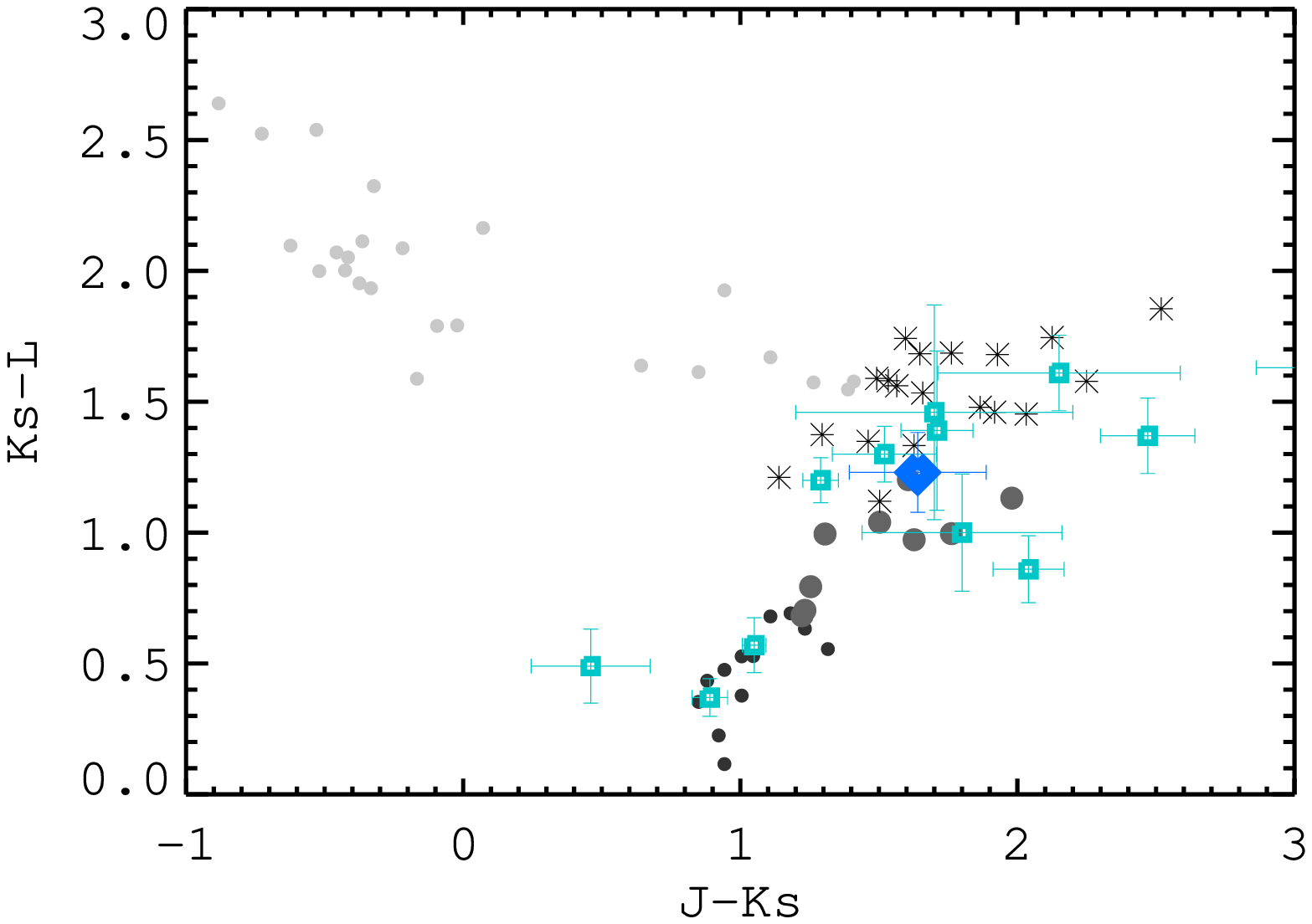}
\\
\includegraphics[scale=0.5,trim = 8mm 0mm 8mm 0mm,clip]{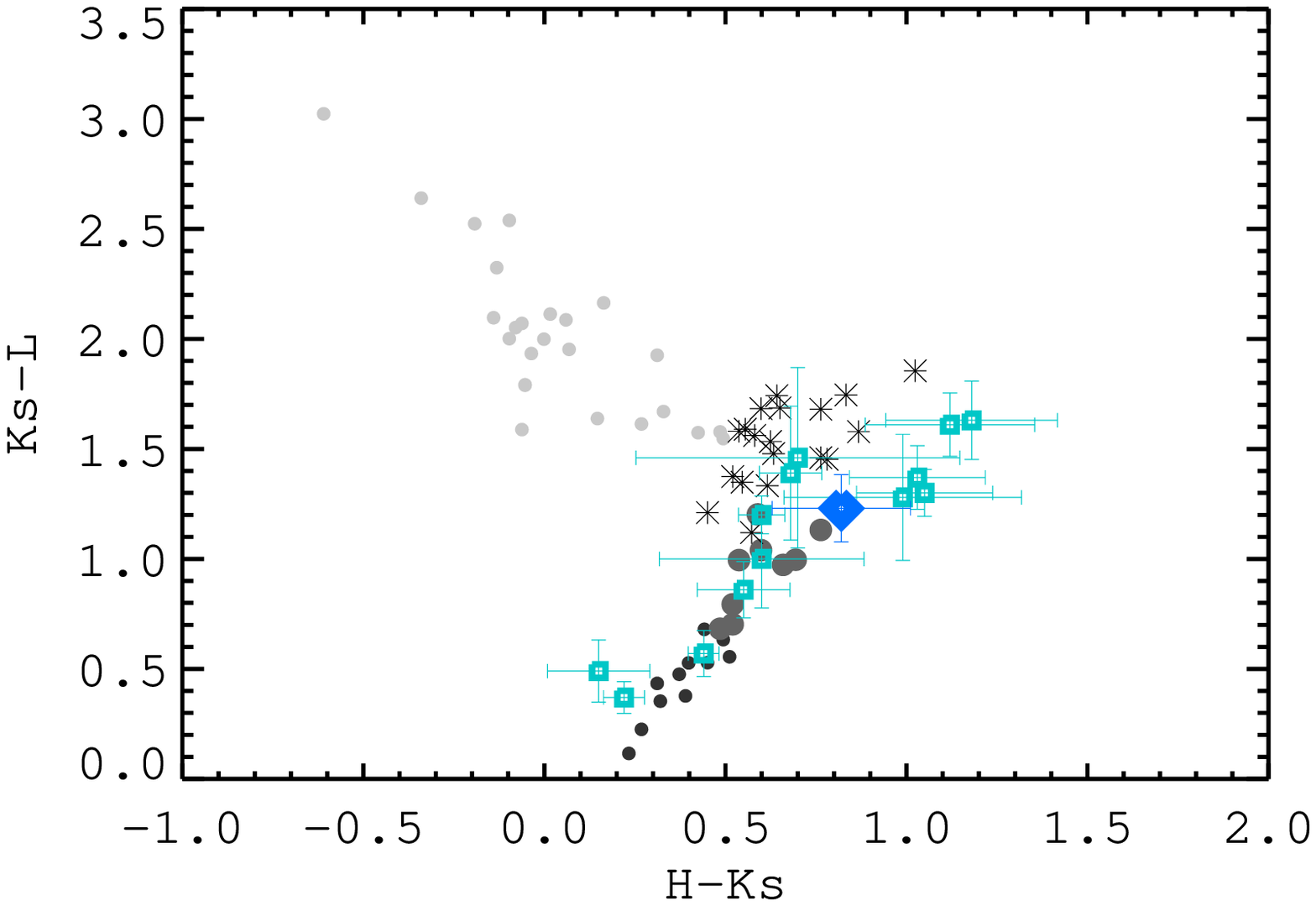}
\includegraphics[scale=0.5,trim = 8mm 0mm 8mm 0mm,clip]{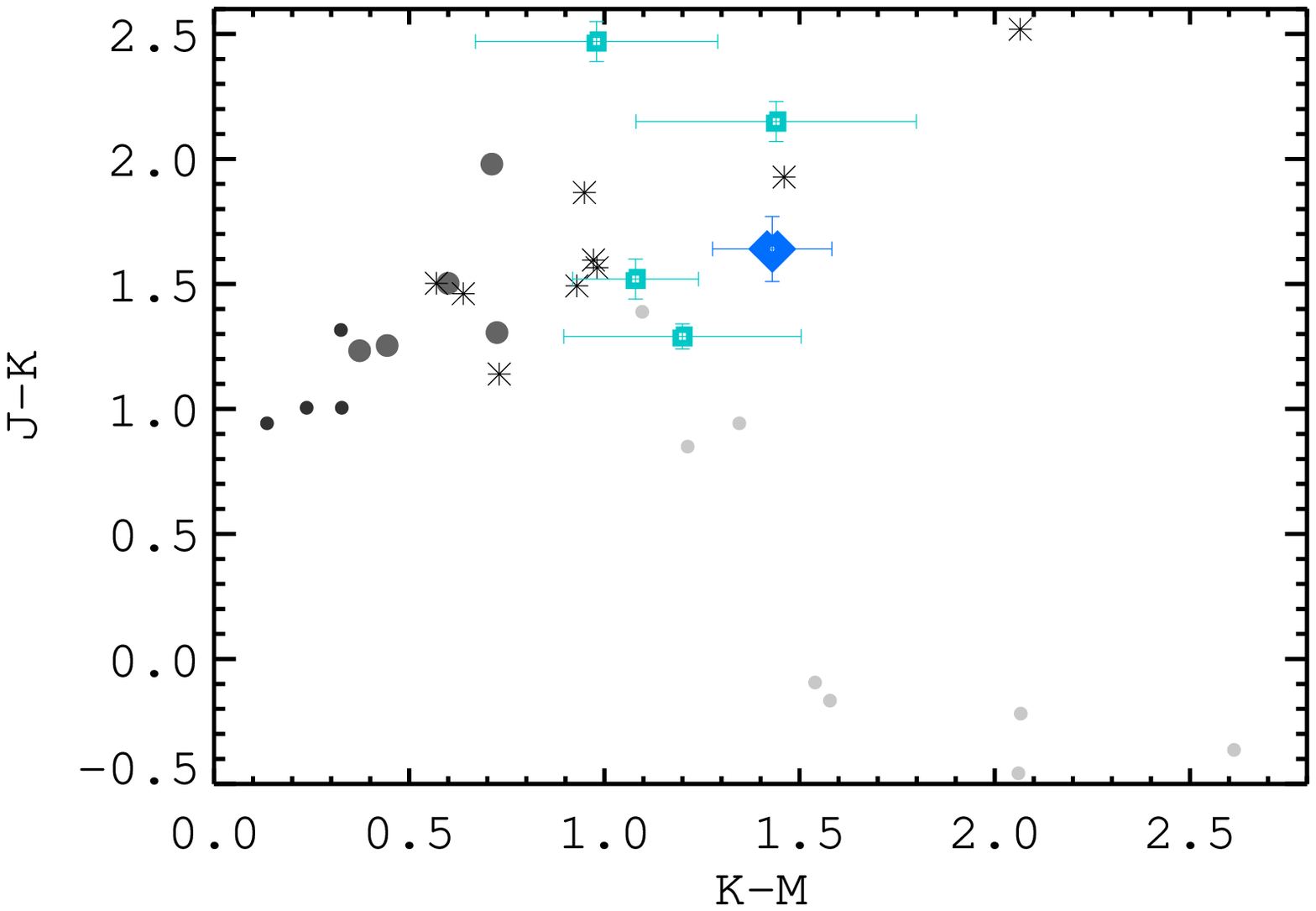}
\caption{Near-to-mid infrared color-color diagrams comparing $\beta$ Pic b's colors (blue diamonds) to those 
of M dwarfs (small dark circles), early L-type brown dwarfs (grey circles), late L dwarfs (asterisks), and T dwarfs 
(light grey dots) from \citet{Leggett2010}.  We also overplot the positions of young, substellar objects/other directly-imaged 
planets (turquoise squares).}
\label{colcol}
\end{figure}

Figure \ref{colcol} compares the IR colors of $\beta$ Pic b (dark blue diamonds) to those for field M dwarfs (small black 
dots), field L0--L5 dwarfs (grey dots), field L5.1-L9 dwarfs (asterisks), T dwarfs (small light-grey dots), 
and planets/low-mass young brown dwarfs (light-blue squares).  
The $J$-$H$/$H$-$K_{s}$ colors for $\beta$ Pic b appear slightly blue in $J$-$H$ and red in $H$-$K_{s}$ compared to 
field L0--L5 dwarfs, though the difference here is not as large as was found in \citet{Bonnefoy2013}.
Other young substellar objects appear to have similar near-IR colors, in particular  
$\kappa$ And b, GSC 06214 B, USco CTIO 108B, 2M 1207A, and Luhman 16 B, whose spectral types 
range between M8 and T0.5.  

The mid-IR colors of $\beta$ Pic b (top-right and bottom panels) show a more complicated situation.  
In $J$-$K_{s}$/$K_{s}$-$L^\prime$ and $H$-$K_{s}$/$K_{s}$-$L^\prime$, $\beta$ Pic b lies along the field L/T dwarf 
locus with colors in between those for L0--L5 and L5.1--L9 dwarfs, overlapping in color with $\kappa$ And b, 1RXJ 1609B, GSC 06214B, HR 8799 d, 
and 2M 1207 B.   Compared to the few field L/T dwarfs from the \citeauthor{Leggett2010} sample with $M^\prime$ photometry, 
$\beta$ Pic b appears rather red, most similar in $K_{s}$-$M^\prime$ color to GSC 06214 B.  

\begin{figure}
\centering
\includegraphics[scale=0.5,trim = 8mm 0mm 8mm 0mm,clip]{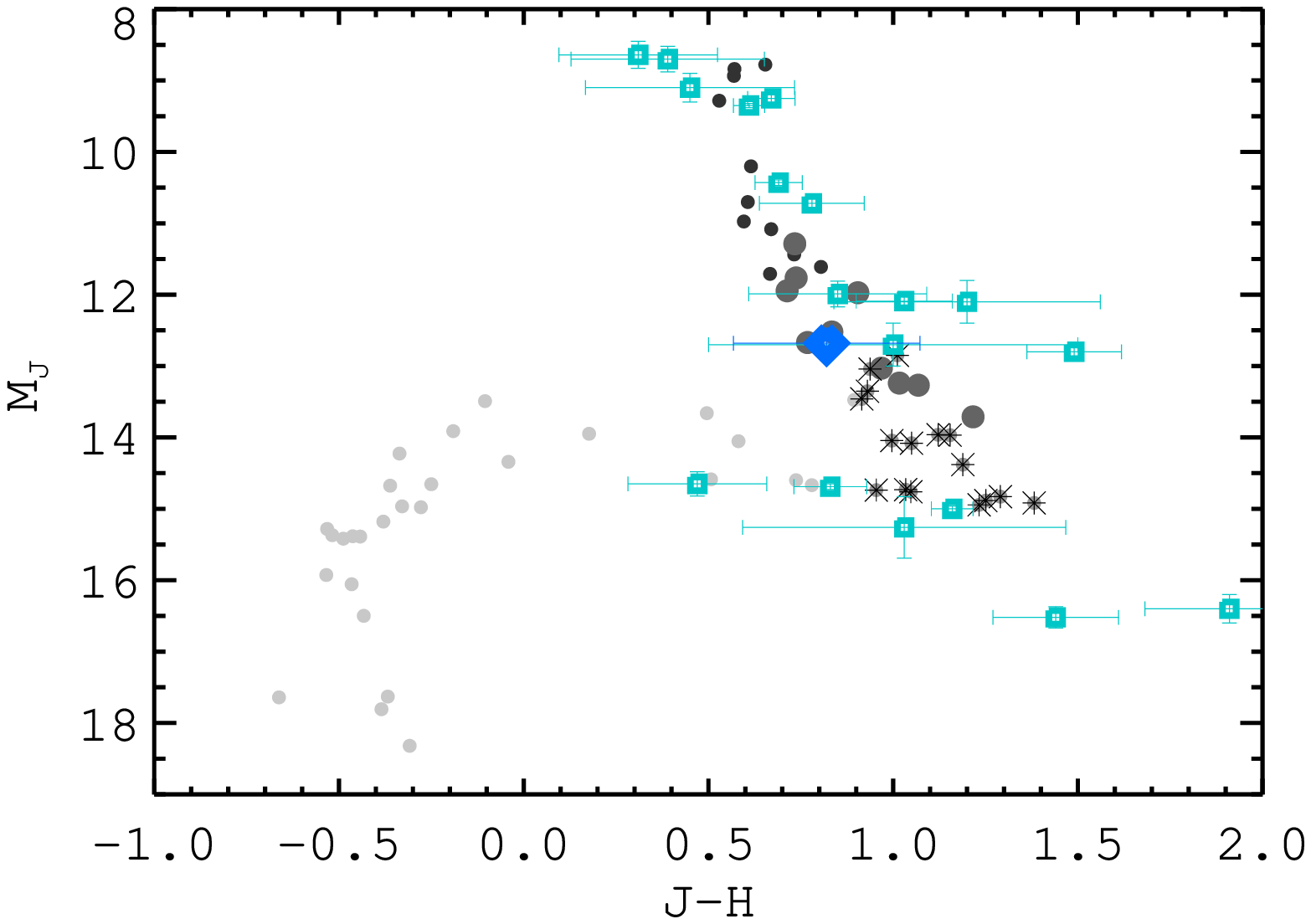}
\includegraphics[scale=0.5,trim = 8mm 0mm 8mm 0mm,clip]{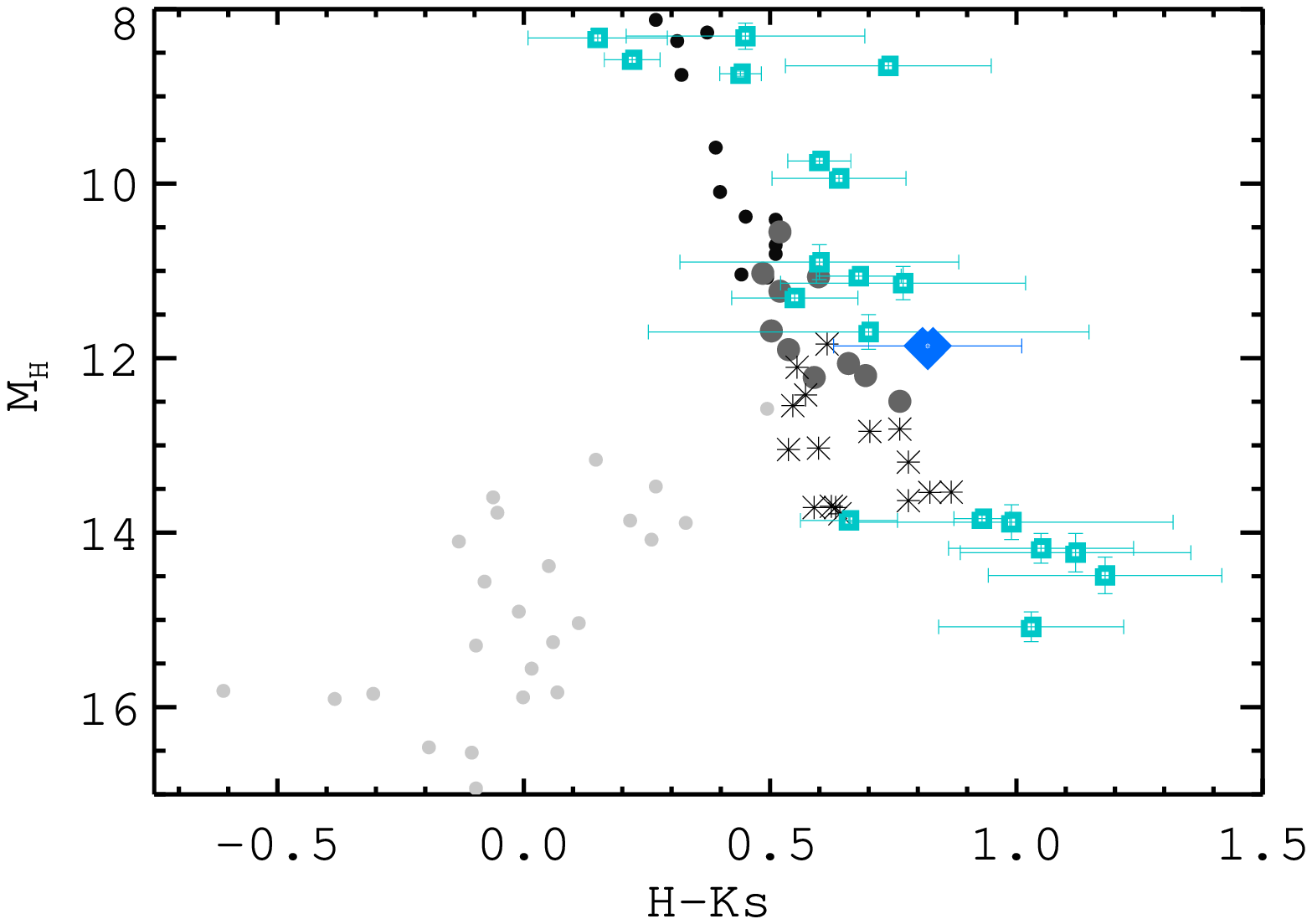}
\\
\includegraphics[scale=0.5,trim = 8mm 0mm 8mm 0mm,clip]{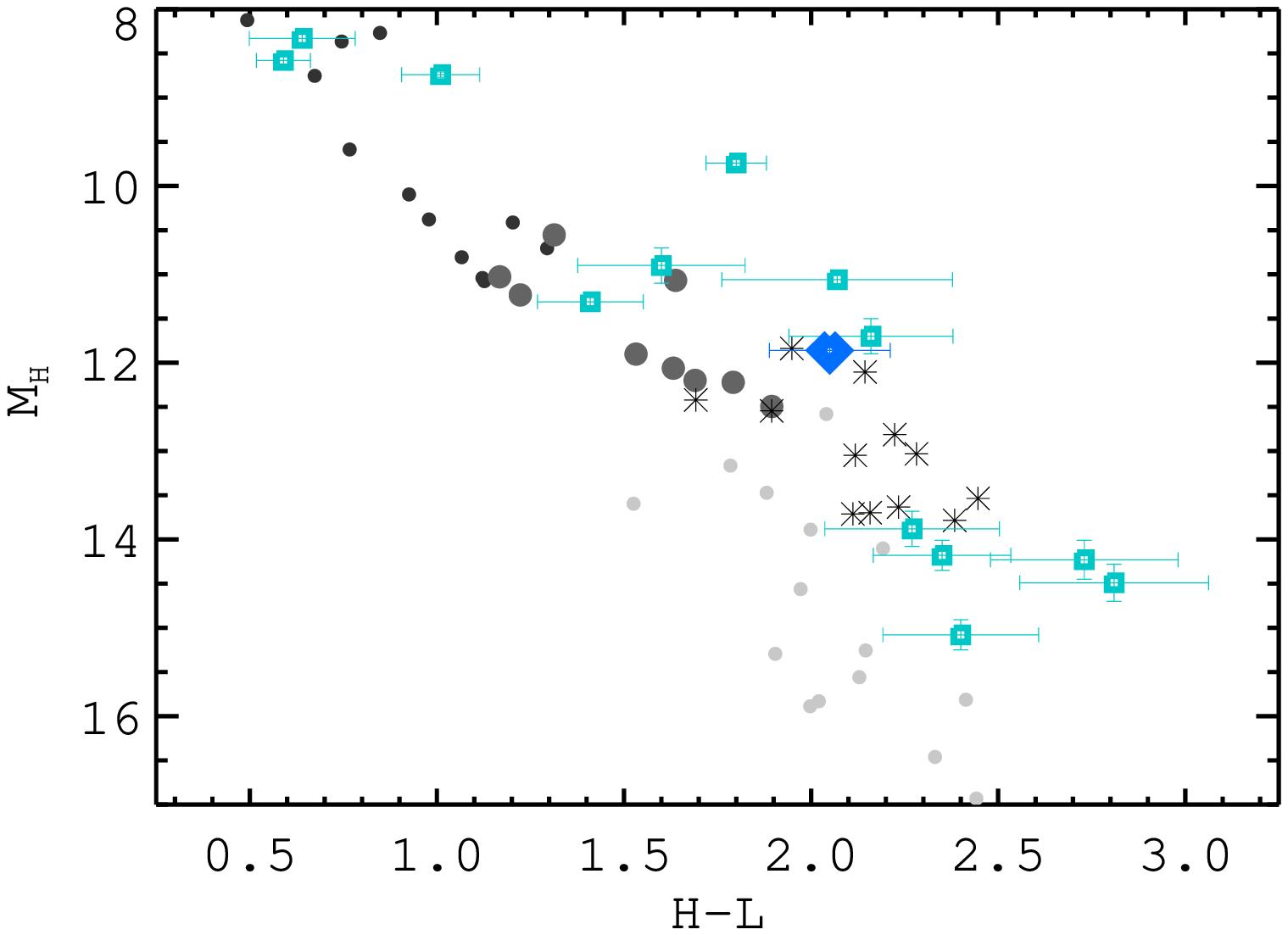}
\includegraphics[scale=0.5,trim = 8mm 0mm 8mm 0mm,clip]{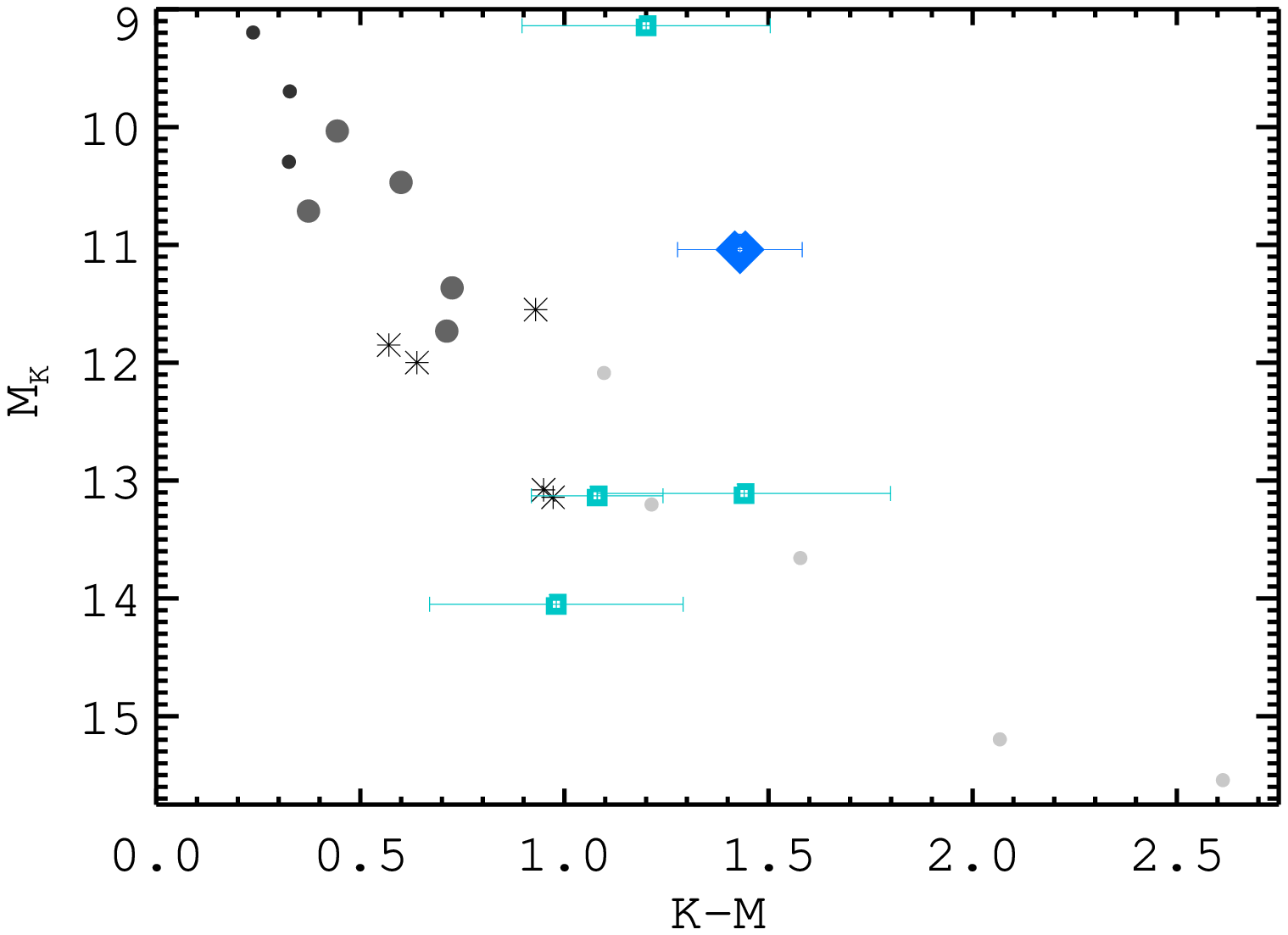}
\caption{Color-magnitude diagrams comparing $\beta$ Pic b to field M, L, and T dwarfs and young, substellar objects.  
Symbols are the same as in Figure \ref{colcol}.}
\label{cmd}
\end{figure}

The color-magnitude diagram positions of $\beta$ Pic b (Figure \ref{cmd}) better clarify how its near-to-mid 
SED compares to the field L/T dwarf sequence and to very low-mass (and gravity?) young substellar objects.
In general, compared to the field L dwarf sequence, $\beta$ Pic b appears progressively redder at 
mid-IR wavelengths.  Similar to the case for GSC 06214 B \citep{Bailey2013}, 
$\beta$ Pic b appears overluminous compared to the entire 
L/T dwarf sequence in the mid-IR.  

\subsection{Comparisons to SEDs of Other Substellar Objects}
To further explore how the SED of $\beta$ Pic b agrees with/departs from the field L/T dwarf sequence 
and other young substellar objects, we first compare its photometry to spectra from the SPeX library 
\citep{Cushing2005,Rayner2009} of brown dwarfs with data overlapping with our narrowband 
mid-IR filters ([3.09] and [4.05]) spanning spectral classes between L1 and L5: 
2MASS J14392836+1929149 (L1), Kelu-1AB (L2), 
2MASS J15065441+1321060 (L3), 2MASS J15074769-1627386 (L5).
To compare the $\beta$ Pic b photometry with cooler L dwarfs, we add combined IRTF/SpeX and 
Subaru/IRCS spectra from 1 to 4.1 $\mu m$ for 
2MASS J08251968+2115521 (L7.5) and DENIS-P J025503.3-470049 (L8) \citep{Cushing2008}.
Finally, we add spectra for the low surface-gravity L4.5 dwarf, 2MASSJ22244381-0158521 \citep{Cushing2008}.
To highlight differences between $\beta$ Pic b and these L dwarfs, we scale the flux densities 
for each of these standards to match $\beta$ Pic b at $\sim$ 2.15 $\mu m$ ($K_{s}$ band).

To convert our photometry derived in magnitudes to flux density units, we use the zeropoint fluxes 
listed in Table \ref{fluxzero}.  The $JHK_{s}$ and $L^\prime$$M^\prime$(4.78 $\mu m$) zeropoints are from \citet{Cohen2003} and \citet{Tokunaga2005}, 
respectively.  We base the other zeropoints off of \citet{Rieke2008}, although alternate sources \citep[e.g.][]{Cohen1995} yield 
nearly identical values.
Because the overlap in wavelengths between $\beta$ Pic and these objects is not uniform, we do not perform a rigorous 
fit between the two, finding the scaling factor that minimizes the $\chi^{2}$ value defined from the planet flux density, 
comparison object flux density, and photometric errors in both.  Rather, we focus on a simple first-order comparison 
between $\beta$ Pic b and the comparison objects to motivate detailed atmospheric modeling later in Section 4. 

\begin{figure}
\centering
\includegraphics[scale=0.4,trim= 0mm 0mm 0mm 0mm,clip]{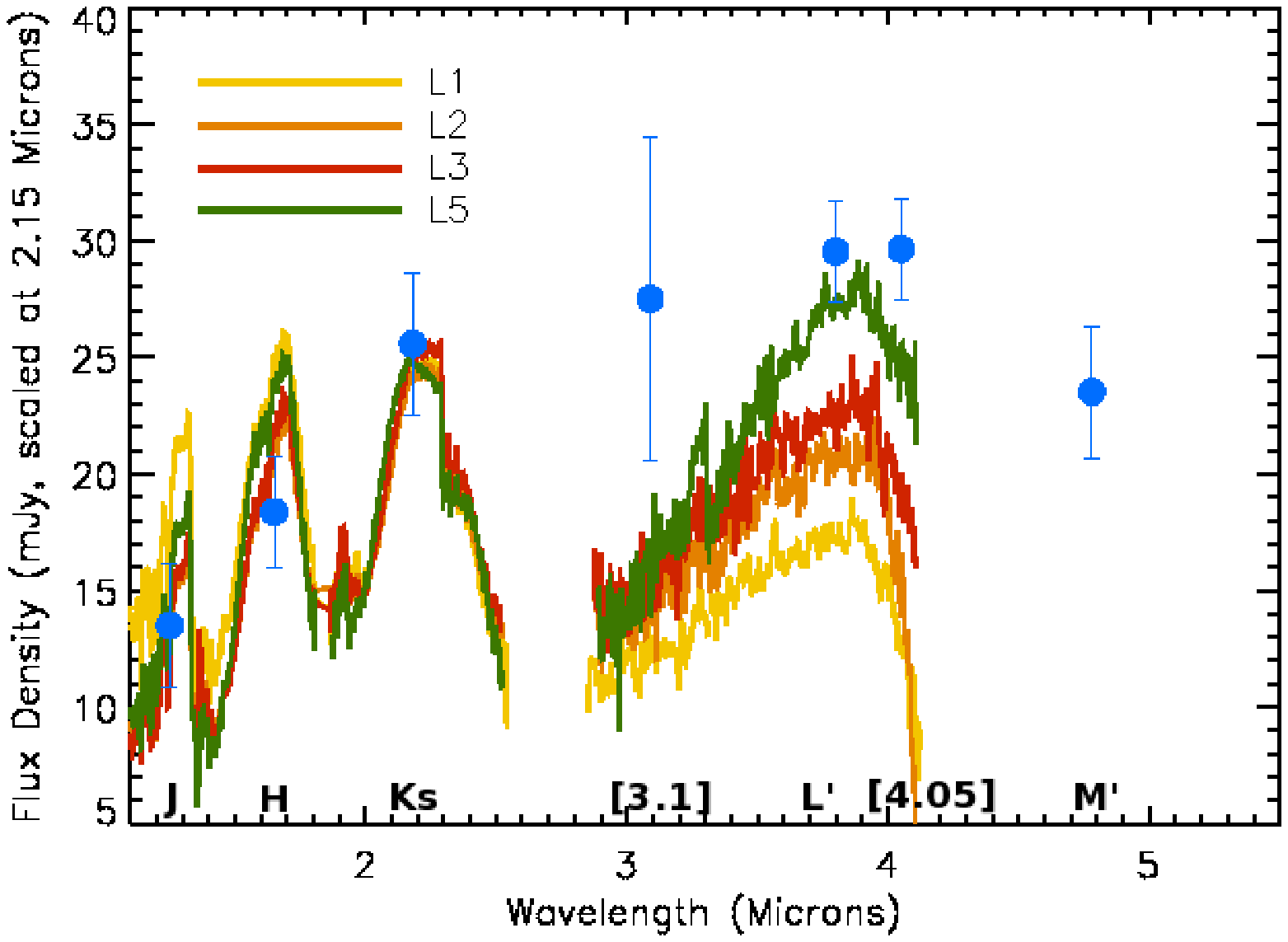}
\includegraphics[scale=0.4,trim= 0mm 0mm 0mm 0mm,clip]{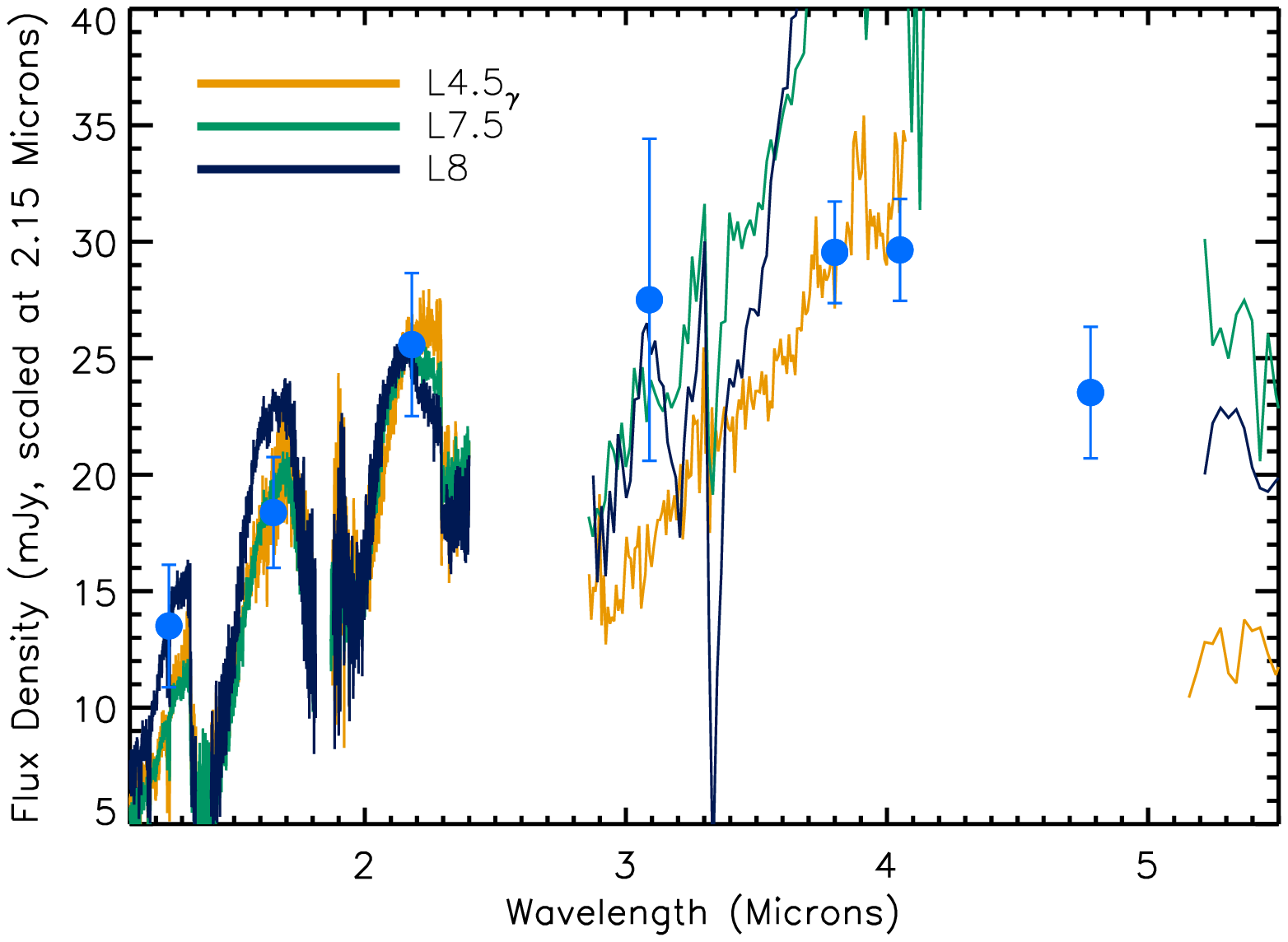}
\caption{Photometric data for $\beta$ Pic b compared to L dwarf standard spectra between L2 and L8 
as well as a low surface gravity L4 dwarf from \citet{Cushing2005,Cushing2008}.  We scale $\beta$ Pic b's 
flux density at $K_{s}$ band to the band-integrated flux density of the standards at 2.15 $\mu m$.  
With the possible exception of the low surface gravity L4 dwarf, none of these standards provide a good match to measurments for 
$\beta$ Pic b.  We identify the passbands along the bottom of the plot (left-hand panel).}
\label{spexcomp}
\end{figure}

Figure \ref{spexcomp} (left panel) compares photometry for $\beta$ Pic b to spectra for field L1--L5 dwarfs.  
While the L1 standard slightly overpredicts the flux density at $J$ band, the other three 
early/mid L standards match the $\beta$ Pic b near-IR SED quite well, indicating a ``near-IR spectral type" 
of $\sim$ L2--L5.  The L7.5 and L8 standards also produce reasonable matches, although they 
tend to underpredict the brightness at $J$ band (right panel).   

However, all standards have difficulty matching the $\beta$ Pic b SED from 3--4 $\mu m$.  
In particular, the $\beta$ Pic b flux density from $\sim$ 3 to $\sim$ 5 $\mu m$ is nearly constant, whereas it 
rises through 4 $\mu m$ and then steeply drops in all six standards depicted here.  Focused on only $\beta$ Pic b photometry 
at 3.8--4.1 $\mu m$, the ``mid-IR spectral type" is hard to define, the low surface gravity
L4.5 dwarf bears the greatest resemblance, although we fail to identify 
good matches at all wavelengths with any of our spectral templates, where the 3.1 $\mu m$, $L^\prime$, 
and [4.05] data points are the most problematic.  While none of our standards have measurements fully overlapping 
with the $M^\prime$ filter, the flux densities at 5.1 $\mu m$ indicate that they may have a very hard time simultaneously 
reproducing our measurements at all four filters between 3 and 5 $\mu m$.  Although non-equilibrium carbon chemistry can 
flatten the spectra of low surface gravity L/T dwarfs \citep{Skemer2012}, its effect 
is to weaken the methane absorption trough at $\sim$ 3.3 $\mu m$ and suppress emission at 
$\sim$ 5 $\mu m$.  Thus, it is unclear whether this effect can explain the enhanced emission at 
$\sim$ 3.1 $\mu m$ (mostly outside of the $CH_{4}$ absorption feature to begin with) \textit{and} 5 $\mu m$.

To understand whether $\beta$ Pic b's SED is unique even amongst other very low-mass 
substellar objects, we compare our photometry to that for companions listed in Table \ref{photcomptable} that 
have photometry from 1 $\mu m$ through $\sim$ 4--5 $\mu m$: HR 8799 bcd, $\kappa$ And b, 1RXJ 1609 B, GSC 06214B, 
HIP 78530 B, 2M 1207A/B, HR 7329B, and AB Pic.  Two objects -- 1RXJ 1609 B and GSC 06214B -- have 3.1 $\mu m$ photometry:
1RXJ 1609 B from \citep{Bailey2013} and $\kappa$ And b has [4.05] from data obtained by T. C. (M$_{[4.05]}$ = 9.45 $\pm$ 0.20) 
(Bonnefoy, Currie et al., 2013 in prep.).

\begin{figure}
\centering
\includegraphics[scale=0.5,clip]{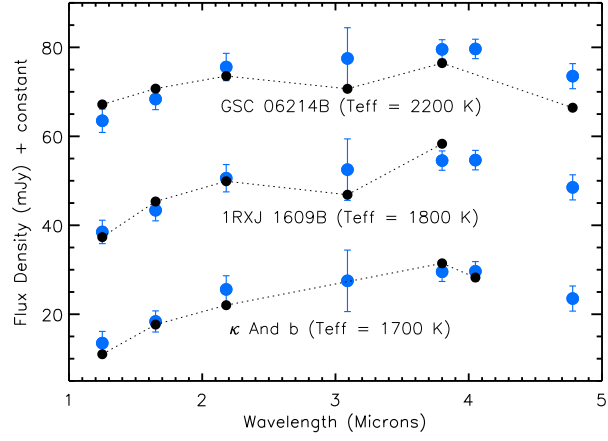}
\caption{Comparisons between the $\beta$ Pic b SED and the closest-matching substellar objects with $JHK_{s}L^\prime$ 
photometry.  Quantitatively, 1RXJ 1609 B and $\kappa$ And b provide the closest matches, although it is as-yet unclear 
whether any known substellar object fully reproduces $\beta$ Pic b's SED at all measured wavelengths.}
\label{empcomp}
\end{figure}

The two far-right columns of Table \ref{photcomptable2} lists the reduced $\chi^{2}$ and goodness-of-fit statistics 
between $\beta$ Pic b's $JHK_{s}L^\prime$ ([3.1],[4.05]) photometry, 
while Figure \ref{empcomp} displays these comparisons for $\kappa$ And b, 1RXJ 1609B, and GSC 06214B, which are all 
thought to be low surface gravity companions with $T_{eff}$ $\sim$ 1700 K, 1800 K, and 2200 K \citep{Carson2013,Lafreniere2010,Bowler2011,Bailey2013}.
Overall, $\kappa$ And b provides the best match to $\beta$ Pic b's photometry, requires negligible flux scaling, and 
is essentially the same within the 68\% confidence limit (C.L.) 
($\chi^{2}$ = 0.946, C.L. = 0.186), although the large photometric uncertainties in the near-IR limit the robustness of these conclusions.
The companion to 1RXJ 1609 likewise produces a very good match ($\chi^{2}$ = 1.369, C.L. = 0.287), while the slightly more luminous (and massive) 
GSC 06214B appears to be much bluer, (relatively) overluminous in $L^\prime$ and $M^\prime$ (or, conversely, overluminous at 
$JHK_{s}$) by $\sim$ 30\%.  In comparison, the cooler ($T_{eff}$ $\approx$ 900-1100 $K$) exoplanets HR 8799 bcd provide far poorer 
matches ($\chi^{2}$ $\sim$ 6--52).

Still, it is unclear whether any object matches $\beta$ Pic b's photometry at all wavelengths: 
both of the objects for which we have [3.1] data, GSC 06214B and 1RXJ 1609B, are still slightly underluminous here.
Moreover, the best-matching companions -- $\kappa$ And b and 1RXJ 1609B -- are still not identical, as the scaling factors between 
$\beta$ Pic b's spectrum and these companions' spectra that minimize $\chi^{2}$ are $\sim$ 0.83 and 0.53, respectively.
While companions with identical temperatures but radii 10\% and 30\% larger than $\beta$ Pic b would achieve this scaling, 
$\kappa$ And b and 1RXJ 1609B are respectively older and younger than $\beta$ Pic b, whereas for a given initial entropy of 
formation planet radii are expected to decrease with time \citep{Spiegel2012}.

In summary, young (low surface gravity?), low-mass objects may provide a better match to $\beta$ Pic b's 
photometry than do field dwarfs, especially those with temperatures well above 1000 $K$ but slightly below 2000 $K$ 
($\kappa$ And b, 1RXJ 1609 B).  However, we fail 
to find a match (within error bars) between the planet's photometry spanning the full range of wavelengths for 
which we have data, especially at $\sim$ 3 $\mu m$.   As the planet spectra depend critically 
on temperature, surface gravity, clouds and (as we shall see) dust particle sizes, our comparisons imply that $\beta$ Pic b may differ from most
young substellar objects in one of these respects.  Next, we turn to detailed atmospheric modeling to 
identify the set of atmospheric parameters that best fit the $\beta$ Pic b data.

\section{Planet Atmosphere Modeling}
To further explore the physical properties of $\beta$ Pic b, we compare its photometry to planet atmosphere models 
adopting a range of surface gravities, effective temperatures, and cloud prescriptions/dust.
For a given surface gravity and effective temperature, a planet's emitted spectrum depends primarily on the atmosphere's composition,
the structure of its clouds, and the sizes of the dust particles of which the clouds are comprised \citep{Burrows2006}.
For simplicity, we assume solar abundances except where noted and leave consideration of anomalous abundances for future work.

Based on $\beta$ Pic b's expected luminosity 
(log(L$_{p}/L_{\odot}$) $\sim$ -3.7 to -4, Lagrange et al. 2010; Bonnefoy et al. 2013) and age, 
it is likely too hot ($T_{eff}$ $\sim$ 1400-1800 K) for 
non-equilibrium carbon chemistry to play a dominant role \citep{HubenyBurrows2007,Galicher2011}.
Therefore, our atmosphere models primarily differ in their treatment of clouds and the dust particles 
 entrained in clouds.  For each model, we explore a range of surface gravities and effective temperatures.

\subsection{Limiting Cases: The \citet{Burrows2006} E60 and A60 Models and AMES-DUSTY Models}
\subsubsection{Model Descriptions}
We begin by applying an illustrative collection of previously-developed atmosphere models to $\beta$ Pic b.
These models will produce limiting cases for the planet's cloud structure and typical dust grain size, which we refine in Section 
\ref{sec-smalldust}.  To probe the impact of cloud thickness, we first adopt a (large) 
modal particle size of 60 $\mu$m and consider three different cloud models: 
the standard chemical equilibrium atmosphere thin-cloud models from \citet{Burrows2006}, which successfully reproduces 
the spectra of field L dwarfs, moderately-thick cloud models from \citet{Madhusudhan2011}, and thick cloud models used in \citet{Currie2011a}.
To investigate the impact of particle size, we then apply the AMES-DUSTY models.  The DUSTY models lack any dust grain sedimentation, 
such that the dust grains are everywhere in the atmosphere, similar to the distribution of dust grains entrained in 
thick clouds.  However, they adopt far smaller dust grains than do the thick cloud models from \citet{Madhusudhan2011} and \citet{Currie2011a}, 
where the grains are submicron in size and follow the interstellar grain size distribution \citep{Allard2001}.
All models described here and elsewhere in the paper assume that the planet is in hydrostatic and radiative equilibrium.
None of them consider irradiation from the star, as this is likely unimportant at $\beta$ Pic b's orbital separation.
  Table \ref{bpicatmosfit} summarizes the range of atmospheric 
properties we consider for each model.

\textbf{The \citet{Burrows2006} E60 Thin Cloud, Large Dust Particle Models} -- As described in \citet{Burrows2006} and later 
works \citep[e.g.][]{Currie2011a,Madhusudhan2011}, the Model E60 case assumes that the clouds are 
confined to a thin layer, where the thickness of the flat part of the cloud encompasses the condensation points 
of different species with different temperature-pressure point intercepts.  Above and below this flat portion, the 
cloud shape function decays as the -6 and -10 powers respectively, so that the clouds have scale heights of 
$\sim$ 1/7th and 1/11th that of the gas.  We adopt a modal particle size of 60 $\mu m$ and a particle 
size distribution drawn from terrestrial water clouds \citep{Deirmendjian1964}.  We consider surface gravities 
with log(g) = 4 and 4.5 and temperatures with a range of $T_{eff}$ = 1400--1800 K in increments of 100 K.

\textbf{The \citet{Madhusudhan2011} AE60 Moderately-Thick Cloud, Large Dust Particle Models} -- 
Described in \citet{Madhusudhan2011}, the Model AE60 case assumes a shallower cloud shape function of 
$s_{u}$ = 1, such that the cloud scale height is half that of the gas as a whole.  We again adopt a 
modal particle size of 60 $\mu m$ and the same particle size distribution.  We consider surface gravities 
 with log(g) = 4 and 4.5 and temperatures between $T_{eff}$ = 1000--1700 K in increments of 100 K.

\textbf{The \citet{Burrows2006} A60 Thick Cloud, Large Dust Particle Models} --
As described in \citet{Currie2011a}, the Model A60 case differs in that it 
assumes that the clouds extend with a scale height that tracks that of the gas as a whole.  
Below the flat part of the cloud, the shape function decays as the -10 power as in the E60 and AE60 models, although 
deviations from this do not affect the emergent spectrum.  Here, we consider surface gravities with 
log(g) = 4 and 4.5 and temperatures with a range of $T_{eff}$ = 1000-1700 K in increments of 100 K.

\textbf{AMES-DUSTY Thick-Cloud, Small Dust Particle Limit} -- The AMES-DUSTY atmosphere models \citep{Allard2001} 
leverage on the PHOENIX radiative transfer code \citep{HauschildtBaron1999} and
explore the limiting case where dust grains do not sediment/rain out in the atmosphere.  
Unlike the \citet{Burrows2006} models and those considered in later works \citep[e.g.][]{Spiegel2012}, the 
AMES-DUSTY models adopt a interstellar grain size distribution favoring far tinier dust grains with 
higher opacities.  The grains' higher opacities reduce the planet's radiation at shorter wavelengths.  
Thus, these models have dramatically different near-IR planet spectra from the E/A/AE60 type models with larger 
modal grain sizes even at the same temperatures and gravities \citep[cf.][]{Burrows2006,Currie2011a}.
Here we consider AMES-DUSTY models with log(g) = 3.5, 4, and 4.5 and $T_{eff}$ = 1000--2000 K ($\Delta T_{eff}$=100 K).  

\subsubsection{Fitting Method}

To transform the DUSTY spectra into predicted flux density measurements (at 10 $pc$), we convolve the spectra 
over the filter response functions and scale by a dilution factor of f = ($R_{planet}$/10 pc)$^{2}$.  
We consider a range of planet radii between 0.9 $R_{J}$ and 2 $R_{J}$.
Likewise, we convolve the E60 and A60 
model spectra over filter response functions.  The E60 models (as do all other \citealt{Burrows2006} and 
\citealt{Madhusudhan2011} models) adopt a mapping between planet radius and surface gravity/temperature set 
by the \citet{Burrows1997} planet evolution models.
To explore departures from these models, we allow the the radius to vary by an additional scale factor 
of 0.7 to 1.7.  For most of our grid, this translates into a radius range of 0.9 to 2 $R_{J}$.

Our atmosphere model fitting follows methods in \citet{Currie2011a,Currie2011b}, where
 we quantify the model fits with the $\chi^{2}$
statistic,
\begin{equation}
\chi^{2} = \sum\limits_{i=0}^{n} (f_{data,i}-
F_{model,i})^{2}/\sigma_{data,i}^{2}.
\end{equation}
We weight each datapoint equally.  Because our photometric calibration fully considers 
uncertainties due to the signal-to-noise ratio, the processing-induced attentuation, and 
the absolute photometric calibration, we do not set a 0.1 mag floor to $\sigma$ for each data point 
as we have done previously.

We determine which models are \textit{formally} consistent with the data
by comparing the resulting $\chi^{2}$ value to that
identifying the 68\% and identify those that can clearly be ruled out 
by computed the 95\% confidence limit.
Note here that these limits are significantly more stringent compared to the ones 
we adopted in \citet{Currie2011a}.
Treating the planet radius as a free parameter, we have five 
degrees of freedom for seven data points, leading to $\chi^{2}_{68\%}$ = 5.87 
and $\chi^{2}_{95\%}$ = 11.06.  `

\subsubsection{Results}
\begin{figure}
\centering
\includegraphics[scale=0.4,clip]{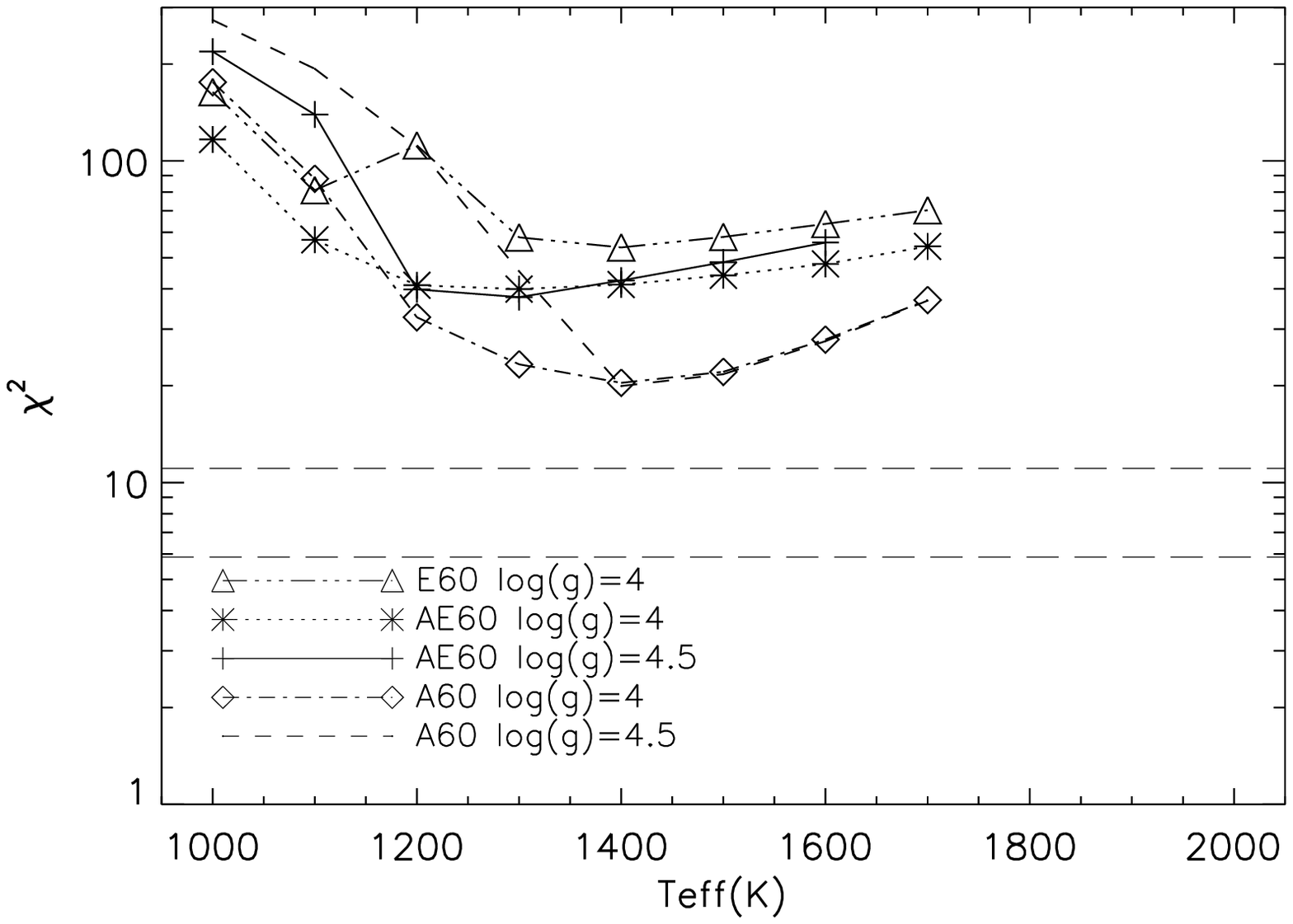}
\includegraphics[scale=0.4,clip]{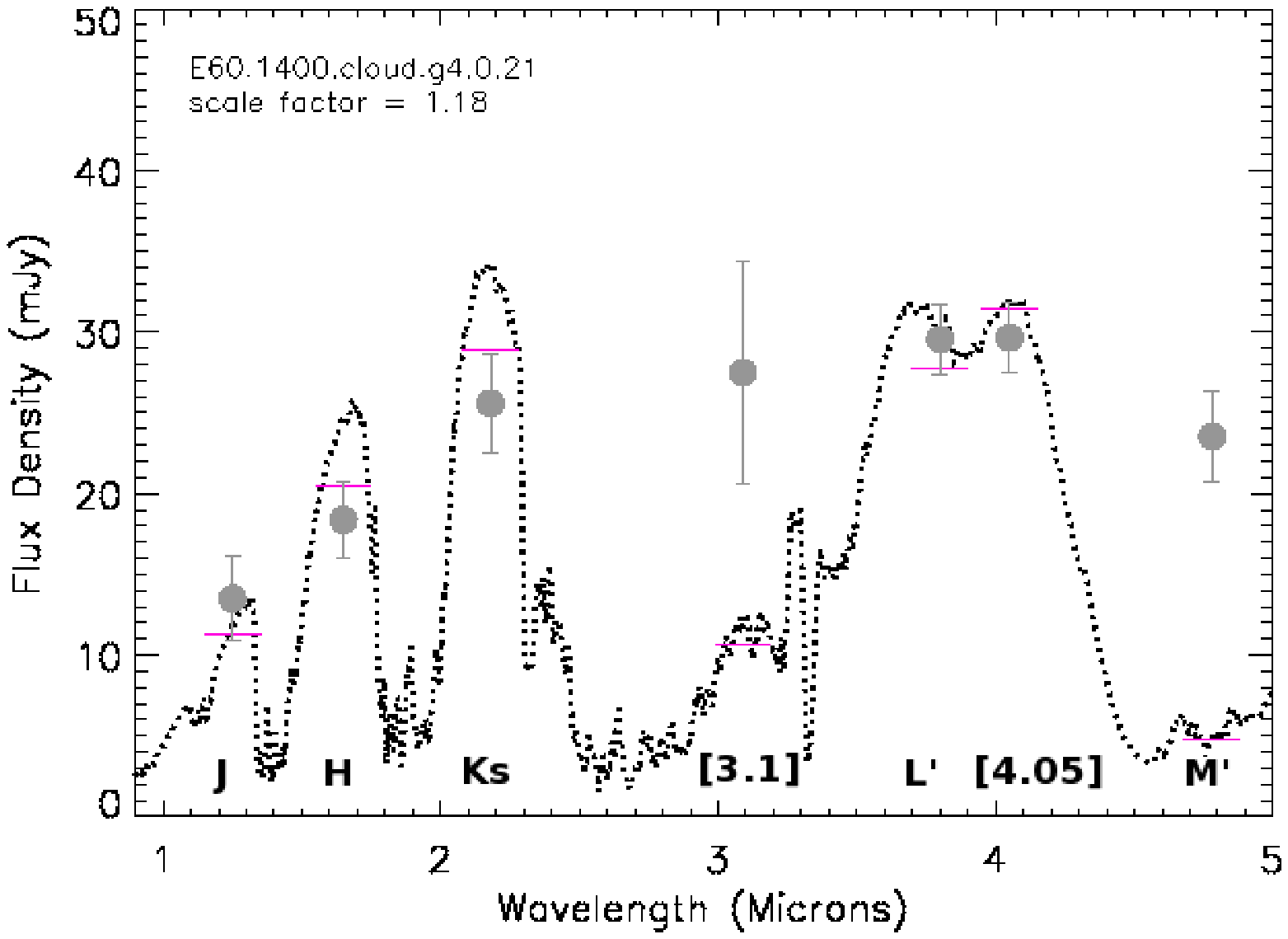}
\\
\includegraphics[scale=0.4,clip]{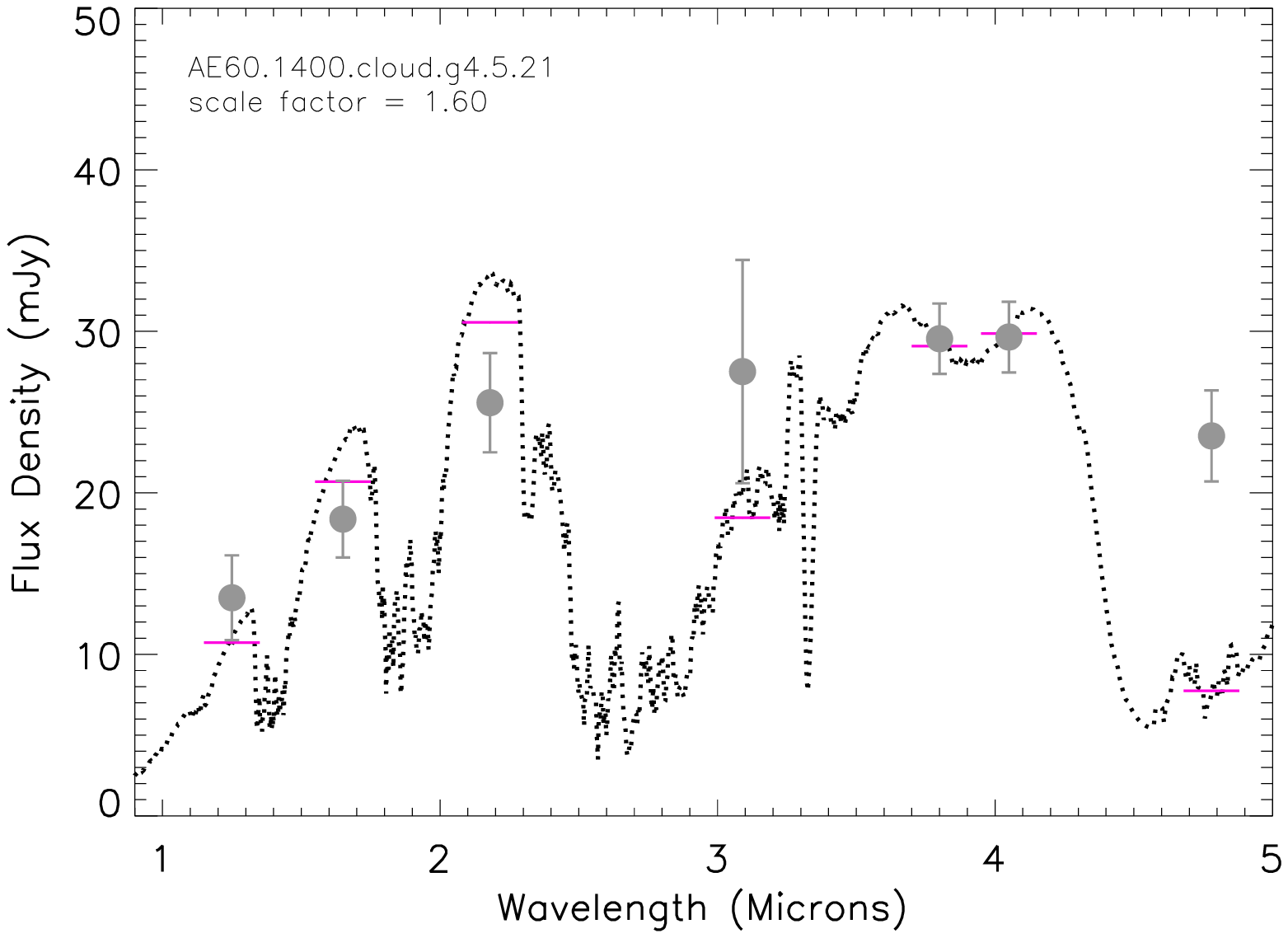}
\includegraphics[scale=0.4,clip]{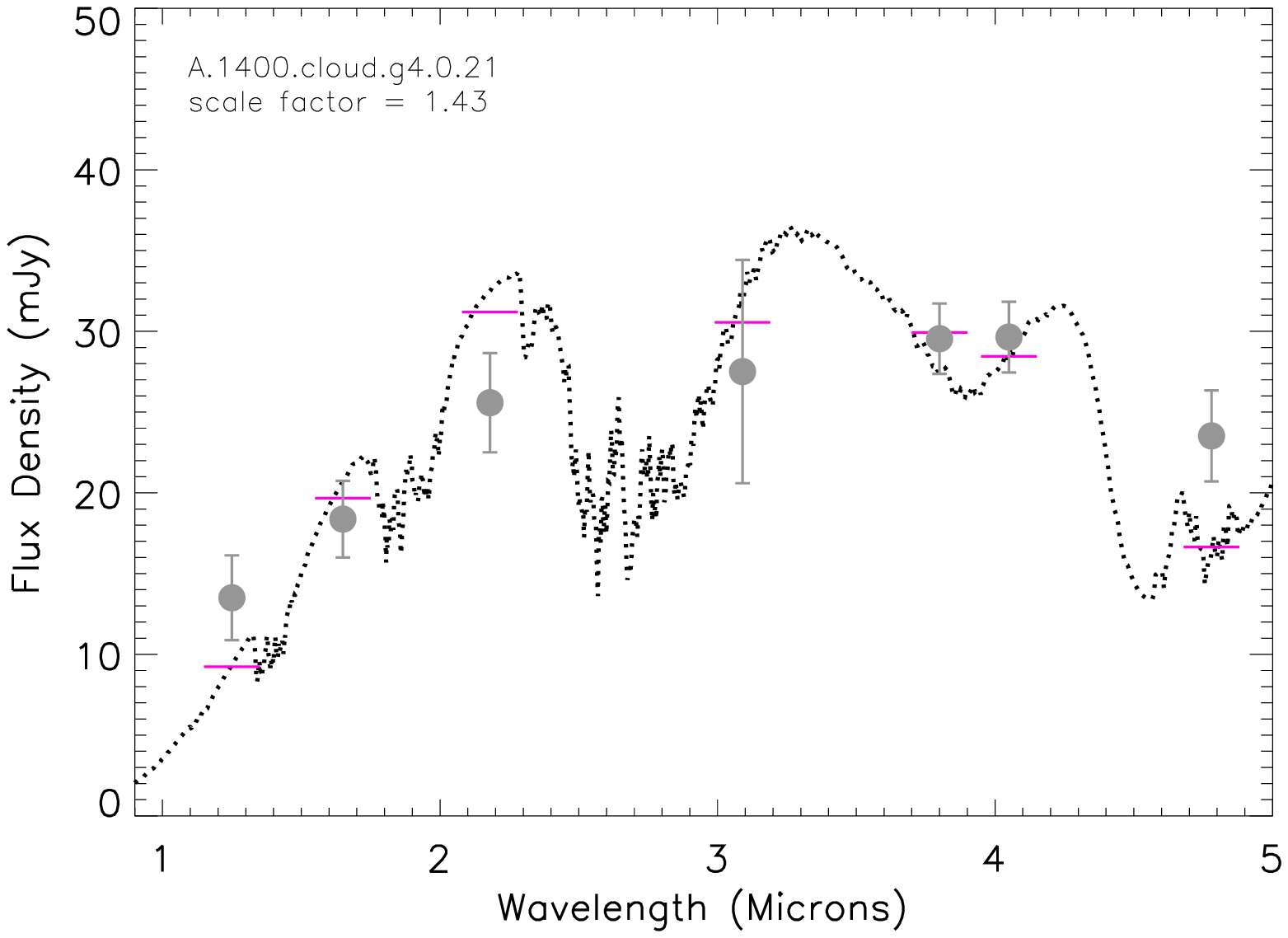}
\\
\includegraphics[scale=0.4,clip]{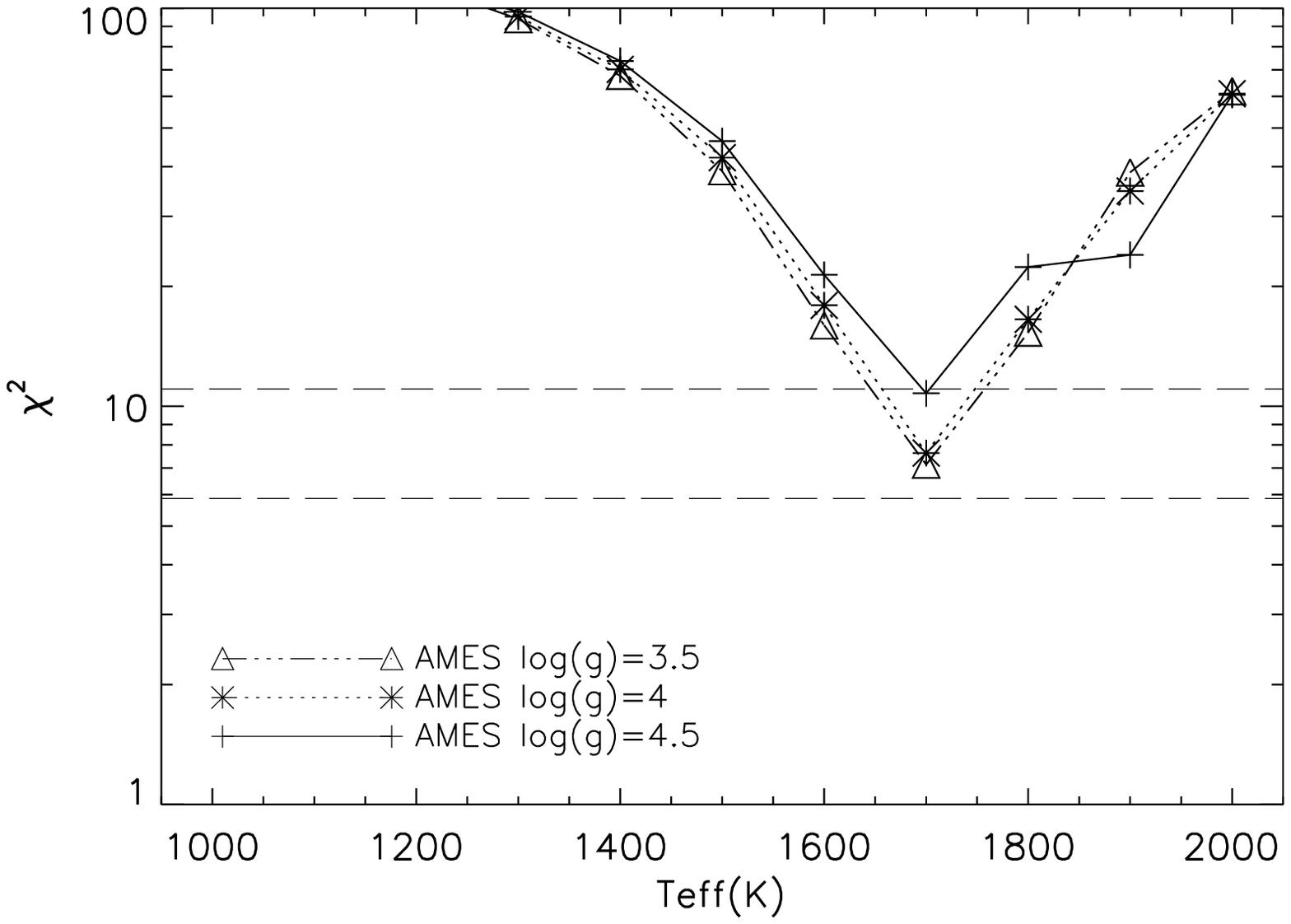}
\includegraphics[scale=0.4,clip]{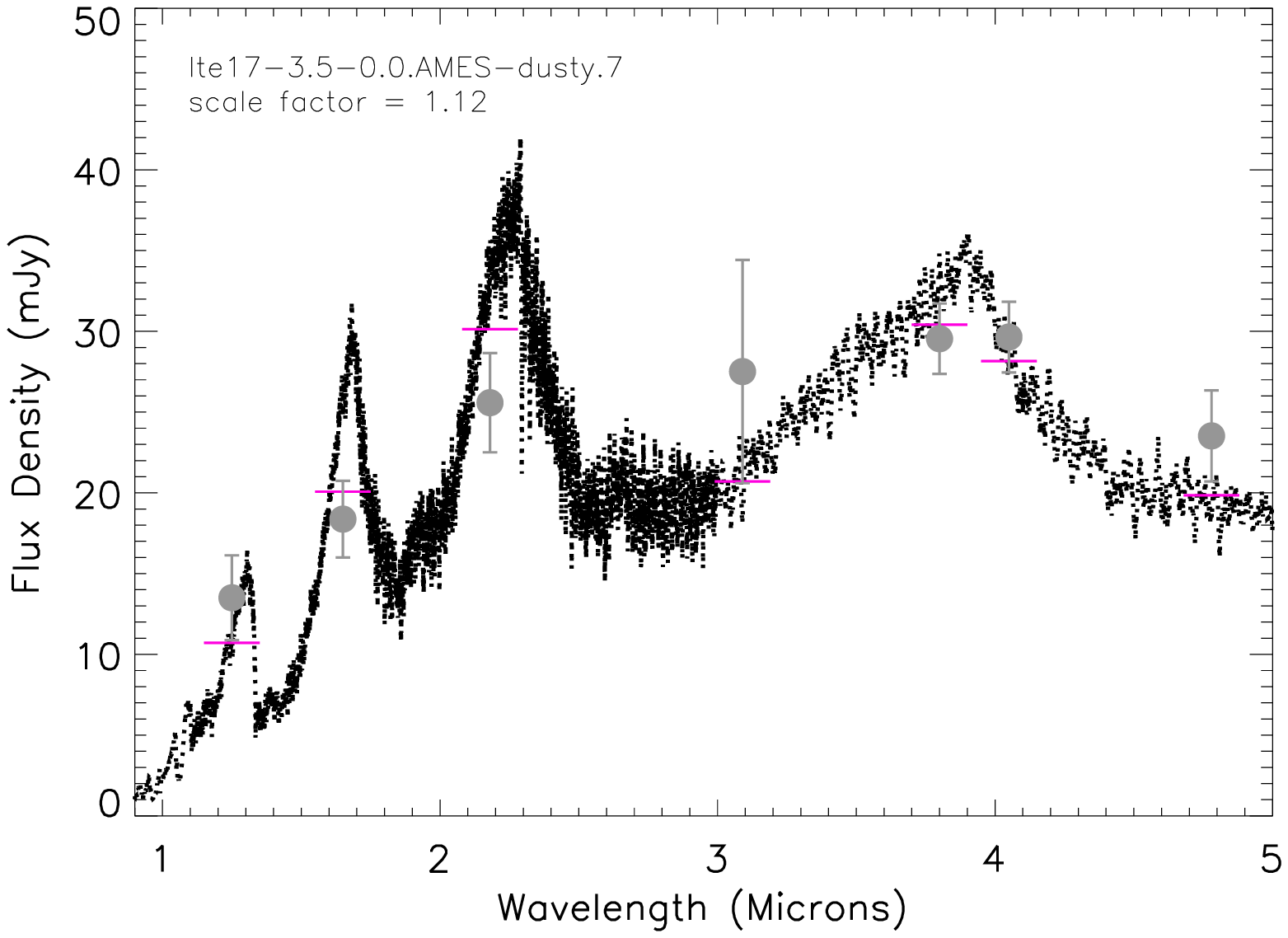}
\caption{The $\chi^{2}$ distributions (top-left) and best-fitting models (top-right, middle panels) for our limiting cases: 
the E60, thin cloud/large dust models the AE60, moderately-thick cloud/large dust models (second row), 
and the A60, thick cloud/large dust models.  (Bottom panels) The $\chi^{2}$ distributions and 
best-fit AMES-DUSTY models which assume ISM-sized dust grains.  The horizontal lines in the 
lefthand panels display the 68\% and 95\% confidence limits.  The pink bars roughly denote the 
model-predicted flux densities at the filters' central wavelength positions.  
We identify the passbands along the bottom of the plot (top-right panel).}
\label{sedfit1}
\end{figure}

Table \ref{bpicatmosfitres} summarizes our fitting results using the E60, AE60, A60, and DUSTY 
models.  Figure \ref{sedfit1} displays some of these fitting results, where the left-hand 
panels show the $\chi^{2}$ distributions with the 68\% and 95\% confidence limits indicated 
by horizontal lines dashed and dotted lines. The right-hand panels and middle-left panel 
show the best-fitting models for each atmosphere prescription.  
A successful model must match three key properties of the observed SED:
(1) At 3--5 $\mu$m, the SED is relatively flat, (2) at 1--3 $\mu$m, 
the spectral slope is relatively shallow, and (3) the overall normalization of the 3--5 $\mu$m flux
 relative to the 1--3 $\mu$m flux must match the data.

For the E60, AE60, and A60 models, we find $\chi^{2}$ minima at log(g) = 4--4.5 and $T_{eff}$ = 1400 $K$ 
in each case with radius scaling factors, the constant we multiple the nominal \citeauthor{Burrows1997} planet radii, 
between 1.185 and 1.680.  For the \citet{Burrows1997} evolutionary models, these scaling factors imply planet radii between $\sim$ 1.8 and 2 $R_{J}$, at 
the upper extrema of our grid in radius.   

 Figure \ref{sedfit1} illustrates the impact on the SED of changing cloud models, 
given a fixed grain size.  The best-fit temperature does not vary dramatically 
because, roughly speaking, the relative fluxes at 1--3 $\mu$m and 3--5$\mu$m are determined by the SED's blackbody envelope.  
However, cloud thickness dramatically  affects the depths of absorption bands superimposed on that envelope.  
Atmosphere models presented here do not feature temperature inversions.   As such, high opacity molecular lines 
have low flux densities because they originate at high altitudes where the temperature is low.  
When clouds are thin, optical depth unity is achieved at very different altitudes in and outside of absorption 
bands such as those at 3.3$\mu$m (methane) and 4.5 $\mu$m (primarily CO), and the bands appear deep.

For a fixed \textit{observed} effective temperature, thicker clouds translate into hotter temperature 
profiles (i.e. at a given pressure in the atmosphere, the temperature is higher) \citep[e.g.][]{Madhusudhan2011}.
The total Rosseland mean optical depth of the atmosphere at a given pressure is higher \citep{Madhusudhan2011}.
As the clouds become thicker, the $\tau$ = 1 surface also is more uniform, such that molecular features wash out and the 
spectrum overall appears flatter and more like a blackbody \citep{Burrows2006}.  Hence, the prominent molecular absorption bands seen in the best-fit 
E60 (thin cloud) model are substantially reduced in the A60 (thick cloud) model, with AE60 lying in between.  
The planet's flat 3--5 $\mu$m SED is best fit by A60.

Although the $\chi^{2}$ minima for all four of the models 
we consider are sharply peaked, none yield fits falling within the 68\% confidence interval.
  The fits from E60 and AE60 are particularly poor, ruled out at a 
greater than 5-$\sigma$ level, whereas the A60 model quantitatively does better but still 
is ruled out as an acceptably-fitting model (C.L. $\sim$ 3.9-$\sigma$).
The best-fit AMES-DUSTY model fits the SED even better than A60, with parameters of  $T_{eff}$ = 1700 and
log(g) = 3.5 and a radius of $r$ = 1.35 $R_{J}$, similar 
parameters to those found in \citet{Bonnefoy2013}.  However, the best-fit DUSTY model 
still falls outside the 68\% confidence limit (C.L. = 0.84).
These exercises suggest that the atmospheric parameters assumed in the models need to 
be modified in order to better reproduce the $\beta$ Pic b photometry.
To achieve this, we restrict ourselves to thick clouds and consider more carefully the impact of dust size.

\subsection{A4, Thick Cloud/Small Dust Models}\label{sec-smalldust}
\subsubsection{The Effect of Small Dust Particles}
Our analyses in the previous section show the extreme mismatch between standard L dwarf atmosphere models 
assuming thin clouds and large dust particles and the data.  While our $\chi^{2}$ values for the Burrows 
thick cloud, large dust particle models are systematically much lower, they likewise are a poor match 
to the data.  In contrast, fits from the AMES-DUSTY models only narrowly lie outside the 68\% confidence interval. 

A closer inspection of the best-fitting models in each case (right-hand panels) illustrates how they fail.  
The main difficulty with matching these models to $\beta$ Pic b spectrum is the planet's flat SED from 2 $\mu m$ 
to 5 $\mu m$, where models tend to underpredict the flux density at 3.1 $\mu m$ and/or $M^\prime$.  The slope from 
$J$ to $K_{s}$ is also a challenge.  Reducing dust sizes can further fill in absorption troughs by 
increasing the opacities of the clouds.  The AMES-DUSTY model, however, appears to overcorrect as its 
spectrum exhibits sharp peaks due to its submicron sized grains that degrade its fit to the data.
Therefore, we consider grain sizes intermediate between those in A60 and AMES-DUSTY (e.g. $\sim$ 1--30 $\mu m$).

\textbf{A4 Thick Cloud, Small Dust Particle Models} -- 
As the primary difference between these models is the typical/modal particle size, we here introduce a 
new set of atmosphere models with the same A-type, thick cloud assumption but with modal particle sizes 
slightly larger than those characteristic of dust in the AMES-DUSTY models but significantly smaller than 
previous Burrows models.  We nominally adopt 4 $\mu m$ as our new modal particle size, comparable 
in wavelength to the peak flux density of $\beta$ Pic b in $F_{\nu}$ units.   Thus, we denote these models 
as ``A4", thick-cloud, small dust particle models.

\begin{figure}
\centering
\includegraphics[scale=0.5,clip]{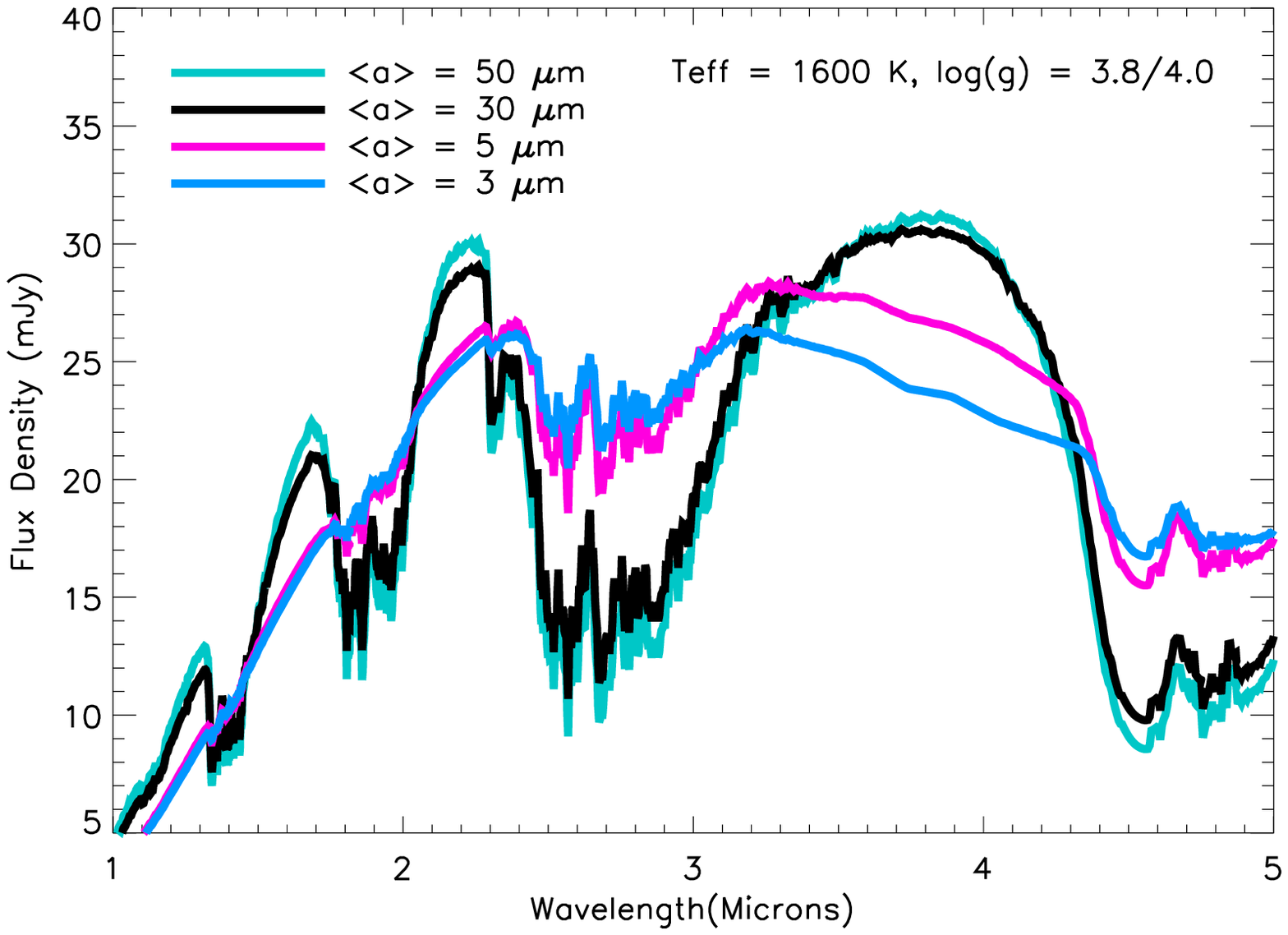}
\caption{The effect of atmospheric dust particle sizes on the shape of a massive planet SED.
 Here, the $<a>$ = 50, 30, and 5 $\mu m$ models depict log(g) = 4, $T_{eff}$ = 1600 $K$ models 
while the $<a>$ = 3 $\mu m$ model assumes log(g) = 3.8 and is scaled to match the luminosity of the $<a>$ = 5 $\mu m$ model.  
For small particle sizes, surface gravity signatures are weak and this parameter's effect is primarily to change the 
flux scaling.}
\label{dustseq}
\end{figure}

Figure \ref{dustseq} illustrates the effect of dust on the planet spectrum for modal particle sizes of 3, 5, 30 and 50 $\mu m$ and 
a temperature and surface gravity consistent with that expected to reflect $\beta$ Pic b based on planet 
cooling models ($T_{eff}$ = 1600 K, log(g)=3.8-4, $r$ $\sim$ 1.5 $R_{J}$) \citep{Burrows1997,Baraffe2003,Lagrange2010,Spiegel2012,Bonnefoy2013}.
As particle sizes decrease, the water absorption troughs at 1.8 $\mu m$ and 2.5 $\mu m$ diminish.  Likewise filled in is the deep absorption trough 
at $\sim$ 3.3 $\mu m$ and 4.5 $\mu m$ that is usually diagnostic of carbon chemistry \citep[e.g.][]{HubenyBurrows2007,Galicher2011}.  
Overall, the spectrum flattens and becomes redder (shorter wavelength emission originates at higher altitudes), with weaker 
emission and a steeper slope at $J$ to $K_{s}$.  This reddening explains the difference in best-fit 
effective temperature between the AMES-DUSTY model and the 60 $\mu$m dust models.

\subsubsection{Model Fitting Procedure}
We follow the steps outlined in \citet{Currie2011a}, where we perform two runs:
one fixing the planet radius to the \citet{Burrows1997} hot-start predictions for a given $T_{eff}$ and log(g) and 
another where we consider a range of planet radii (as in the previous section).   For the fixed-radii modeling, the 
68\% and 95\% confidence limits now lie at $\chi^{2}$ = 7.01 and 12.6, respectively, whereas they are at 5.87 and 11.06 
for the varying-radii fits as before.  
Similar to the Burrows A/E60 model runs, we consider a range of temperatures between 1400 K and 1900 K.  To explore whether 
or not the fits are sensitive to surface gravity, we consider models with log(g) = 3.6, 3.8, 4, and log(g) = 4.25.  For 
the age of $\beta$ Pic (formally, 8 to 20 Myr), this surface gravity range fully explores the masses (in the hot-start 
formalism) allowed given the radial-velocity dynamical mass limits \citep{Lagrange2012a}.

To further explore the effect that carbon chemistry may have on our planet spectra, we take the best-fitting model 
from the above exercise, significantly enhance the methane abundances over solar and re-run a small 
grid of temperatures based on that, to determine if departures from solar abundances may yield a wider range of 
acceptable atmosphere parameters. 
 Because variations in molecular abundances affect the depths of molecular absorption bands, we 
expect that such variations may improve our fit.

\subsubsection{Results}
\begin{figure}
\centering
\includegraphics[scale=0.4,clip]{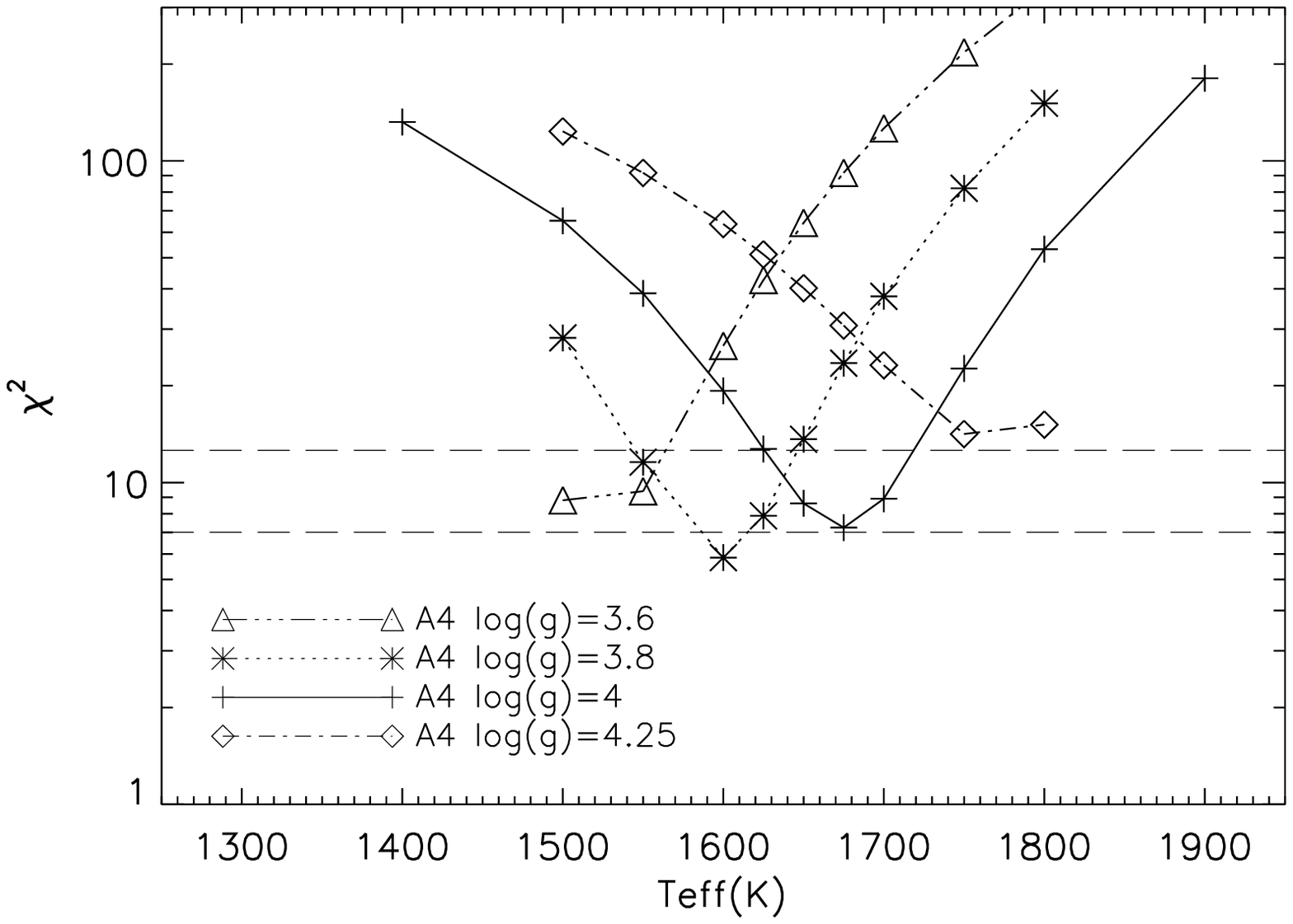}
\includegraphics[scale=0.4,clip]{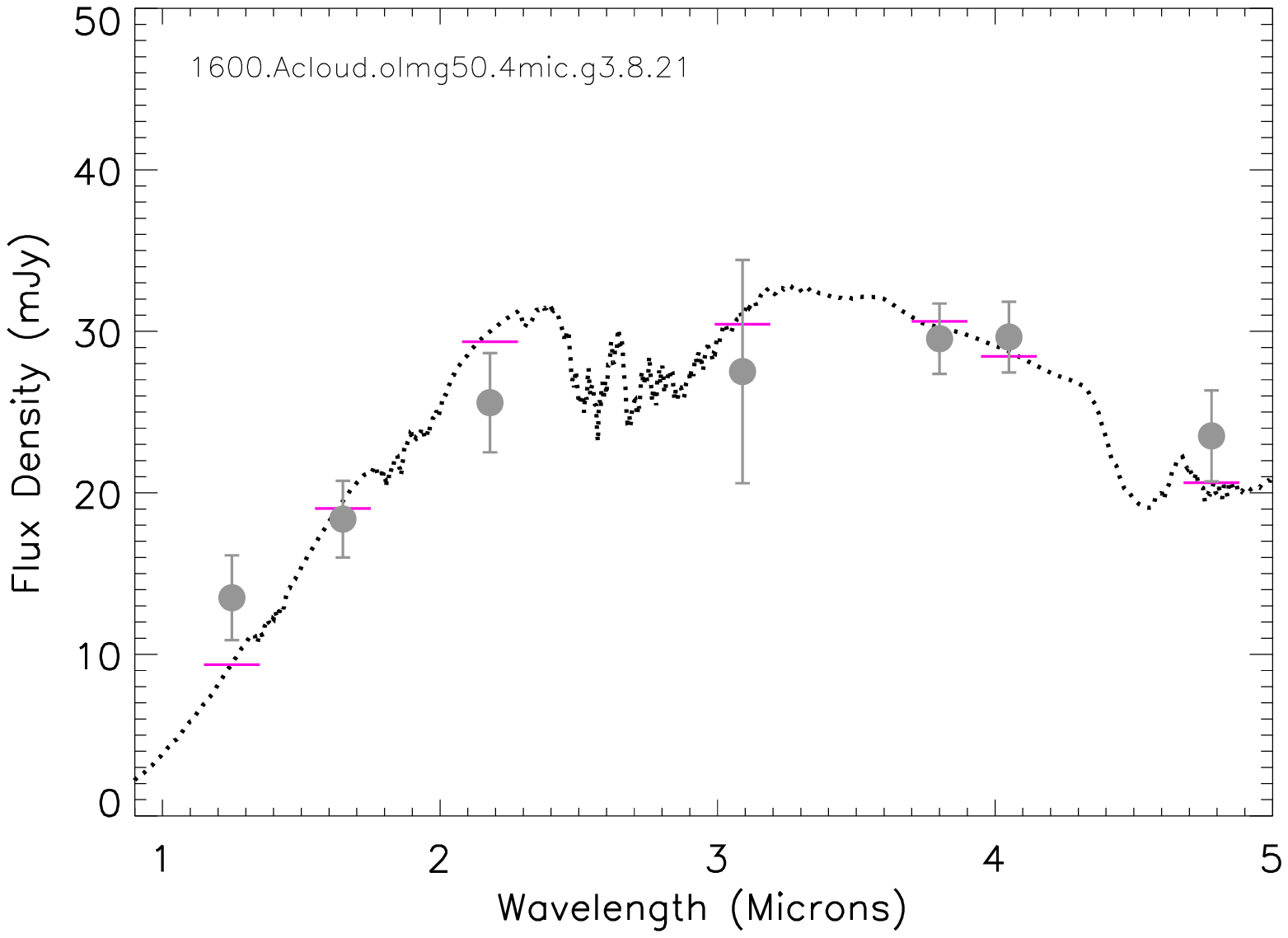}
\caption{The $\chi^{2}$ distribution (left) and best-fitting model (right) for the thick cloud, small dust A4 models, showing 
that we can achieve statistically significant fits to the data provided that the clouds are thick and the atmospheric 
dust particles are significantly smaller than those we have previously assumed in matching L dwarf spectra \citep[c.f.][]{Burrows2006}.
This model fitting ties the planet radius to predictions for hot-start models from \citet{Burrows1997}.
The horizontal dashed lines identify the 95\% confidence limit (top) and 68\% confidence limit (bottom).
}
\label{sedfit2}
\end{figure}

\begin{figure}
\centering
\includegraphics[scale=0.4,clip]{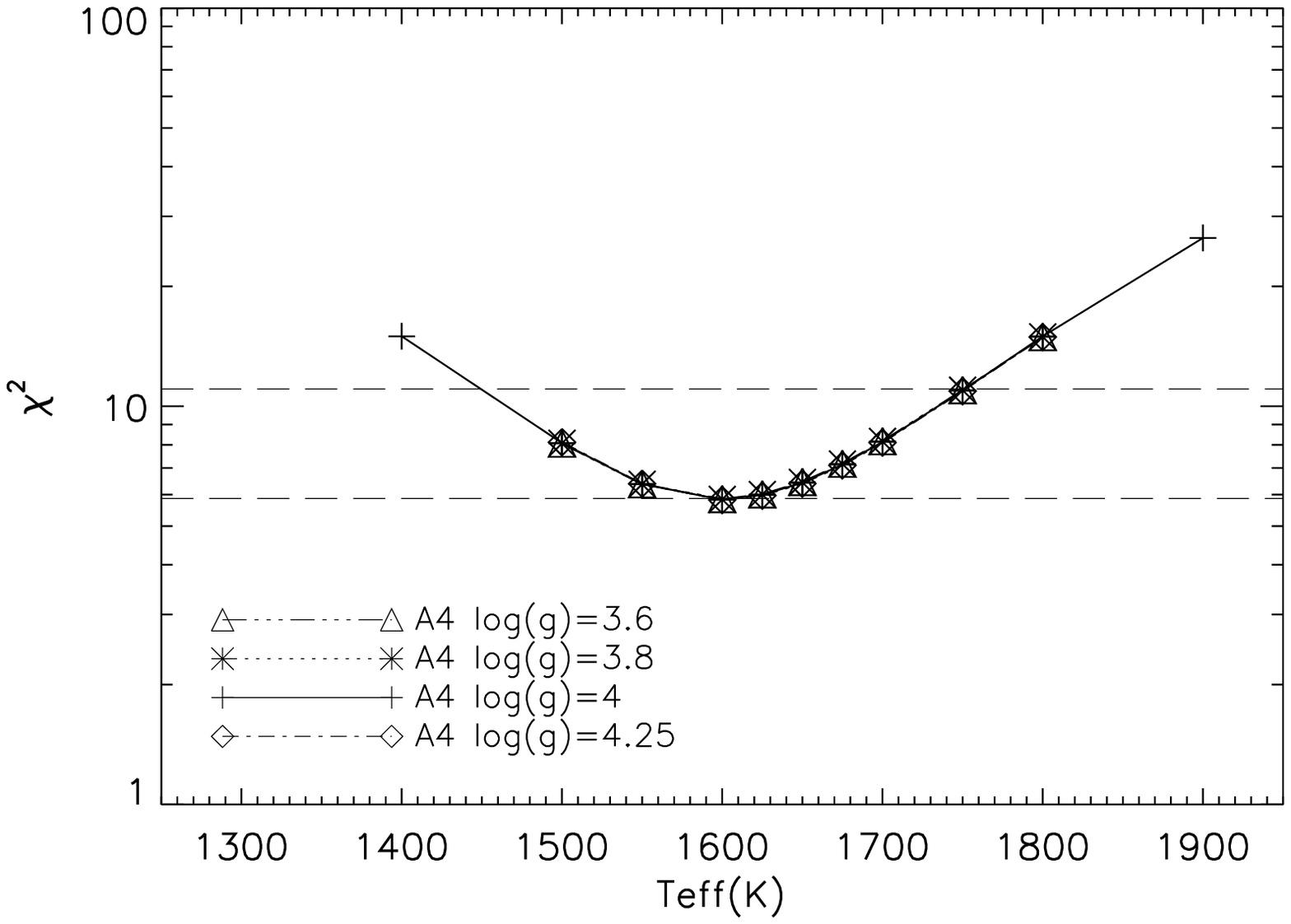}
\includegraphics[scale=0.4,clip]{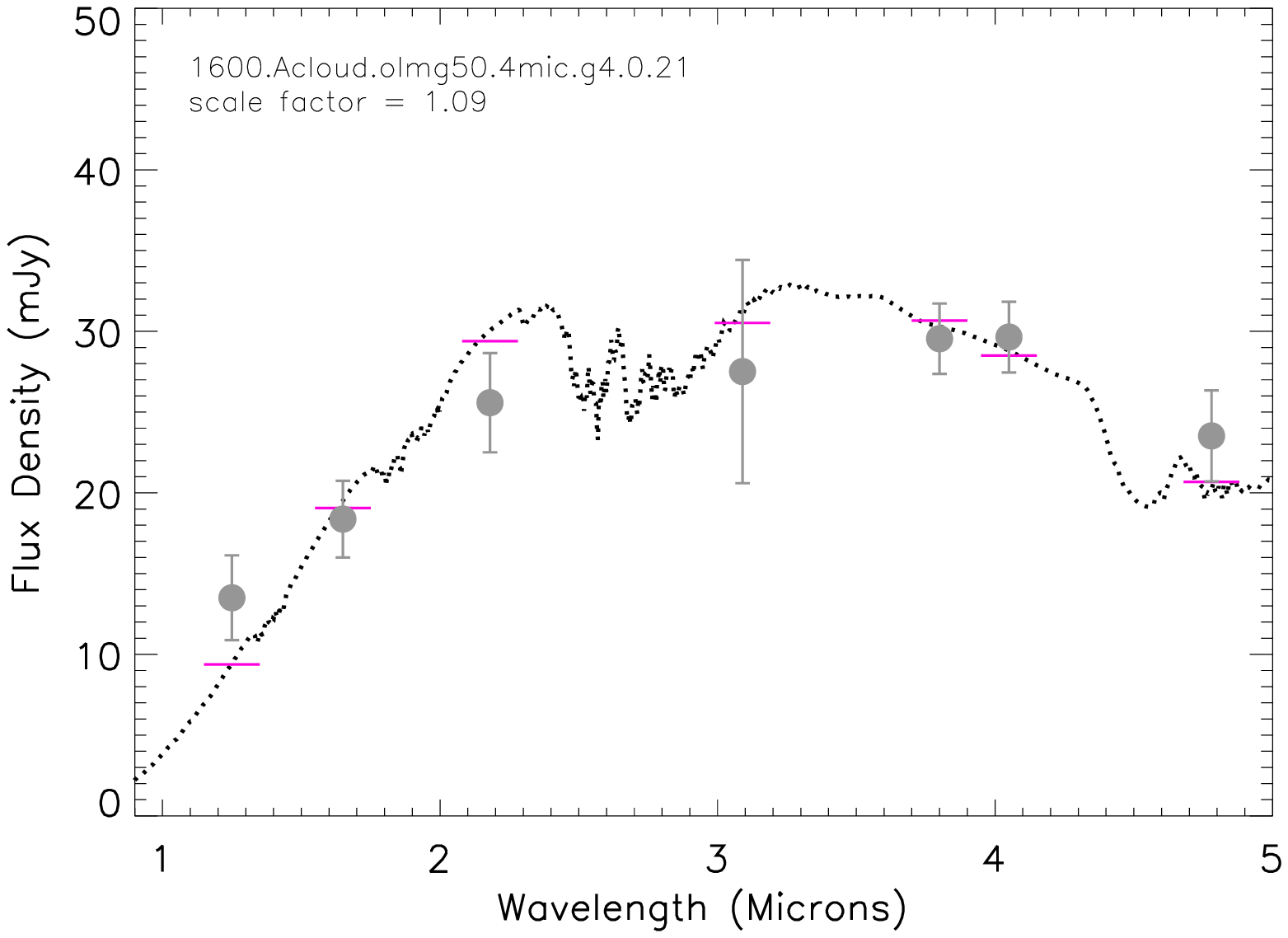}
\caption{Same as Figure \ref{sedfit2} except treating the planet radius as a free parameter.  Here we show the 
log(g) = 4, $T_{eff}$ = 1600 $K$, though the synthetic spectrum's appearance and its agreement with the data 
at neighboring gridpoints in surface gravity is nearly identical.
}
\label{sedfit3}
\end{figure}

Figures \ref{sedfit2} and \ref{sedfit3} and Table \ref{bpicatmosfitres} present our results for fitting 
the $\beta$ Pic b data with the A4, thick cloud/small dust models.  Quantitatively, these models 
better reproduce the $\beta$ Pic b SED.  Fixing the planet radius to values assumed in the \citet{Burrows1997} 
planet cooling model, we find one atmosphere model -- log(g) = 3.8, $T_{eff}$ = 1600 $K$ -- consistent 
with the data to within the 68\% confidence interval.  A wide range of models are consistent with the data
at the 95\% confidence limit, covering $\pm$ 0.2 dex in surface gravity and $\pm$ 100 $K$ in temperature.  

We can slightly improve upon these fits if we allow the planet radius to freely vary.  In this case, 
the best-fitting models yield a slightly higher surface gravity of log(g) = 4--4.25 but the same temperature of 
1600 $K$.   But in contrast to the fixed-radius case above, a wide range of models are consistent with the 
data at the 68\% confidence limit.  In particular, all surface gravities considered in our model grid 
are consistent with the data provided that the temperature is 1600 $K$ and the radius is rescaled accordingly: 
log(g) = 3.6--4.25, $T_{eff}$ = 1600 $K$.  
Another set of models with the full range of surface gravities 
and 250 $K$ spread in temperature (1500--1750 $K$) are marginally consistent with the data.

\begin{figure}
\centering
\includegraphics[scale=0.3,clip]{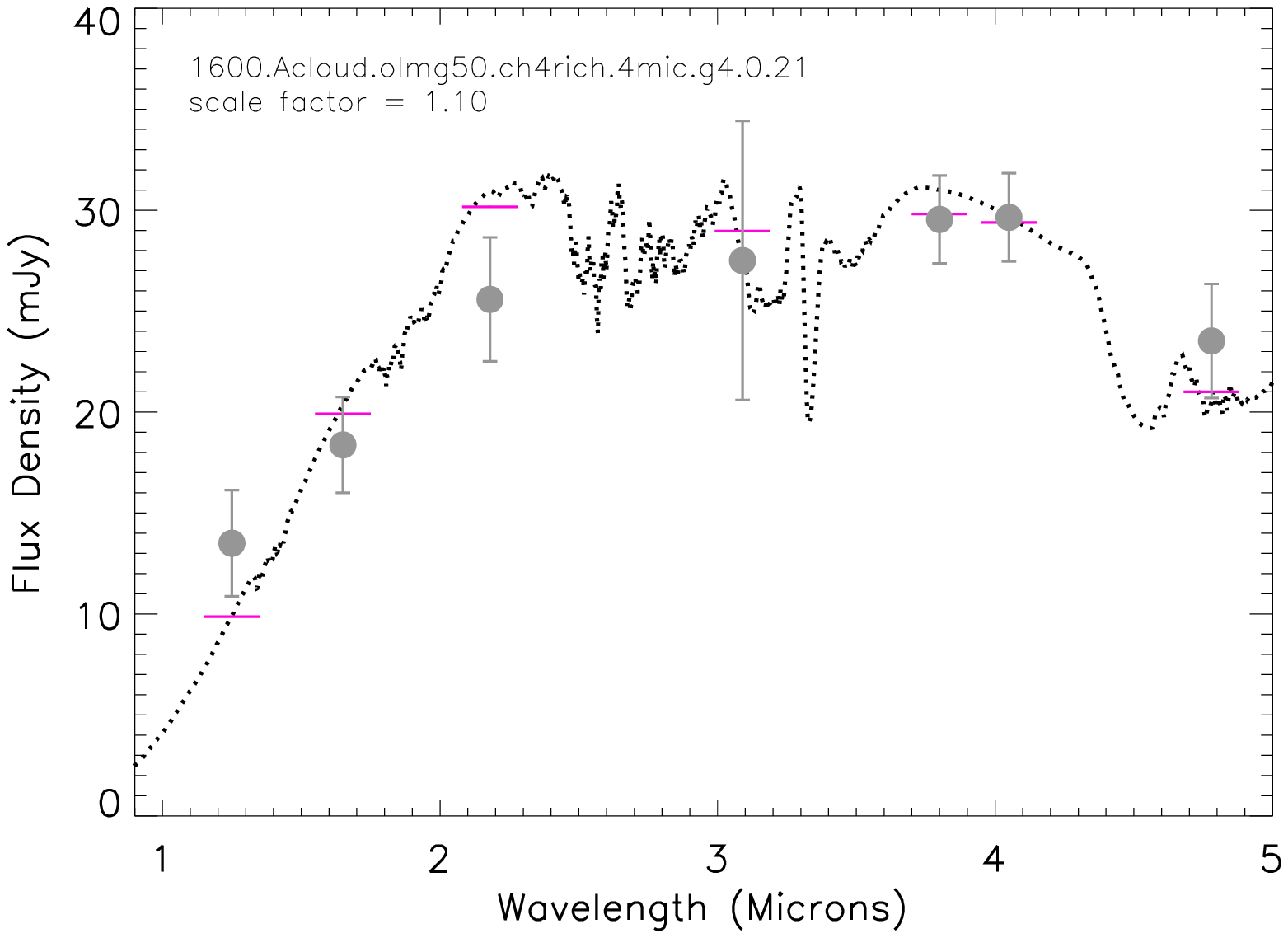}
\includegraphics[scale=0.3,clip]{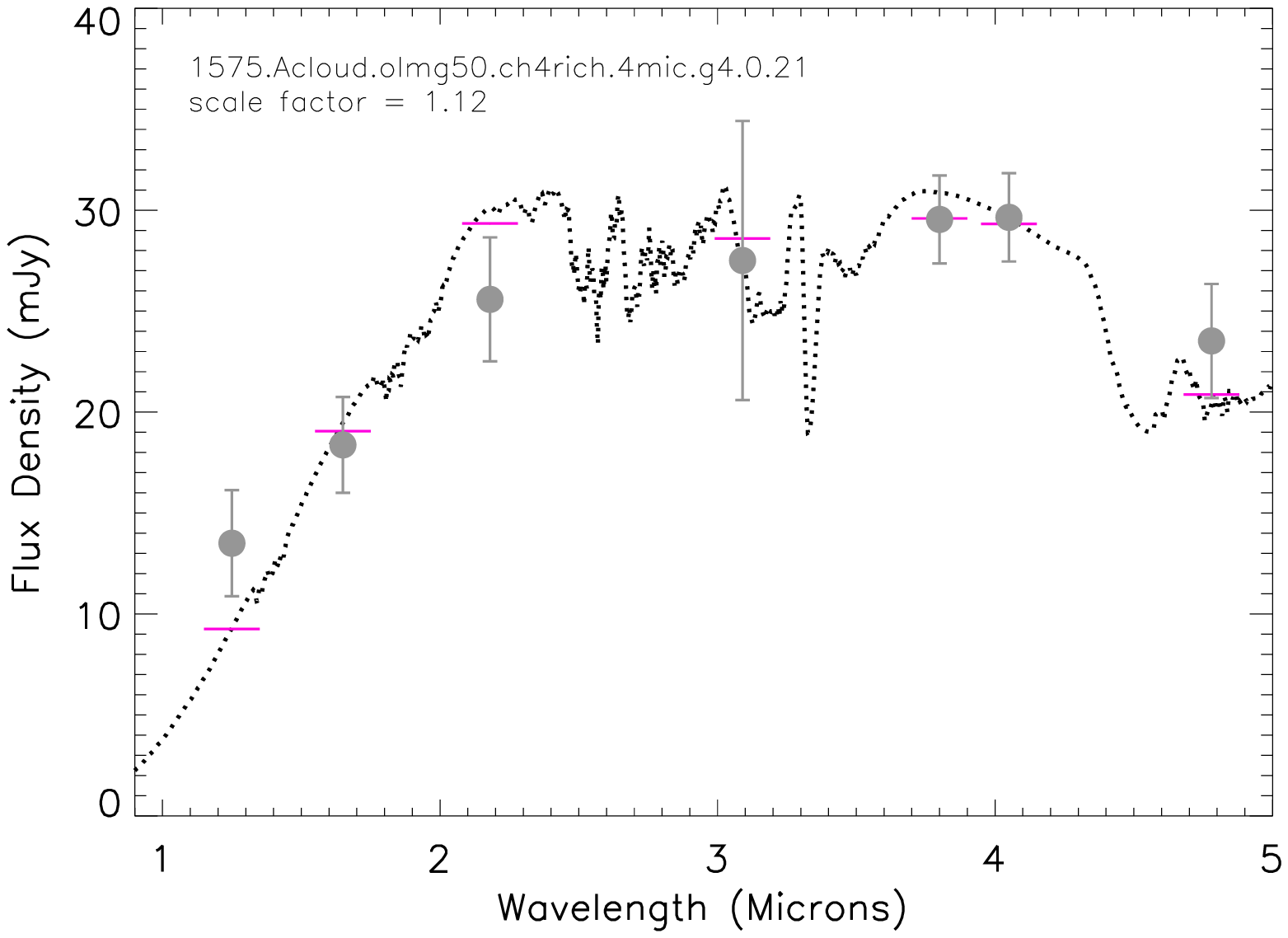}
\includegraphics[scale=0.3,clip]{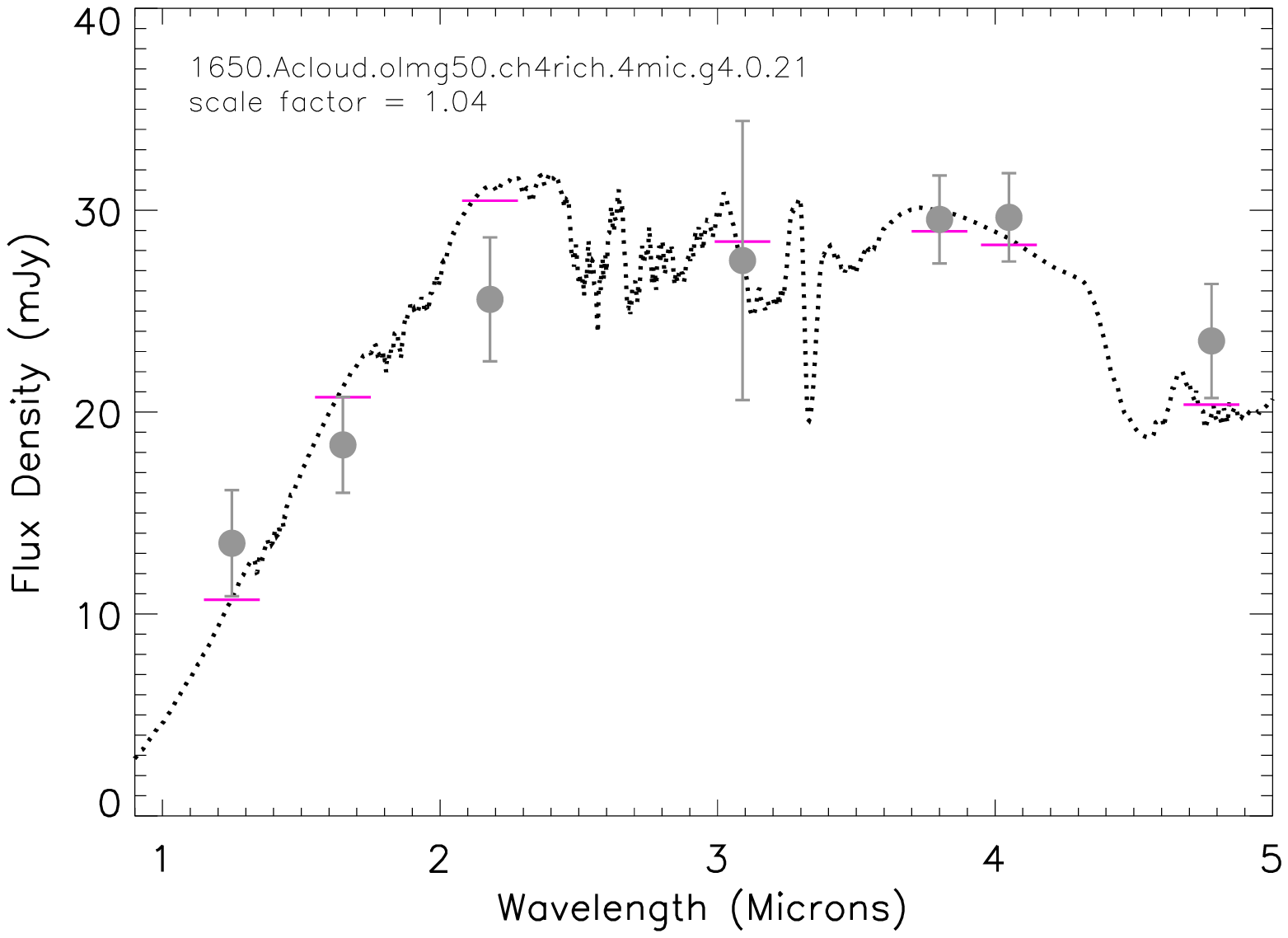}
\caption{SED fits adopting the nominally good-fitting atmosphere parameters depicted in Figure \ref{sedfit3} 
but enhancing the atmosphere of methane.}
\label{sedfit4}
\end{figure}

The methane-enhanced models are shown in Figure \ref{sedfit4} for log(g)=4 and $T_{eff}$ = 1575--1650 $K$.
The 1575 $K$ and 1600 $K$ models (Figure \ref{sedfit4}) likewise produce good fits to the data ($\chi^{2}$ = 5.13--5.3), 
where the 1650 $K$ model barely misses the 68\% cutoff.  
Thus, while best-fitting solar abundance models appear narrowly peaked at $T_{eff}$ = 1600 $K$, the range
in temperature enclosing the 68\% confidence interval is larger when non-solar abundances are considered, 
at least extending from 1575 $K$ to almost 1650 $K$.  
Changes in molecular abundances, as expected, allow us to very slightly improve the SED fit.
However, thick clouds and small dust grains are likely still needed to match the emission 
from $\beta$ Pic b, since given molecules (i.e. $CH_{4}$) by themselves do not change fluxes comparably 
at 1--3 $\mu$m and 3--5 $\mu$m.  

In summary, adopting the \citet{Burrows1997} hot-start models to set our planet radii and the A4 thick cloud/small 
dust atmosphere models, we derive log(g) = 3.8 and $T_{eff}$ = 1600 $K$ for $\beta$ Pic b.  Allowing the radius to 
vary and considering non-solar carbon abundances we derive log(g) = 3.6--4.25 and $T_{eff}$ = 1575--1650 $K$, meaning 
that the planet temperature is well constrained but the surface gravity is not.  However, in Section 5
we narrow the range of surface gravities to log(g) = 3.8 $\pm$ 0.2, as higher surface gravities imply planet masses 
ruled out by dynamical estimates.

\subsubsection{Varying Grain Sizes and Fits Over Other Model Parameter Space} 
The models considered in the previous subsections assume thick clouds, dust grains with 
a modal size of 4 $\mu m$, and (in most cases) solar abundances.  Although we achieve statistically 
significant fits to the $\beta$ Pic b photometry with these models, our exploration of 
model parameter space is still limited.  While an exhaustive parameter space search
is beyond the scope of this paper, here we argue that models either 
thick clouds or small dust grains are unlikely to produce good-fitting models.  
Thus, small grains and thick clouds are likely important components of $\beta$ Pic b's atmosphere 
required in order to fit the planet's spectrum.

To consider the robustness of our results concerning the modal grain size, we also ran some 
model fits for modal particle sizes of 3 $\mu m$, 5 $\mu m$, 10 $\mu m$, and 30 $\mu m$.  
The models with 3 and 5 $\mu m$ modal sizes yielded fits slightly worse than those with 
modal sizes of 4 $\mu m$.  For example,  models with modal sizes of $<a>$ = 3 and 5 $\mu m$, $T_{eff}$ = 1600 K, 
log(g) = 3.8 and a freely-varying planet radius yield $\chi^{2}$ 6.31 and 6.28, respectively.
These values lie slightly outside the 68\% confidence interval, although they are still smaller than 
those from the best-fit DUSTY models.  
In contrast, models with $<a>$ = 10 $\mu m$ and 30 $\mu m$ fit the data significantly worse 
($\chi^{2}$ = 10.0 and 19.6, respectively).

Similarly, our investigations show that small dust grains do not obviate the need to assume 
thick, A-type clouds in our atmosphere models.  For example, adopting the AE-type cloud prescription,
modal particle sizes of 5 $\mu m$, a temperature of $T_{eff}$ = 1600 K, and a surface gravity 
of log(g) = 3.8--4, our model fits are substantiailly worse than the A4-type models and even 
the AMES-DUSTY models and are easily ruled out ($\chi^{2}$ $\sim$ 15--40).
The AE-type cloud prescription fails to reproduce the $\beta$ Pic b spectrum because by 
confining clouds to a thinner layer the $\tau$ = 1 surface varies too much in and out of 
molecular absorption features such as $CH_{4}$ and $CO$.  In disagreement with the 
$\beta$ Pic b SED, the AE model spectra thus have suppressed emission at $\approx$ 3 $\mu m$ 
and 5 $\mu m$ and an overall shape looking less like a blackbody.

In contrast, non-solar abundances may slightly widen the range of parameter space (in radius, temperature, 
gravity, etc.) yielding good fits.  The methane-rich model from the previous section adopting $<a>$ = 5 $\mu m$ 
instead of 4 $\mu m$, log(g) = 3.8, and $T_{eff}$ = 1600 K still yields a fit in agreement with the data 
to within the 68\% confidence limit ($\chi^{2}$ = 5.59).
Thus, within our atmosphere modeling approach we need 1) grains several microns in size, 
comparable to the typical sizes of grains in debris disks, 
and 2) thick clouds to yield fits consistent with the data to 
within the 68\% confidence limit.  These results are not strongly sensitive to chemical abundances 
although varying the range of abundances may slightly widen the corresponding range of other parameter space 
(in temperature, gravity, etc.) yielding good-fitting models.

\section{Planet Radii, Luminosities, Masses, and Evolution}
From the set of models that reproduce the $\beta$ Pic b SED to the 68\% confidence interval, we 
derive a range of planet radii, luminosities and inferred masses.  The planet radii for each model 
run are given in Table \ref{bpicatmosfitres}.  Interestingly, all of our 1-$\sigma$ solutions fall on or 
about $R$ $\sim$ 1.65 $R_{J}$ with very little dispersion ($\pm$ $\sim$ 0.05 dex).  If we consider the 
range of radii for a given atmosphere model consistent with the data to within the 68\% (or 95\%) confidence 
interval regardless of whether the given radius is the best-fit one, then the range in acceptable radii 
marginally broadens: $r$ = 1.65 $\pm$ 0.06.  Note that these 
radii are larger than those inferred for HR 8799 bcde based either on its luminosity and hot-start 
cooling tracks \citep{Marois2008,Marois2011} or from atmosphere modeling, 
where in \citet{Currie2011a} and \citet{Madhusudhan2011} 
our best-fit models typically had $R$ $\sim$ 1.3 $R_{J}$.
The range in inferred planet luminosities is even narrower,  The values inferred from our best-fit 
models center on log($L$/$L_{\odot}$)= -3.80 with negligible intrinsic dispersion ($\pm$ 0.01 dex).
The uncertainty in $\beta$ Pic's distance affects both our radius and luminosity 
determinations.  Treating the distance uncertainty ($\pm$ 1 $pc$) as a separate, additive source of error, 
$\beta$ Pic b's range in radii is 1.65 $\pm$ 0.06 $R_{J}$ and its luminosity 
is log($L$/$L_{\odot}$)= -3.80 $\pm$ 0.02.

From our best-fit surface gravities and inferred radii, we can derive the mass of the planets 
inferred from our modeling.  Adopting the hot-start formalism without rescaling the radius, 
our modeling implies a best-fit planet mass of $\sim$ 7 $M_{J}$; the range covering the 95\% 
confidence limit of 5--9 $M_{J}$.  If we allow the radius to freely vary, we derive a 
range of 4 $M_{J}$ to 18.7 $M_{J}$, where the spread in mass reflects primarily the 
spread in surface gravity from best-fitting models (log(g) = 3.6--4.25).  However, RV data limits $\beta$ Pic b's 
mass to be less than 15 $M_{J}$ if its semimajor axis is less than 10 $AU$, which appears 
to be the case \citep{Lagrange2012a,Chauvin2012,Bonnefoy2013}.  Thus, limiting the atmosphere models 
to those whose implied masses do not in violate the RV upper limits (ones with log(g) = 3.6--4), 
our best-estimated (68\% confidence limit) planet masses are $\sim$ 7$^{+4}_{-3}$ $M_{J}$.

\begin{figure}
\centering
\includegraphics[scale=0.4,clip]{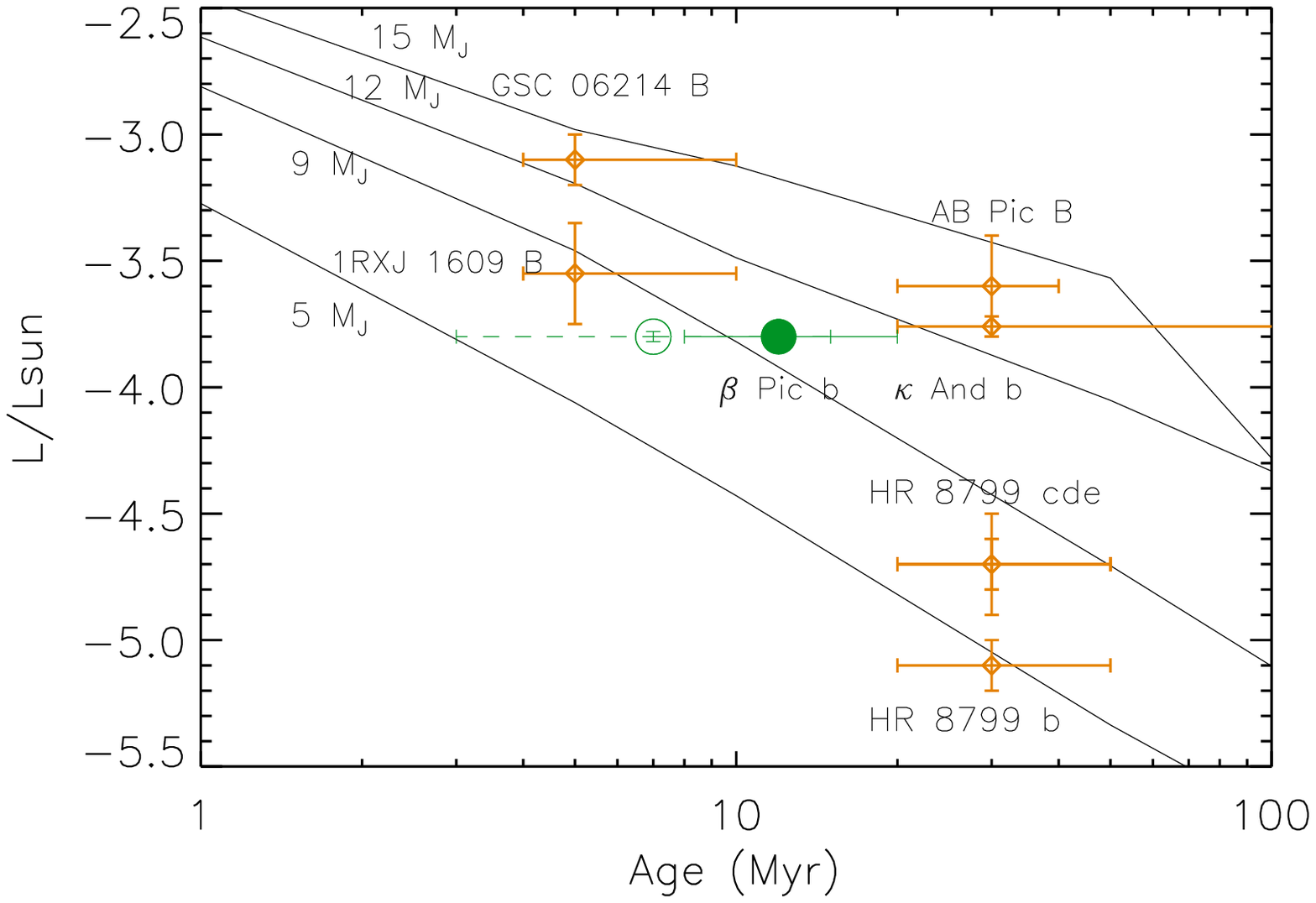}
\includegraphics[scale=0.4,clip]{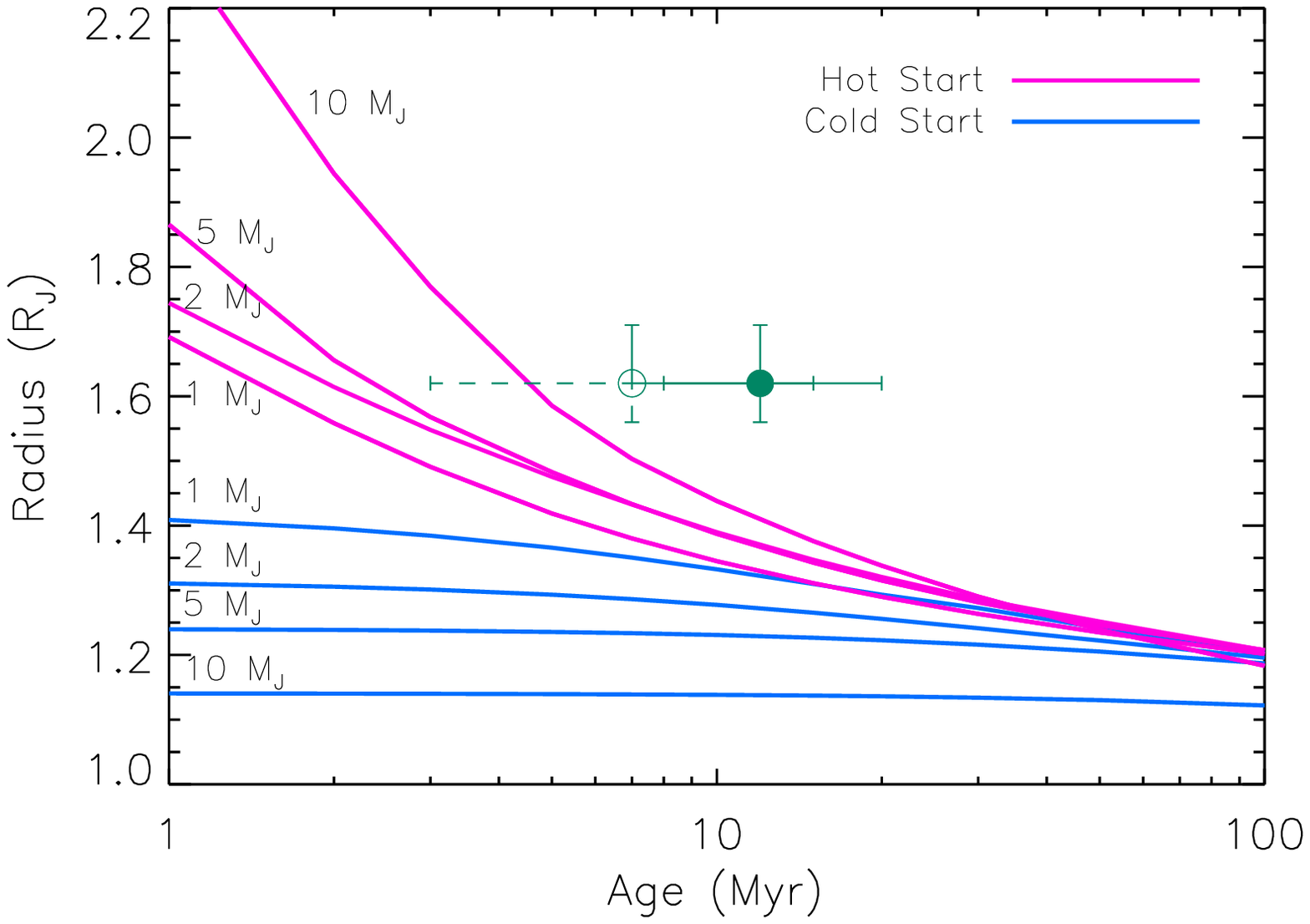}
\caption{(left) Luminosity evolution for hot-start models from \citet{Baraffe2003}, comparing $\beta$ Pic b's 
luminosity as derived in this work to that for the directly-imaged planets HR 8799 bcde, the planet/low-mass brown 
dwarf $\kappa$ And b, and other very low-mass brown dwarf companions.  The solid circle denotes $\beta$ Pic 
b's nominal positions, whereas the open circle identifies its effective position on this plot if it 
formed after 5 Myr.  (Right) Evolution of the radius for planets with masses of 1--10 $M_{J}$ for the 
``hot-start" and ``cold-start" planet cooling models from \citet{Spiegel2012}.  The radius error bars 
define the range of planet radii from models consistent with the data to within the 68\% confidence limit.}
\label{lumevo}
\end{figure}

Planets cool and contract as a function of time, and we can compare our inferred luminosities and 
radii to planet cooling models.  Figure \ref{lumevo} compares the inferred planet luminosity 
to the hot-start planet evolution models from \citet{Baraffe2003}.  For context, we also show 
the luminosities of other 5--100 Myr old companions with masses that (may) lie below 15 $M_{J}$: GSC 06214 B, 
1RXJ 1609 B, HR 8799 bcde, AB Pic B, and $\kappa$ And b.  From our revised luminosity estimate, 
the \citet{Baraffe2003} hot-start models imply a mass range of $\sim$ 8--12 $M_{J}$ if 
the planet's age is the same as the star's inferred age (12$^{+8}_{-4}$ Myr; \citealt{Zuckerman2001}).
If we use the \citet{Burrows1997} hot-start models, we obtain nearly identical results of
9--13 $M_{J}$.  These masses are slightly higher than most of the implied masses from our atmosphere modeling 
but still broadly consistent with them and with the dynamical mass upper limits of 15 $M_{J}$ 
from \citet{Lagrange2012a}.  Note also that the luminosities and planet radii are completely inconsistent 
with predictions from low-entropy, cold-start models for planet evolution.

Still, the righthand panel of Figure \ref{lumevo} highlights one possible complication with our results, namely 
that our best-estimated planet radii are near the upper end of the predicted range for 5--10 $M_{J}$ companions 
in the hot-start formalism.  For the hot-start models presented in \citet{Burrows1997} and \citet{Baraffe2003}, 
5--10 $M_{J}$ companions are predicted to have radii of $\sim$ 1.5--1.6 $R_{J}$.  
For the hot-start models presented in \citet{Spiegel2012}, the predicted range for 5--10 $M_{J}$ planets covers 
$\approx$ 1.4--1.5 $M_{J}$\footnote{This mismatch does not mean that the AMES-DUSTY models, whose fits 
to the data imply planet radii of $\approx$ 1.3 $R_{J}$ and lie just outside the 68\% confidence limit, 
are preferable.  The best-fit AMES-DUSTY radii lie \textit{below} the 
radii predicted for 5--10 $M_{J}$ objects at $\beta$ Pic b's age and are only consistent for `warm-start' models 
that imply lower luminosities and colder temperatures than otherwise inferred from the AMES-DUSTY fits.}.  

To reduce the planet radius of $\sim$ 1.65 $R_{J}$ by $\sim$ 10\% while yielding the same luminosity 
requires raising the effective temperature from $\approx$ 1600 $K$ to $\sim$ 1700 $K$.  This is a small change and 
atmospheric modeling of $\beta$ Pic b and similar substellar objects is still in its early stages.  Thus, it is 
quite plausible that future modeling efforts, leveraging on additional observations of $\beta$ Pic b and those of 
other planets with comparable ages and luminosities, will find quantitatively better fitting solutions that 
imply smaller planet radii and higher temperatures.  We consider this to be the most likely 
explanation.

Alternatively, we can bring the atmosphere modeling-inferred radius into more comfortable agreement with hot-start 
evolutionary models if $\beta$ Pic b is $\approx$ 7 Myr old or less.  For a system age of $\approx$ 12 Myr, 
this is consistent with it forming late in the evolution of the protoplanetary disk that initially surrounded the primary.
Even adopting the lower limit on $\beta$ Pic's age (8 Myr), $\beta$ Pic b may still need to be younger than the star.
While most signatures of protoplanetary disks around 1--2 $M_{\odot}$ stars disappear within 3--5 Myr, 
some $\sim$ 10--20\% of such stars retain their disks through 5 Myr \citep{Currie2009,CurrieSiciliaAguilar2011,Fedele2010}.  
Several 1--2 $M_{\odot}$ members of Sco-Cen and h and $\chi$ Persei apparently have even retained their disks for 
more than 10 Myr \citep{Pecaut2012,Bitner2010,Currie2007c}, 
comparable to or greater than the age of $\beta$ Pic.  Models for even rapid planet formation by core accretion predict that 
several Myrs elapse before the cores are massive enough to undergo runaway gas accretion at $\beta$ Pic-like separations 
\citep{KenyonBromley2009,Bromley2011}.  

In Figure \ref{lumevo} the open circles depict a case where 
$\beta$ Pic b formed after 5 Myr, effectively making the planet 5 Myr younger than the star, where 
the implied masses and radii overlap better with our atmospheric modeling-inferred values.
The overlap is even better for some hot-start models such as COND, which predict larger planet radii at $\approx$ 
5--10 Myr than depicted here.  Note that a young $\beta$ Pic b as depicted in 
Figure \ref{lumevo} with an implied mass mass of $M$ $\ge$ 5 $M_{J}$ is still consistent with a scenario
where the planet produces the warped secondary disk \citep[c.f.][]{Dawson2011}.

\section{Discussion}
\subsection{Summary of Results}
This paper presents and analyzes new/archival VLT/NaCo and Gemini/NICI 1--5 $\mu m$ photometry for $\beta$ Pictoris b, 
  These data allow a detailed comparison between 
$\beta$ Pic b's SED and that of field brown dwarfs and other low-mass substellar objects such as 
directly imaged planets/candidates around HR 8799 and $\kappa$ And.  Using a range of planet atmosphere models, we 
then constrain $\beta$ Pic b's temperature, surface gravity and cloud properties.  Our study yields the following 
primary results.

\begin{itemize}
\item \textbf{1.} - The near-IR ($JHK_{s}$) colors of $\beta$ Pic b appear fairly consistent with the field 
L/T dwarf sequence.  Compared to other young, low-mass substellar objects, $\beta$ Pic b's near-IR colors 
bear the most resemblance to late M to early T dwarfs such as Luhman 16B and $\kappa$ And b.  From its near-IR colors 
and color-magnitude positions, $\beta$ Pic b's near-IR properties most directly resemble those of a L2--L5 dwarf.

\item \textbf{2.} - $\beta$ Pic b's mid-IR properties identify a significant departure from the field L/T dwarf sequence.  
The planet is slightly overluminous at $L^\prime$ and significantly overluminous at $M^\prime$, with deviations from the field L dwarf 
sequence matched only by GSC 06214B and $\kappa$ And b.  The mid-IR portion of $\beta$ Pic b's SED appears more like 
that of a late L dwarf or low surface gravity mid L dwarf.  The broadband $JHK_{s}L^\prime$ photometry for 
$\beta$ Pic b also closely resembles that of $\kappa$ And b.  However, it is unclear whether any object matches $\beta$ Pic 
b's SED at all wavelengths for which we have measurements.  Its 3.1 $\mu m$ brightness and 3.8--5 $\mu m$ spectral 
shape are particularly difficult to match.

\item \textbf{3.} -- Compared to limiting-case atmosphere models E60 (large dust confined to very thin clouds), 
AE60/A60 (large dust confined to moderately-thick/thick clouds) and DUSTY (copious 
small dust everywhere in the atmosphere), $\beta$ Pic b appears to have evidence for thick clouds consistent 
with a high $T_{eff}$ and low surface gravity.  We fail to find any E60/AE60/A60 model providing statistically significant fits 
over a surface gravity range of log(g) = 4--4.5 and any $T_{eff}$.  The DUSTY models come much closer to yielding 
statistically significant fits but mismatch the planet 
flux at $J$, $K_{s}$, [3.1], and $M^\prime$.  From these fiducial comparisons, we infer that $\beta$ Pic b's atmosphere 
shows evidence for clouds much thicker than those assumed in the E60 models but is slightly less dusty 
than the DUSTY models imply.

\item \textbf{4.} -- Using thick cloud models with particle sizes slightly larger than those found in the 
interstellar medium ($<a>$ = 4 $\mu m$), we can match $\beta$ Pic b's SED in both the near and mid IR.  
Assuming planet radii appropriate for the \citet{Burrows1997} `hot-start' models, we derive 
log(g) = 3.80 and $T_{eff}$ = 1600 $K$ for $\beta$ Pic b.  Allowing the 
radius to freely vary, leaves the surface gravity essentially 
unconstrained, where models consistent with the data at the 68\% confidence limit include 
log(g) = 3.6--4.25 and $T_{eff}$ = 1600 $K$.  Considering departures from solar abundances 
and eliminating models that imply masses ruled out by dynamical estimates, 
the acceptably fitting range of atmosphere parameters cover log(g) = 3.6--4 and 
$T_{eff}$ = 1575--1650 $K$.   

\item \textbf{5.} -- Using our best-fit atmosphere models and eliminating models inconsistent 
with $\beta$ Pic b's dynamical mass upper limit, within the hot-start formalism 
we derive a mass of 7 $M_{J}$ for a fixed radius and 7$^{+4}_{-3}$ $M_{J}$ for a scaled radius.  
Our best-fit planet radius is $\sim$ 1.65 $\pm$ 0.06 $R_{J}$ and luminosity of log(L/L$_{\odot}$) = -3.80 $\pm$ 0.02.

\item \textbf{6.} -- While our derived luminosity and radius for $\beta$ Pic b rules out cold start models, 
the radius is near the upper end of predicted radii for hot start-formed planets with $\beta$ Pic's age.  
As the planet only needs to be $\sim$ 100 $K$ hotter to easily eliminate this discrepancy, it 
likely identifies a limitation of the atmosphere models.  Alternatively, if $\beta$ Pic b has a significantly 
younger age than the star's age consistent with it forming late in the protoplanetary disk stage
 our derived radius is comfortably within the range predicted by hot start models.

\end{itemize}
\subsection{Comparisons to Other Recent $\beta$ Pictoris b Studies}
\subsubsection{Currie et al. 2011b}
In our first-look analysis of the atmosphere of $\beta$ Pictoris b \citep{Currie2011b}, 
we compared its $K_{s}$, $L^\prime$, and [4.05] photometry to an array of atmosphere models, from
atmospheres completely lacking clouds to those with the Model A-type thick clouds that extend to the 
visible surface of the atmosphere.  In that paper, we found that the AE thick cloud models from \citet{Madhusudhan2011} 
yielded the smallest $\chi^{2}$ value.  The fits degraded at about the same level for 
the Model A thick cloud and Model E ``normal" L dwarf atmosphere prescriptions, while the cloudless case fared the worst.
\citet{Currie2011b} conclude that while the AE thick cloud model quantitatively produced the best fit, the existing data 
were too poor to say whether the clouds in $\beta$ Pictoris b were any different in physical extent, in 
mean dust particle size, etc. from those for field L dwarfs with the same range of temperatures.

Our present study greatly improves upon the analyses in \citet{Currie2011b}.  First, our photometry covers seven 
passbands, not three, at 1.25--4.8 $\mu m$, not 2.18--4.05 $\mu m$.  This expanded coverage allows far firmer 
constraints on $\beta$ Pic b's atmospheric properties.  In particular, our new photometry strongly favors 
the Model A thick-cloud prescription over AE, largely due to the relatively low planet flux densities at 1.25--1.65 $\mu m$ 
and the relatively high flux densities at 3.1 $\mu m$ and $M^\prime$/4.8 $\mu m$, trends that the Model A cases consistently 
reproduce better.  While all models considered in \citet{Currie2011b} assumed a modal particle size of 60 $\mu m$ for dust 
entrained in clouds, our fits improve if we use smaller sizes.  The combined effect of thicker clouds and smaller particle sizes 
favor atmosphere models with a slightly higher surface gravity and temperature than the best-fit model in \citet{Currie2011b}.
Our new data more clearly demonstrate the failure of the E models successful in fitting most of the field L dwarf 
sequence and thus better distinguish $\beta$ Pic b's atmosphere from that of a typical cloudy field L dwarf.

\subsubsection{Bonnefoy et al. (2013)}
\citet{Bonnefoy2013} presented new photometry for $\beta$ Pictoris b in the $J$, $H$, and $M^\prime$ filters from 
data taken in 2011 and 2012.  The $J$ and $H$ detections are firsts and greatly expand the wavelength coverage 
for $\beta$ Pic b's SED.  Their $M^\prime$ detection is first \textit{well-calibrated} detection, building upon and 
following the detection presented in \citet{Currie2011b}, which lacked a contemporaneous flux-calibration data to provide 
precise photometry.
They then combined these measurements with their previously published $K_{s}$ and $L^\prime$ photometry and [4.05] from \citet{Quanz2010}.  


In general, our study clarifies and modifies, instead of contradicting, 
the picture of $\beta$ Pictoris b constructed in \citet{Bonnefoy2013}.  
On the same datasets, the SNR of our detections is slightly higher 
but our photometry agrees within theirs derived from their CADI, RADI, and LOCI reductions
within their adopted photometric uncertainties ($\sim$ 0.2--0.3 mag).  
We derive smaller photometric uncertainties, owing to a more uniform 
throughput as a function of azimuthal angle, probably due to our pixel masking technique 
and SVD cutoff in A-LOCI \citep[see also][]{Marois2010b}.  
We concur that the planet's mid-IR colors are unusually red and highlight a potentially strong, 
new disagreement with field L dwarfs at 3.1 $\mu m$.

We agree with \citeauthor{Bonnefoy2013}'s general result that the best-fitting atmosphere models 
are those intermediate between the AMES-DUSTY models (submicron-sized dust everywhere) and 
the COND or BT-Settl models (no dust/clouds or very thin clouds).  Quantitatively, the $\chi^{2}$ 
values we derive are much larger than the best-fitting models in \citeauthor{Bonnefoy2013} because 
our photometric uncertainties are significantly smaller (e.g. 7 vs. 3 for AMES-DUSTY).
Our analyses point to thick clouds and particle sizes small compared to the range typically 
used in the \citet{Burrows2006} models but larger than the ISM-like grains in the AMES-DUSTY 
models.  The temperatures, surface gravities, and luminosities they derive are generally consistent 
with our best-fit values.  

 While they derive a lower limit to the initial entropy of 
9.3 $kB$/baryon, we do not provide a detailed similar analysis since the inferred entropy range
depends on the planet radius which, considering our studies together, is very model dependent.  Similarly, 
it depends on the planet mass (for which there still is some range) and the planet's age (which is very poorly constrained).
Still, we agree that cold start models are ruled out for $\beta$ Pic b as they fail to reproduce the inferred 
luminosity and radii of the planet determined from both our studies.

\subsection{Future Work to Constrain $\beta$ Pic b's Properties}
Deriving $\beta$ Pic b's mass and other properties is difficult 
since they are based on highly uncertain parameters such as the planet's age and its entropy at formation.
However, dynamical mass limits can be derived from continued radial-velocity measurements \citep{Lagrange2012a}.  
As these limits depend on $\beta$ Pic b's orbital parameters, future planet astrometry may be particularly 
important in constraining $\beta$ Pic b's mass.  If $\beta$ Pic b is responsible for the warp observed in 
the secondary debris disk \citep{Golimowski2006}, planet-disk interaction modeling can likewise yield 
a dynamical mass estimate \citep{Lagrange2009,Dawson2011} provided the planet's orbit is known.

Finally, while our models nominally assume solar abundances, we showed that changing the methane abundance 
might yield marginally better fits to the data.  Near-infrared 
spectroscopic observations of $\beta$ Pic b as can be done soon with $GPI$ and $SPHERE$ may clarify 
its atmospheric chemistry.  Future observations with GMTNIRS on the \textit{Giant Magellan Telescope} should 
be capable of resolving molecular lines in $\beta$ Pic b's atmosphere \citep{Jaffe2006}, providing a more detailed look 
at its chemistry, perhaps even constraining its carbon to oxygen ratio and formation history \citep[e.g][]{Oberg2011,Konopacky2013}.
\acknowledgements 
We thank Christian Thalmann, France Allard, and the anonymous
referee for helpful comments and discussions and
Michael Cushing for providing IRTF/SPeX and Subaru/IRCS spectra of field L dwarfs.
We are grateful to the 
telescope staffs at ESO Paranal Observatory and Gemini-South Cerro Pachon Observatory for support 
for our observations, all of which were obtained with ``delegated visitor mode" or ``eavesdropping mode".
Finally, we thank Christian Marois for very detailed discussions on image processing techniques and 
extensive helpful suggestions that improved this manuscript. 
T. C. acknowledges support from a McLean Postdoctoral Fellowship.
R. D. acknowledges NSF-GRFP grant DGE-1144152.
{}
\input{tab_obs.tex}
\input{tab_photbpic.tex}
\input{tab_photerror.tex}
\input{tab_photcomp.tex}
\input{tab_photcomp2.tex}
\input{tab_fluxzero.tex}
\input{tab_atmosfit.tex}
\input{tab_atmosfitres.tex}

\vspace{-2in}
\vspace{-2in}
\clearpage

\end{document}

%% file: tab_obs.tex
\begin{deluxetable}{lcllccccccc}
\setlength{\tabcolsep}{0pt}
\tablecolumns{8}
\tablecaption{Observing Log}
\tiny
\tablehead{{UT Date}&{Telescope/Instrument}&{Mode}&{Pixel Scale}&{Filter}&{$t_{int}$}&{$N_{images}$}&{$\Delta$PA} &\\
{} & {} &{} &{(mas pixel$^{-1}$)} &{}&{(s)}&{}&{(degrees)}}
\startdata
2011-12-16 & {VLT/NaCo} & Direct& 13.22&$J$ & 50 & 48 & 26\\
2012-01-11 & {VLT/NaCo} & Direct& 13.22&$H$ & 40 & 92 & 34.6\\
\\
2012-12-15 & {VLT/NaCo} & Direct& 27.1&{$M^\prime$}&{20}&{176}&{70.2}\\
2012-12-16 & {VLT/NaCo} & Direct& 27.1&{$L^\prime$}&{30}&{112}&{67.7}\\
2012-12-16 & {VLT/NaCo} & Direct& 27.1&{[4.05]}&{30}&{64}&{57.4}\\
2012-12-23 & {Gemini/NICI} & Direct & 17.97&{[3.09]}&{38}&{60}&{30.2}\\
2012-12-26 & {Gemini/NICI} & Direct & 17.97&{[3.09]}&{38}&{60}&{31.7}\\
2013-01-09 & {Gemini/NICI} & {0\farcs{}22 mask}& 17.97/17.94&{$H$/$K_{s}$}&{11.4}&{117}&{41.1}
 \enddata
\label{bpiclog}
\end{deluxetable}

%% file: tab_photbpic.tex
\begin{deluxetable}{llcllccccccc}
\setlength{\tabcolsep}{0pt}
\tablecolumns{8}
\tablecaption{$\beta$ Pictoris b Detections and Photometry}
\tiny
\tablehead{{UT Date}&{Telescope/Instrument }&{ Filter }&{ Wavelength ($\mu m$)}&{ SNR }&{ Apparent Magnitude } &{Absolute Magnitude}}
\startdata
2011-12-16&VLT/NaCo & J & 1.25 & 9.2 & 14.11 $\pm$ 0.21 & 12.68 $\pm$ 0.21\\
2012-01-01&VLT/NaCo & H & 1.65 & 30 & 13.32 $\pm$ 0.14 & 11.89 $\pm$ 0.14\\
\\
2013-01-09&Gemini/NICI & H & 1.65 & 6.4 & 13.25 $\pm$ 0.18 & 11.82 $\pm$ 0.18\\
2013-01-09&Gemini/NICI & K$_{s}$ & 2.16 & 10 & 12.47 $\pm$ 0.13 & 11.04 $\pm$ 0.13\\
2012-12-23&Gemini/NICI& [3.09] & 3.09 & 4.6 & 11.71 $\pm$ 0.27 & 10.28 $\pm$ 0.27\\
2012-12-26&Gemini/NICI& [3.09] & 3.09 & 11 & -- & --\\
\\
2012-12-16&VLT/NaCo &  L$^\prime$ & 3.8 & 40 & 11.24 $\pm$ 0.08 & 9.81 $\pm$ 0.08\\
2012-12-16&VLT/NaCo &  [4.05] & 4.05& 20 & 11.04 $\pm$ 0.08 & 9.61 $\pm$ 0.08\\
2012-12-15&VLT/NaCo & M$^\prime$ & 4.78& 22 & 10.96 $\pm$ 0.13 & 9.54 $\pm$ 0.13\\
 \enddata
\label{bpicphot}
\end{deluxetable}

%% file: tab_photerror.tex
\begin{deluxetable}{llcllccccccc}
\tablecolumns{8}
\tablecaption{Sample Photometric NaCo and NICI Error Budgets} 
\tiny
\tabletypesize{\scriptsize}
\tablehead{{Telescope/Instrument}&{Filter}&{Apparent Magnitude}&{$\sigma_{det}$}&{$\sigma_{atten}$}&{$\sigma_{fluxcal}$}}
\startdata
Gemini/NICI & $K_{s}$ & 12.47 $\pm$ 0.13&0.11 & 0.06 & 0.03\\
Gemini/NICI & [3.09] & 11.71 $\pm$ 0.27&0.24 & 0.11 & 0.08\\
VLT/NaCo & $L^\prime$& 11.24 $\pm$ 0.08&0.03 & 0.07 & 0.03\\
VLT/NaCo & $M^\prime$ & 10.96 $\pm$ 0.13&0.05 & 0.11 & 0.06\\
 \enddata
\tablecomments{The photometric uncertainty from the intrinsic SNR ($\sigma_{det}$) scales as 1.0857/SNR.}
\label{bpicphoterror}
\end{deluxetable}

%% file: tab_photcomp.tex
\begin{deluxetable}{llllllllllllll}
 \tiny
\tabletypesize{\small}
\tablecolumns{11}
\tablecaption{Young Directly Imaged Planets and Very Low-Mass Brown Dwarfs Used for Comparison}
\tiny
\tablehead{{Companion}&{D (pc)}&{Assoc.}&{Age}&{ST(Primary)}&{ST(Companion)}&{Sep.}&{Mass}&{References}}
\startdata
\textit{Planets/Planet Candidates}\\
HR 8799 b & 39.4 $\pm$ 1&Columba& 30& A5&??&67.5--70.8&4--5&1,2,3,4,5\\
HR 8799 c & 39.4 $\pm$ 1&Columba& 30& A5&L/T?&42.1--44.4&$\sim$ 7&1,2,3,4,5\\
HR 8799 d & 39.4 $\pm$ 1&Columba& 30& A5&L/T?&26.4--28.1&$\sim$ 7&1,2,3,4,5\\
HR 8799 e & 39.4 $\pm$ 1&Columba& 30& A5&L/T?&$\sim$15&$\sim$ 7&2,4,6\\
$\kappa$ And b & 51.6 $\pm$ 0.5 & Columba & 30 & B9IV & L2--L8?&55$\pm$ 2&11.8-14.8&7\\
\\
\textit{Low Mass}\\
\textit{Brown Dwarfs}\\
1RXJ 1609 B & 145 $\pm$ 14 & US & 5--10 & K7 & L4 $\pm$ 2 & 330 & 6--12 & 8,9\\
GSC 06214 B & 145 $\pm$ 14 & US & 5--10 & K7 & L0 $\pm$ 1 & 320 $\pm$ 30 & 14 $\pm$ 2 & 10,11\\
USco CTIO 108 B& 145 $\pm$ 14 & US & 5--10 & M7 & M9.5 & 670 $\pm$ 64 & 6-16 & 12\\
HIP 78530 B & 156.7 $\pm$ 13.0 & US & 5--10 & B9V & M8 $\pm$ 1 & 710 $\pm$ 60 & 22 $\pm$ 4&11,13\\
2M 1207 B & 52.4 $\pm$ 1.1 & TWA& $\sim$8& M8 & ??&40.8 $\pm$ 9 & 8$\pm$ 2&14\\
2M 1207 A & 52.5 $\pm$ 1.1 & TWA& $\sim$8& M8 & M8& 40.8 $\pm$ 9 & 24$\pm$ 6&14\\
TWA 5B & 44.4 $\pm$ 4 & TWA & $\sim$ 8 & M2Ve &M8--M8.5&$\sim$ 98&$\sim$ 20&15\\
HR 7329 B & 47.7 $\pm$ 1.5 & $\beta$ Pic & 12 & A0 & M7.5 & 200 $\pm$ 7 & 26 $\pm$ 4&16\\
PZ Tel B & 51.5 $\pm$ 2.6 & $\beta$ Pic & 12 & K0 & M7 $\pm$ 2 & 17.9 $\pm$ 0.9 & 36 $\pm$ 6&17\\
2M0103AB B & 47.2 $\pm$ 3.1 & Tuc-Hor? & 30 & M5/M6 & L? & 84 & 12--14 & 18\\
AB Pic B & 45.5 $\pm$ 1.8 & Carina & 30 & K1Ve & L0$\pm$1& 248$\pm$ 10& 13--14 & 19, 20\\
Luhman 16 B&2.02 $\pm$ 0.15& Argus? &40?&L7.5 & T0.5&3.12 $\pm$ 0.25&40--65&21,22\\
Luhman 16 A&2.02 $\pm$ 0.15& Argus?&40?&L7.5 & L7.5 & 3.12 $\pm$ 0.25&40--65&21,22\\
CD-35 2722 B & 21.3 $\pm$ 1.4 & AB Dor & $\sim$ 100 & M1Ve & L4$\pm$1 & 67.4$\pm$4 & 31$\pm$8 &23\\
\enddata
\tablecomments{References: 1) \citet{Marois2008}, 2) \citet{Currie2011a}, 3) \citet{Galicher2011}, 4) \citet{Skemer2012}, 
5) \citet{Currie2012b}, 6) \citet{Marois2011}, 7) \citet{Carson2013}, 8) \citet{Lafreniere2008a}, 9) \citet{Lafreniere2010}, 
10) \citet{IrelandKraus2011}, 11) \citet{Bailey2013}, 12) \citet{Bejar2008}, 13) \citet{Lafreniere2010}, 14) \citet{Chauvin2004}, 
15) \citet{Lowrance1999}, 16) \citet{Lowrance2000}, 17) \citet{Biller2010}, 18) \citet{Delorme2013}, 19) \citet{Chauvin2005}, 
20) \citet{Bonnefoy2010}, 21) \citet{Luhman2013}, 22) \citet{Burgasser2013}, 23) \citet{Wahhaj2011}}
\label{photcomptable}
\end{deluxetable}

%% file: tab_photcomp2.tex
\begin{deluxetable}{llllllllllllll}
 \tiny
\tabletypesize{\small}
\tablecolumns{11}
\tablecaption{Photometry for Young Directly Imaged Planets and Very Low-Mass Brown Dwarfs}
\tiny
\tablehead{{Companion}&{J}&{H}&{K$_{s}$}&{[3.09]}&{$L^\prime$}&{$M^\prime$}& {$\chi^{2}_{\beta Pic b}$}&{C.L.}}
\startdata
\textit{Planets/Candidates}\\
HR 8799 b &16.52$\pm$0.14&15.08 $\pm$ 0.17 & 14.05 $\pm$ 0.08 & --&12.68 $\pm$ 0.12 & 13.07 $\pm$ 0.30&52.8 & $\sim$ 1\\
HR 8799 c &14.65$\pm$0.17&14.18 $\pm$ 0.17 & 13.13 $\pm$ 0.08 & --&11.83 $\pm$ 0.07 & 12.05 $\pm$ 0.14&6.098&0.893\\
HR 8799 d &15.26$\pm$0.43&14.23 $\pm$ 0.22 & 13.11 $\pm$ 0.08 & --&11.50 $\pm$ 0.12 & 11.67 $\pm$ 0.35&8.351&0.961\\
HR 8799 e &-- & 13.88 $\pm$ 0.20 & 12.89 $\pm$ 0.26 & --&11.61 $\pm$ 0.12& $>$ 10.09&--&--\\
$\kappa$ And b & 12.7 $\pm$ 0.30 & 11.7 $\pm$ 0.20 & 11.0 $\pm$ 0.4 & --&9.54 $\pm$ 0.09 & --&0.946&0.186\\
\\
\textit{Low-Mass}\\
\textit{Brown Dwarfs}\\
1RXJ 1609 B & 12.09 $\pm$ 0.12 & 11.06 $\pm$ 0.07 & 10.38 $\pm$ 0.05 & 9.84 $\pm$ 0.21 & 8.99 $\pm$ 0.30&--&1.369 & 0.287\\
GSC 06214 B & 10.43 $\pm$ 0.04 & 9.74 $\pm$ 0.04 & 9.14 $\pm$ 0.05 & 8.60 $\pm$ 0.08 & 7.94 $\pm$ 0.07 & 7.94 $\pm$ 0.30&7.001&0.864\\
USco CTIO 108 B & 10.72 $\pm$ 0.09 & 9.94 $\pm$ 0.08 & 9.30 $\pm$ 0.11 & -- & -- & --&--&--\\
HIP 78530 B & 9.25 $\pm$ 0.05 & 8.58 $\pm$ 0.04 & 8.36 $\pm$ 0.04 & -- & 7.99 $\pm$ 0.06 & --&88.086&$\sim$ 1\\ 
2M 1207 B & 16.40 $\pm$ 0.2 & 14.49 $\pm$ 0.21 &13.31 $\pm$ 0.11&-- & 11.68 $\pm$ 0.14&--&20.601 & $\sim$ 1\\
2M 1207 A & 9.35 $\pm$ 0.03 & 8.74 $\pm$ 0.03 & 8.30 $\pm$ 0.03 & -- & 7.73 $\pm$ 0.10 & --&48.044 & $\sim$ 1\\
TWA 5B & 9.1 $\pm$ 0.2 & 8.65 $\pm$ 0.06 & 7.91 $\pm$ 0.2 & -- & -- & --\\
HR 7329 B & 8.64 $\pm$ 0.19 & 8.33 $\pm$ 0.1 & 8.18 $\pm$ 0.1 & -- & 7.69 $\pm$ 0.1 & --&59.455 & $\sim$ 1\\
PZ Tel B &  8.70$\pm$ 0.18 & 8.31 $\pm$ 0.15 & 7.86 $\pm$ 0.19 & -- & -- & --&--&--\\
2M0103AB B & 12.1 $\pm$ 0.3 & 10.9 $\pm$ 0.2 & 10.3 $\pm$ 0.2 & -- & 9.3 $\pm$ 0.1&--&2.666&0.736\\
AB Pic B & 12.80 $\pm$ 0.10 & 11.31 $\pm$ 0.10 & 10.76 $\pm$ 0.08 & -- & 9.9 $\pm$ 0.1 & --&11.231&0.996\\ 
Luhman 16 B& 14.69 $\pm$ 0.04 & 13.86 $\pm$ 0.04 & 13.20 $\pm$ 0.09& -- & -- & --&--\\
Luhman 16 A& 15.00 $\pm$ 0.04 & 13.84 $\pm$ 0.04 & 12.91 $\pm$ 0.04&--&--&--&--\\ 
CD-35 2722 B & 11.99 $\pm$ 0.18 & 11.14 $\pm$ 0.19 & 10.37 $\pm$ 0.16 & -- & --& --&--\\ 
\enddata
\tablecomments{We only quantitatively compare the photometry between $\beta$ Pic b and those objects with full $JHK_{s}L^\prime$ photometry.}
\label{photcomptable2}
\end{deluxetable}

%% file: tab_fluxzero.tex
\begin{deluxetable}{llllccccccc}
\setlength{\tabcolsep}{0pt}
\tablecolumns{3}
\tablecaption{Adopted Flux Density Zero-Points}
\tiny
\tablehead{{Filter}&{$\lambda_{o}$ ($\mu m$)}&{F$_{\nu,o}$ (Jy)}}
\startdata
J & 1.25 & 1594\\
H & 1.65 & 1024\\
K$_{s}$ & 2.15 & 666.20\\
$[3.09]$ & 3.09 & 356\\
L$^\prime$&3.78 & 248\\
$[4.05]$ & 4.05 & 207\\
M$^\prime$&4.78 & 154
\enddata
\label{fluxzero}
\end{deluxetable}

%% file: tab_atmosfit.tex
\begin{deluxetable}{lcllccccccc}
\setlength{\tabcolsep}{0pt}
\tablecolumns{7}
\tablecaption{$\beta$ Pictoris b Atmosphere Modeling Grid}
\tiny
\tablehead{{\textbf{Model}}&{}&{\textbf{Range}}&{}&{References}&{}\\
{}&{$T_{eff}$}&{log(g)}&{$R_{p}$ ($R_{J}$)} & {}}
\startdata
\textit{Limiting Cases}\\
E60& 1000-1800 & 4--4.5 & 0.9--2 & 1\\
AE60 & 1000-1700 & 4--4.5 & 0.9--2 & 2\\
A60& 1000-1700 & 4--4.5 & 0.9--2 & 1,3\\
AMES-DUSTY  &1000-2000& 3.5--4.5&0.9--2& 4\\
\\
\textit{New Models}\\
A4 & 1400-1900 & 3.6--4.25 & 0.9--2$^{a}$ & 5\\
 \enddata
\tablecomments{a) In our modeling, we perform two runs for the A4 models: one where we fix the radius to values adopted in the 
\citet{Burrows1997} hot-start models and one where we allow the radius to freely vary between 
0.9 $R_{J}$ and 2 $R_{J}$.  References: 1) \citet{Burrows2006}, 2) \citet{Madhusudhan2011}, 3) \citet{Currie2011a}, 4) \citet{Allard2001}, and 5) this work.}
\label{bpicatmosfit}
\end{deluxetable}

%% file: tab_atmosfitres.tex
\begin{deluxetable}{llcccc}
 \tiny
\tabletypesize{\tiny}
\tablecolumns{11}
\tablecaption{Model Fitting Results}
\tablehead{{Model}&{$\chi^{2}_{min}$}&{log(g), T$_{eff}$, R(R$_{J}$) (for $\chi^{2}_{min}$)}&
{log(g), T$_{eff}$, R(R$_{J}$) (68\%)} & {log(g), T$_{eff}$, R(R$_{J}$) (95\%)}}
\startdata
E60 & 53.80 & 4.0, 1400 $K$, 1.79 & -- & --\\
AE60 & 37.70 & 4.5, 1400 $K$, 1.96&--&--\\
A60 & 19.99 & 4.5, 1400 $K$, 2.05&--&--\\
AMES-DUSTY & 7.14& 3.5, 1700 $K$, 1.35 & -- & 3.5--4, 1700 $K$, 1.32--1.35 \\
\\\\
A4 (fixed radius) & 5.85 & 3.8, 1600 K, 1.65 & 3.8, 1600 K, 1.65 & 3.6, 1500--1550 K, 1.79--1.80\\
                                   &   &  &              & 3.8, 1550--1625 K, 1.65\\
                                   &   &  &             & 4.0, 1650--1700 K, 1.54\\
\\
A4 (scale)     & 5.82 & 4--4.25, 1600 K, 1.65 & 3.6--4.25, 1600 K, 1.64--1.66 & 3.6--4.25, 1500--1750 K, 1.44--1.82\\
\\
A4 (1\% $CH_{4}$, scale) & 5.13 & 4, 1575 K, 1.71 & 4, 1575--1650 K, 1.59--1.71 & 4, 1575--1650 K, 1.59--1.71\\ 
 \enddata
\tablecomments{The $\chi^{2}$ values quoted here do \textit{not} refer to $\chi^{2}$ per degree of freedom.  
The columns for A4 (fixed radius) do not have a range in radius in some columns because only 
one model (with a fixed radius) identifies the $\chi^{2}$ minimum}.
\label{bpicatmosfitres}
\end{deluxetable}